\documentclass[11pt,a4paper,twoside,headings=small,numbers=noenddot,abstracton,fleqn,titlepage,openright]{scrreprt}

\usepackage [english]{babel}
\usepackage[utf8]{inputenc}
\usepackage{pifont,fancyhdr,gensymb,geometry,amsfonts,amssymb,amsmath,amsopn,amsthm,caption,footnote,colortbl,color,lscape,pdflscape,pdfpages}
\usepackage{subfigure}
\usepackage{setspace} %

\usepackage{rotating}

\usepackage[acronym,nonumberlist]{glossaries}
\newacronym{adrl}{ADRL}{Agile & Dexterous Robotics Lab}

\newcommand{\sref}[1]{Section~\ref{#1}}
\newcommand{\fref}[1]{Fig.~\ref{#1}}
\newcommand{\eref}[1]{Eq.~(\ref{#1})}

\newglossaryentry{x}{
  name = $x$ ,
  description = Position of robot,
}

\newcommand{\argmin}{\mathop{\mathrm{argmin}}}
\newcommand{\argmax}{\mathop{\mathrm{argmax}}}

\newcommand{\vep}{\mbox{\boldmath $\epsilon$}}

\newcommand{\vth}{\mbox{\boldmath $\theta$}}

\newcommand{\vmu}{\mbox{\boldmath $\mu$}}

\newcommand{\vvep}{\mbox{\boldmath $\varepsilon$}}

\newcommand{\vTh}{\mathbf \Theta}

\newcommand{\vUpsilon}{\mathbf \Upsilon}

\newcommand{\vf}{\mathbf f}
\newcommand{\vg}{\mathbf g}

\newcommand{\vq}{\mathbf q}
\newcommand{\vr}{\mathbf r}
\newcommand{\vs}{\mathbf s}

\newcommand{\vu}{\mathbf u}

\newcommand{\vw}{\mathbf w}
\newcommand{\vx}{\mathbf x}
\newcommand{\vy}{\mathbf y}
\newcommand{\vz}{\mathbf z}

\newcommand{\vdx}{\delta \mathbf x}
\newcommand{\vdu}{\delta \mathbf u}

\newcommand{\vA}{\mathbf A}
\newcommand{\vB}{\mathbf B}
\newcommand{\vC}{\mathbf C}

\newcommand{\vG}{\mathbf G}
\newcommand{\vH}{\mathbf H}
\newcommand{\vI}{\mathbf I}
\newcommand{\vJ}{\mathbf J}
\newcommand{\vK}{\mathbf K}

\newcommand{\vP}{\mathbf P}
\newcommand{\vQ}{\mathbf Q}
\newcommand{\vR}{\mathbf R}
\newcommand{\vS}{\mathbf S}

\newcommand{\vU}{\mathbf U}
\newcommand{\vV}{\mathbf V}
\newcommand{\vW}{\mathbf W}
\newcommand{\vX}{\mathbf X}

\newcommand{\cN}{\mathcal{N}}

\newcommand\numberthis{\addtocounter{equation}{1}\tag{\theequation}} 

\makeglossaries

\makesavenoteenv{tabular}
\makesavenoteenv{figure}
\makesavenoteenv{itemize}

\providecommand{\norm}[1]{\lVert#1\rVert}

\usepackage{beamerarticle}
\usepackage{tikz}
\usepackage{verbatim}
\usetikzlibrary{arrows,shapes}
\usepackage{wrapfig}
\usepackage{tabularx}

\usepackage{mathrsfs,amsmath,amsfonts,mathtools,empheq}

\usepackage{cancel}
\usepackage{breqn}
\usepackage{xcolor}
\usepackage{float}

\usepackage{array,multirow,graphicx}
\usetikzlibrary{decorations.pathreplacing,calc}

\usepackage{enumitem}
\newcommand{\vect}[1]{\boldsymbol{\mathbf{#1}}} %

\newcommand{\tikzmark}[2][-3pt]{\tikz[remember picture, overlay, baseline=-0.5ex]\node[#1](#2){};} %

\newlength{\bracewidth}

\newcommand{\highlight}[1]{%
 \colorbox{red!20}{$\displaystyle#1$}}

\newcounter{arrow}
\newcommand{\drawcurvedarrow}[3][]{%
\refstepcounter{arrow}
\tikz[remember picture, overlay]\draw (#2.center)edge[#1]node[coordinate,pos=0.5, name=arrow-\thearrow]{}(#3.center);
}

\newcommand{\annote}[3][]{%
\tikz[remember picture, overlay]\node[#1] at (#2) {#3};
}

\usepackage{xfrac}

\usepackage{algpseudocode}
\usepackage{algorithm}

\algrenewcommand\algorithmicrequire{\textbf{Input:}}
\algrenewcommand\algorithmicensure{\textbf{Output:}}

\algnewcommand\algorithmicforeach{\textbf{for each:}}
\algnewcommand\ForEach{\item[ \algorithmicforeach]}

\usepackage{amsmath, amsfonts, mathtools, xfrac, empheq}
\usepackage{mathrsfs}
\usepackage{xcolor}
\usepackage{algpseudocode}
\usepackage{algorithm}

\usepackage{wrapfig}
\usepackage{graphicx}
\usepackage{tikz}

\usepackage{array}

\algrenewcommand\algorithmicensure{\textbf{Return:}}

\begin{document}

\thispagestyle{empty}
\begin{titlepage}

\begin{center}

\begin{picture}(0,60)(0,0) 
\put(-255,30){\includegraphics[height=1.8cm]{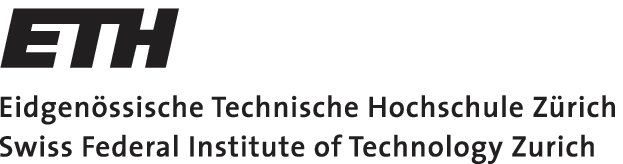}} 
  
\put(145,30){\includegraphics[height=1.8cm]{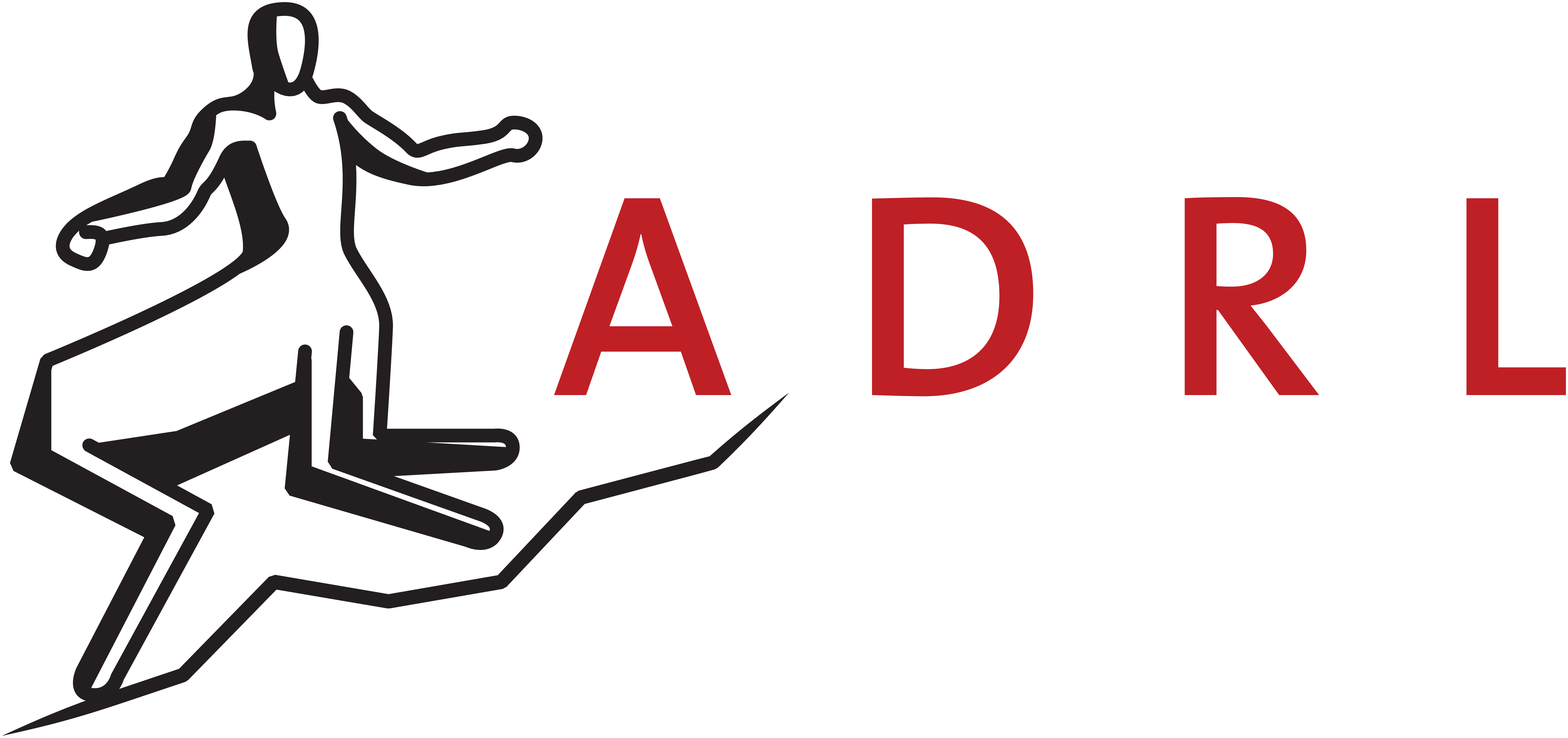}}

\put(-255,5){\line(1,0){510}}
\end{picture}

\vspace{15mm}
{ \bfseries\huge{Optimal and Learning Control\\ for Autonomous Robots}} \\
\vspace{15mm}
{\bfseries \large Lecture Notes} \\
\vspace{5mm}
{\bfseries \large \today}
\vspace{10mm}


\vfill

\begin{picture}(0,0)(0,0) 

\put(100,10){\bfseries{Authors}}
\put(100,-6){Jonas Buchli}
\put(100,-18){Farbod Farshidian}
\put(100,-30){Alexander Winkler}
\put(100,-42){Timothy Sandy}
\put(100,-54){Markus Giftthaler}
\end{picture}

\end{center} 
\end{titlepage}

\cleardoublepage
\newgeometry{left=3cm, width=16cm,top=2cm, includehead}
\pagestyle{fancy} 

\fancyhead{} %
\fancyhead[RO,LE]{}	
\fancyhead[CO,CE]{}
\fancyhead[LO,RE]{}

\fancyfoot[RO,LE]{\thepage}
\fancyfoot[CO,CE]{}
\fancyfoot[LO,RE]{} 
\renewcommand{\headrulewidth}{0pt} 
\renewcommand{\footrulewidth}{0pt}

\renewcommand{\chaptermark}[1]{\markboth{#1}{}}
\renewcommand{\sectionmark}[1]{\markright{\thesection\ #1}}

\renewcommand*{\chapterpagestyle}{fancy}

\pagenumbering{Roman} \setcounter{page}{1}

\section*{Foreword - ArXiv Edition} 

\emph{Optimal and Learning Control for Autonomous Robots} has been taught in the Robotics, Systems and Controls Masters at ETH Z\"urich with the aim to teach optimal control and reinforcement learning for closed loop control problems from a unified point of view. The starting point is the formulation of of an optimal control problem and deriving the different types of solutions and algorithms from there.

These lecture notes aim at supporting this unified view with a unified notation wherever possible, and a bit of a translation help to compare the terminology and notation in the different fields. There are a number of outstanding and complementary text books which inspired some of the material in here. We highly recommend to peruse these textbooks to deepen the knowledge in the specific topics.

Thus, some chapters follow relatively closely some well known text books (see below) with adapted notation, while others section have been originally written for the course.

\begin{itemize}
\item{Section 1.3} - Robert F Stengel. Stochastic Optimal Control: Theory and Application. J. Wiley \& Sons, New
York, NY, USA, 1986 \cite{Stengel1986}. \\

\item{Chapter 2} - Richard S. Sutton and Andrew G. Barto. Introduction to Reinforcement Learning. MIT Press,
Cambridge, MA, USA, 1st edition, 1998 \cite{Sutton1998}. \\
\end{itemize}

Slides and additional information on the course can be found at:\\
\url{http://www.adrl.ethz.ch/doku.php/adrl:education:lecture}

The course assumes basic knowledge of Control Theory, Linear Algebra and Stochastic Calculus.

\vspace{1cm}
Jonas Buchli \\
Z\"urich, July 2017

\tableofcontents
\thispagestyle{empty} 
\cleardoublepage

\renewcommand{\headrulewidth}{1pt}
\fancyhead[RO,LE]{\rightmark}
\fancyhead[LO,RE]{\leftmark}

\pagenumbering{arabic} \setcounter{chapter}{0} 

\graphicspath{{pics/}}
\chapter{Optimal control} \label{ch:optimal_control}
\section{Principle of optimality}

This section describes two different approaches of finding an optimal path
through a graph, namely \emph{forward} and \emph{backward} search. Backwards
search introduces the \emph{Principle of Optimality} which lies at the core of
\emph{dynamic programming}. This section closely follows \cite{Bertsekas1995}, pages 18-19.

\subsection{Graph Search Problem}
\begin{figure}[htb]
\centering
  \includegraphics[width=0.75\textwidth]{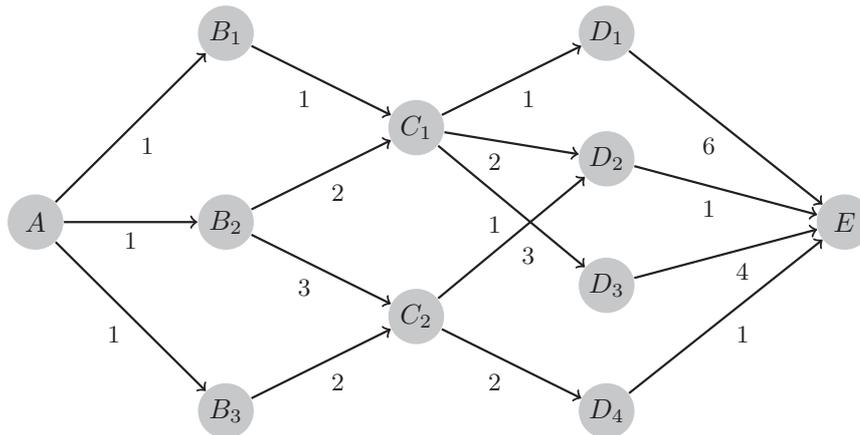}
  \caption{Example of a directed graph with the cost at each edge. The goal is passing from 
           $A \rightarrow E$, while accumulating the least costs.}
  \label{fig:shortest_path}
\end {figure}

Consider the directed graph shown in \fref{fig:shortest_path}. The goal
is to find the path with the lowest cost from node $A$ to node $E$, with the values at the 
edges corresponding to costs. Two different approaches, namely \emph{forward} and \emph{backward search}, are discussed
in the following.  
 
\subsection*{Forward Search}
\begin{figure}[htb]
\centering
  \includegraphics[width=0.8\textwidth]{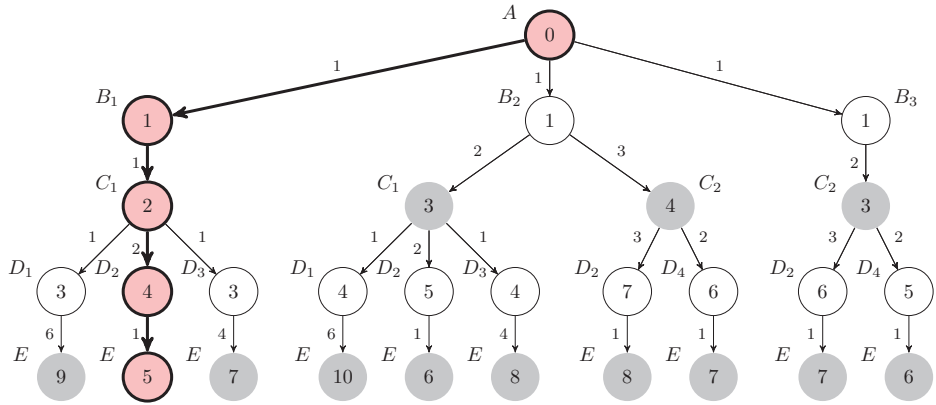}
  \caption{Forward search tree. The accumulated cost to reach each node from node $A$ is shown 
  inside the nodes. The optimal path with the lowest accumulated cost when reaching 
  node $E$ is shown in bold.}
  \label{fig:forward_search}
\end {figure}

A forward search through this graph consists of computing the cost for
\textit{all possible paths} starting at node $A$ until node $E$. The 10 possible
paths can be seen in
\fref{fig:forward_search}. The value inside each node corresponds to the 
\emph{accumulated cost} from node $A$ up to the specific node.  The optimal path is the one with the lowest accumulated
cost after reaching node $E$, namely path $A-B_1-C_1-D_2-E$ (bold path).

Now consider that we reformulate the problem and want to find the shortest path
from node $C_2$ to $E$, after unintentionally reaching state $C_2$. We cannot
reuse our calculated values of the \textit{accumulated} costs (node $A \rightarrow
E$), because we do not care anymore about the cost to reach $C_2$. To find the
optimal path we must rebuild the tree, starting at $C_2$, and then again choose the
lowest cost path. This rebuilding of the tree for every change of the initial state
is computationally expensive. The following approach allows us to store the information in a
more reusable way.

\subsection*{Backward Search}
\begin{figure}[htb]
\centering
  \includegraphics[width=0.8\textwidth]{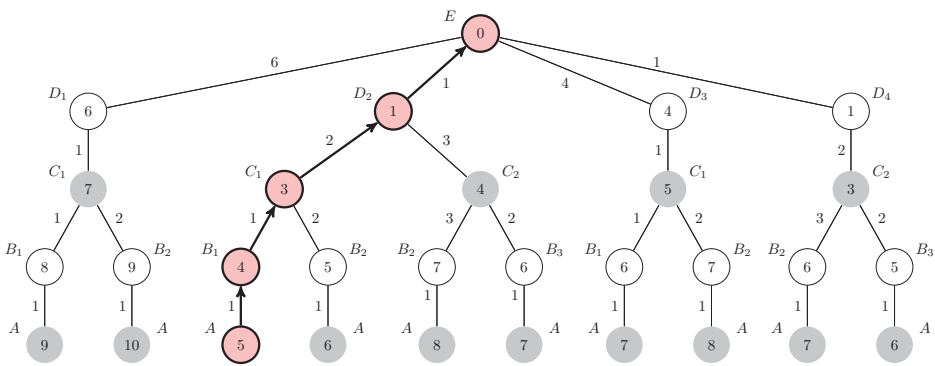}
  \caption{Backward search tree. The graph is constructed by starting at the goal node $E$
           and evaluating the possible paths towards the start node $A$. The values 
           inside each node refers to the cost to go from this node to the goal node $E$.}
  \label{fig:tree_backward}
\end {figure}

A backward search tree is constructed by starting at the goal node and
evaluating all possible paths towards the start node. The resulting tree can be
seen in \fref{fig:tree_backward}. The values in the nodes describe the
``cost-to-go''(``value function'') from this specific node to the goal node $E$. 
The cost-to-go for each node is equal to the cost-to-go of the node one level above
plus the cost of the connecting edge. The highlighted path in \fref{fig:tree_backward} starting
at node $A$ is the optimal path from $A \rightarrow E$ since it has the lowest cost-to-go,
with the cost being $5$.

Storing the information about the graph as the cost-to-go, instead of
the \textit{accumulated} cost for each possible path, yields one main advantage:
The optimal ``control strategy'' at each step is the one which leads to the node
with the lowest cost-to-go. There is no dependency on how this node has been
reached as in the forward search, so there is no need to build a new tree when
deviating from our initially planned path. In terms of computational effort, the
construction of both trees is comparable (considering one unit of computation for each node of the trees: $28$
for the forward search tree,and $30$ for the backward search tree). The relative cost for the two methods is only dependent on the structure of the directed graph. This is easily seen by considering what the computational cost of each method would be if the direction of all edges in the example were switched (ie reverse the notion of backwards and forwards). The real value in considering the cost-to-go for each node, though, is that we can easily \textit{reuse} this information when deviating from our original plan.

\subsection{Principle of optimality}
\label{subsec:principle_of_optimality}
The reason why the backward search tree in the previous example can be constructed
as described is due to the \textit{Principle of optimality}, which states:
\begin{center}
\fbox{\parbox[c][1.5\height]{15cm}{\centering
If path $ABCDE$ is optimal, then $BCDE$ is  optimal for the truncated problem.\\
In "control" terms: $\vU_0^* \ldots    ,\vU_n^*,\vU_{n+1}^*,   \ldots   \vU_N^* 
                   \Rightarrow        \vU_n^*,\vU_{n+1}^*,   \ldots   \vU_N^*$}}
\end{center} 

This means it is possible to construct an optimal path in a piecemeal fashion:
First an optimal path is found for only the last stage of the problem (e.g. $D
\rightarrow E$). The problem can then be solved for the next stage, reusing the
already solved ``tail subproblem''. This iterative process reduces overall
complexity and computational time by decomposing the problem into solvable
sub-tasks. This principle is the foundation of the dynamic programming (DP) algorithms
presented in the following sections.

\section{Bellman equation}
The Bellman equation introduces a functional for the value function (cost-to-go)
of a given policy in \emph{discrete time} systems. It expresses a relationship
between the value function of a policy in a specific time and state to its
successor states in the next time step. By applying the principle of optimality,
the optimal Bellman equation can be derived for any discrete time optimal control
problem. This section derives the Bellman equation in four different problem settings. The optimal control problem for discrete time
deterministic systems is presented, first over a finite time horizon and then over an infinite time horizon. The optimal control problem for discrete time stochastic systems is then presented, again for both finite and infinite time horizons. For each problem setting, the notions of value function and optimal value
function are introduced and the Bellman equation is then derived.

\subsection{Finite time horizon, deterministic system} 
\paragraph*{Problem definition} \label{subsec:prob_fin_det}
The problem definition consists of two components. The first is a model that describes
the system dynamics by the function $\vf$. This corresponds to the nodes and
arrows of the graph in Figure~\ref{fig:shortest_path}. The second is the cost function
$J$, which captures the cost associated with a certain path taken over the whole
time interval considered.

In the discrete time case, time steps $t = t_n$ are indexed by integer values $n \in \{0,1,...,N\}$. In this section, we consider a finite time interval $t \in \left[ t_0, t_N \right]$. For the system state at time $t_{n}$, we use the short hand notation $\vx_n = \vx(t_n)$. 
The discrete time deterministic system is described as
\begin{equation}
\vx_{n+1}=\vf_n(\vx_n,\vu_n), \hspace{10mm} n \in \{0,1,...,N-1\}
\label{eq:discrete_time_deterministic_system}
\end{equation}
where \\
\begin{tabular}[t]{ll}
$n$ & is the discrete time index, \\
$\vx_n$ & is the state of the system at time $n$, \\
$\vu_{n}$ & is the control input at time $n$ and \\
$\vf_{n}$ & is the state transition equation. \\
\end{tabular} \\

The cost function $J$ gives the cost associated with a particular trajectory, 
i.e. a time sequence of states $\vx(t)$,
starting from state $\vx_{0}$ at time step $n=0$ up to the final state $\vx_N$ at
time step $n=N$ under a specific control policy $\mu = \{ \vu_0,\vu_1,...,\vu_{N-1}
\}$. 
\begin{equation}
J=\alpha^N \Phi(\vx_{N})+\sum_{k=0}^{N-1} \alpha^k L_k (\vx_k,\vu_k)
\label{eq:discrete_time_deterministic_system_cost_function}
\end{equation}
$L(\vx_n,\vu_n)$ defines the intermediate cost incurred at each time step, as a function of the state and control input applied at that time step. $\Phi(\vx_{N})$ defines the terminal cost, which depends only on the state at the final time. $\alpha$ is a parameter $0 \leq \alpha \leq 1$ called the discount or decay rate,
that continuously reduces the effect of costs further away in time (in the finite
time horizon case this discount factor is usually set to 1). 

The goal of the optimal control problem is to find
the optimal control policy $\mu^*$, that minimizes the cost function, and thus gives
to optimal cost $J^*$. This goal can equivalently be written as
\begin{equation}
\mu^* = \arg \min_{\vu} J  
\end{equation}

\paragraph*{Value function} \label{subsec:val_fun_dis_det}  
As motivated in the previous chapter, it is more useful to consider the
``cost-to-go'' from a specific state to a goal state, than the ``accumulated cost''
$J$. The cost-to-go from a specific state $\vx$ at a specific time step $n$ when
following a policy $\mu$ is described by the \emph{value function}%
\begin{align} \label{eq:val_fun}
    &V^{\mu}(n,\vx) = \alpha^{N-n} \Phi(\vx_{N}) + \sum_{k=n}^{N-1} \alpha^{k-n}  L_k(\vx_k,\vu_k) \\
    &\vx_n = \vx \nonumber \\
    &\vx_{k+1} = \vf_k(\vx_k, \vu_k)   \hspace{2cm}  k=n, \dots, N-1 \nonumber\\
    &\vu_k = \mu(k,\vx_k) \nonumber    
\end{align}
Note that the value function depends on both time $n$ and state $\vx$, which
indicates the initial condition at time $n$ for integrating the system dynamics.
The value function evaluated at the final stage $N$ corresponds to the terminal
cost $\Phi(\cdot)$ of the cost function.
\begin{equation}
    V^{\mu}(N,\vx) = \Phi(\vx)
    \label{eq:val_fun_final_cost}
\end{equation}

The value function with the lowest value, e.g. the minimum cost, is called the
\emph{optimal value function} and is denoted as
\begin{equation}
\begin{split}
    V^*(n,\vx) &= \min_{\mu} V^{\mu}(n,\vx) \\ 
             &= \min_{\vu_n,\ldots, \vu_{N\!-\!1}} 
                \{\alpha^{N-n} \Phi(\vx_{N}) + \sum_{k=n}^{N-1} \alpha^{k-n}  L_k(\vx_k,\vu_k)\} \\
\end{split}
\label{eq:val_fun_opt}
\end{equation}
Notice that in general a control policy, $\mu$, that minimizes the right-hand
side of equation~\eqref{eq:val_fun_opt} is a function of both time and state.
Furthermore, due to the Principle of Optimality
(\sref{subsec:principle_of_optimality}), the control sequence that minimizes the
value function in time $n$ should be same as the tail sequence of a policy that
minimizes the value function for a time step before the time step $n$.

The corresponding \emph{optimal policy} for the optimal value function is defined as
\begin{equation}
\begin{split}
    \mu^* = \{ \vu_n^*,\ldots, \vu_{N\!-\!1}^* \} &= \arg \min_{\mu} V^{\mu}(n,\vx)    \hspace{1cm}  \forall n:0, \dots, N-1    \\
          &= \arg \min_{\vu_n,\dots,\vu_{N\!-\!1}} \{\alpha^{N-n} \Phi(\vx_{N}) + \sum_{k=n}^{N-1} \alpha^{k-n}  L_k(\vx_k,\vu_k)\} \\
\end{split}
\label{eq:val_fun_opt_policy}
\end{equation}

Note that the optimal value function $V^*(\cdot,\cdot)$ at time step $0$  corresponds to the optimal accumulated cost $J^*$
\begin{equation}
J^* = V^*(0,\vx_0)
\label{eq:val_fun_cost_fun_relation}
\end{equation}

\paragraph*{Bellman equation} \label{subsec:bell_dis_det}
The Bellman equation is derived starting with the definition of the value function (equation~\ref{eq:val_fun}). Taking the intermediate cost at time $n$ out of the summation and using the fact that $\vx_n=\vx$ leads to the following equation
\begin{equation}
    V^{\mu}(n,\vx) = L_n (\vx,\vu_n) + \alpha^{N-n} \Phi(\vx_{N}) + \sum_{k=n+1}^{N-1} \alpha^{k-n}  L_k(\vx_k,\vu_k)
\end{equation}
Factoring $\alpha$ out of the last terms leads to
\begin{equation}
    V^{\mu}(n,\vx) = L_n (\vx,\vu_n) + \alpha \left[ \alpha^{N-n-1} \Phi(\vx_{N}) + \sum_{k=n+1}^{N-1} \alpha^{k-n-1}  L_k(\vx_k,\vu_k) \right]
\end{equation}
The terms inside the brackets are equal to $V^{\mu}(n+1,\vx_{n+1})$, where $\vx_{n+1}=f(\vx,\vu_n)$:
\begin{equation}
    \label{eq:bellman_det}
    V^{\mu}(n,\vx) = L_n (\vx,\vu_n) + \alpha V^{\mu} \left( n+1,f_n \left( \vx,\vu_n \right) \right)
\end{equation}
with final condition $V^{\mu}(N,\vx)=\Phi(\vx)$.

As previously discussed, the optimal value function corresponds to the policy that
minimizes the right-hand side of \eqref{eq:bellman_det} which is known as 
the \emph{optimal Bellmann equation} for deterministic systems
\begin{equation}
    \label{eq:bellman_det_opt}
    V^{*}(n,\vx) = \min_{\vu_n} \left[ L_n (\vx,\vu_n) + \alpha V^* \left( n+1,\vf_n \left( \vx,\vu_n \right) \right) \right]
\end{equation}
and thus the optimal control policy at time n is computed as
\begin{equation}\label{eq:Bellman_control_det}
    \vu^*(n,\vx) = \arg \min_{\vu_n} \left[ L_n (\vx,\vu_n) + \alpha V^* \left( n+1,\vf_n \left( \vx,\vu_n \right) \right) \right]
\end{equation}
To find the optimal value function as defined in
equation~\eqref{eq:bellman_det_opt}, we search for the control input $\vu_n$ that minimizes the sum of the instantaneous cost $L_n$ and the optimal
value function at the next time step, $(n+1)$, considering the state
which would be reached by applying $\vu_n$. To solve this equation, one should first find the optimal control input for time instance $n=N-1$ and then proceed backwards in time, finding the optimal control input at each step, until the first time instance $n=0$.
The optimal state trajectory will be obtained if the calculated optimal policy
$\mu^{*} = \{\vu_0^{*}, \vu_1^{*},...,\vu_{N-1}^{*} \}$ is applied to the system
described by equation~\eqref{eq:discrete_time_deterministic_system}.

The advantage of using the optimal Bellman equation compared to solving
\eqref{eq:val_fun_opt} is problem decomposition: whereas in
equation~\eqref{eq:val_fun_opt} the entire sequence of control inputs (policy)
must be optimized for at once, the Bellman equation allows us to optimize for a
single control input $\vu_n$ at each time. Therefore the computational effort in
the latter case increases only linearly with the number of time steps, as opposed to exponentially for
equation~\eqref{eq:val_fun_opt}.

\subsection{Infinite time horizon, deterministic system} 
\paragraph*{Problem definition} \label{subsec:prob_infin_det} 
In the infinite horizon case the cost to minimize resembles the finite horizon
case except the accumulation of costs is never terminated, so $N \to \infty$ and
\eqref{eq:discrete_time_deterministic_system_cost_function} becomes
\begin{equation}
    J=\sum_{k=0}^{\infty} \alpha^k L(\vx_k,\vu_k)  ,   \hspace{10mm} \alpha \in [0,1]    \label{eq:cost_infin_det}
\end{equation}
Since the evaluation of the cost never ends, there exists no terminal cost
$\Phi(\cdot)$. In contrast to \sref{subsec:prob_fin_det}, the discount factor
$\alpha$ is usually chosen smaller than  $1$, since otherwise it is a summation
over a infinite numbers which can lead to an unbounded cost. Additionally for the
sake of convenience, the system dynamics $\vf(\cdot)$ and the instantaneous costs
$L(\cdot)$ are assumed time-invariant.

\paragraph*{Value function} \label{subsec:val_infin_det}
The optimal value function for \eqref{eq:cost_infin_det} at time step $n$
and state $\vx$ can be defined as
\begin{equation}
V^*(n,\vx)=\min_\mu \left[ \sum\limits_{k=n}^{\infty} \alpha^{k-n} L(\vx_k,\vu_k) \right].
\label{eq:v_n}
\end{equation}

Calculating the optimal value function for the same state $\vx$, but at a
different time $n+\Delta n$ leads to
\begin{equation}
\begin{split}
V^*(n+\Delta n,\vx) 
&=\min_\mu \left[ \sum\limits_{k=n+\Delta n}^{\infty} \alpha^{k-n-\Delta n} L(\vx_k,\vu_k) \right] \\
&=\min_\mu \left[ \sum\limits_{k'=n        }^{\infty} \alpha^{k'-n         } L(\vx_{k'+\Delta n},\vu_{k'+\Delta n}) \right]
\end{split}
\label{eq:v_ndn}
\end{equation}
The only difference between \eqref{eq:v_n} and \eqref{eq:v_ndn} is the state trajectory over which the cost is calculated. However, since the system dynamics are time invariant and the initial state for the both paths is the same, these two trajectories are identical except for the shift in time by $\Delta n$. It follows that
\begin{equation}
V^*(n,\vx) = V^*(n+\Delta n,\vx) = V^*(\vx).
\end{equation}
This shows that the optimal value function for the infinite time
horizon is time-invariant and therefore only a function of the state.

\paragraph*{Bellman equation} \label{subsec:ball_infin_det} 
Using the results from the finite time horizon Bellman equation and the knowledge that the value function for the infinite time horizon case is time-invariant, we can simplify \eqref{eq:bellman_det_opt} to give
\begin{equation}
V^*(\vx)=\min\limits_{u}\{L(\vx,\vu)+\alpha V^*(f(\vx,\vu)) \},
\label{eq:bell_inf_det}
\end{equation}
which is the optimal Bellman equation for the infinite time horizon problem.

\subsection{Finite time horizon, stochastic system} 
\paragraph*{Problem definition} \label{subsec:prob_fin_stoch}
We model stochastic systems by adding a random variable $\vw_n$ to the
deterministic system's dynamics.
\begin{equation}
\vx_{n+1}=\vf_n(\vx_n,\vu_n)+\vw_n
\label{eq:discrete_time_ stochastic}
\end{equation}
$\vw_n$ can take an arbitrary conditional probability distribution given $\vx_n$ and $\vu_n$. 
\begin{equation}
\vw_n\sim P_w(\cdot \mid \vx_n,\vu_n)
\label{eq:discrete_time_ noise_probability}
\end{equation}
We can now introduce a new random variable $\vx'$, which incorporates the deterministic and random parts of the system:
\begin{equation}
\vx_{n+1} = \vx'
\label{eq:discrete_time_probability_1}
\end{equation}
where $\vx'$ is distributed according to a new conditional distribution given  $\vx_n$ and $\vu_n$.
\begin{equation}
\vx' \sim P_f(\cdot \mid \vx_n,\vu_n)
\label{eq:discrete_time_probability_2}
\end{equation}
Although any system described by equations~\eqref{eq:discrete_time_ stochastic} and \eqref{eq:discrete_time_ noise_probability} can be uniquely formulated by equations~\eqref{eq:discrete_time_probability_1} and \eqref{eq:discrete_time_probability_2}, the opposite is not correct. Therefore, equations~\eqref{eq:discrete_time_probability_1} and \eqref{eq:discrete_time_probability_2} describe a more general class of discrete systems. We will consider this class of systems in the remainder of this subsection. 

Once again, the goal is to find the control policy $\mu^* = \{\vu_0^{*}, \vu_1^{*},...,\vu_{N-1}^{*} \}$
which results in the path associated with the lowest cost defined by
equation~\eqref{eq:discrete_time_stochastic_system_cost_function}.
\begin{equation}
J= E \left[ \alpha^N \Phi(\vx_{N})+\sum_{k=0}^{N-1} \alpha^k L_k (\vx_k,\vu_k) \right]
\label{eq:discrete_time_stochastic_system_cost_function}
\end{equation}

\paragraph*{Value function} \label{subsec:val_fin_stoch}
The value function for a given control policy in a stochastic system is identical to that which was used in the deterministic case, expect for the fact that we must take the expectation of the value function to account for the stochastic nature of the system.
\begin{equation}\label{eq:val_fun_stoch}
    V^{\mu}(n,\vx) = E \left[ \alpha^{N-n} \Phi(\vx_{N}) + \sum_{k=n}^{N-1} \alpha^{k-n}  L_k(\vx_k,\vu_k) \right]
\end{equation}
The optimal value function and optimal policy similarly follow as
\begin{equation}
    V^*(n,\vx) = \min_{\mu} E \left[ \alpha^{N-n} \Phi(\vx_{N}) + \sum_{k=n}^{N-1} \alpha^{k-n}  L_k(\vx_k,\vu_k) \right]
\label{eq:val_fun_opt_stoch}
\end{equation}
\begin{equation}
    \mu^* = \arg \min_{\mu} E \left[ \alpha^{N-n} \Phi(\vx_{N}) + \sum_{k=n}^{N-1} \alpha^{k-n}  L_k(\vx_k,\vu_k) \right]
\label{eq:val_fun_opt_policy_soch}
\end{equation}

\paragraph*{Bellman equation} \label{subsec:bell_fin_stoch}
Through the same process as in the deterministic case, we can show that the Bellman equation for a given control policy will be as follows
\begin{equation}
    V^{\mu}(n,\vx) = L_n (\vx,\vu_n) + \alpha E_{x' \sim P_f(.\mid \vx,\vu_n)} \left[ V^\mu \left( n+1,\vx' \right) \right]
\end{equation}
and the optimal Bellman equation is
\begin{equation}\label{eq:Bellman_stoch}
    V^{*}(n,\vx) = \min_{\vu_n} \left[ L_n (\vx,\vu_n) + \alpha E_{\vx' \sim  P_f ( \cdot \mid \vx,\vu_n)} \left[ V^* \left( n+1,\vx' \right) \right] \right]
\end{equation}
and thus the optimal control at time n is computed as
\begin{equation}\label{eq:Bellman_control_stoch}
    \vu^*(n,\vx) = \arg \min_{\vu_n} \left[ L_n (\vx,\vu_n) + \alpha E_{\vx' \sim  P_f ( \cdot \mid \vx,\vu_n) } \left[ V^* \left( n+1,\vx' \right) \right] \right].
\end{equation}

Note that it is possible to convert Equation~\eqref{eq:Bellman_stoch} to the deterministic
case by assuming $P(\cdot \mid \vx,\vu)$ as a Dirac delta distribution.
\begin{equation}
P_f(\vx'\mid \vx,\vu_n)=\delta(\vx' - \vf(\vx,\vu_n))
\end{equation}

\subsection{Infinite time horizon, stochastic system} 
\paragraph*{Problem Definition} \label{subsec:prob_infin_stoch} 
The Bellmann equation for a stochastic system over an infinite time horizon is a
combination of \sref{subsec:prob_infin_det} and \sref{subsec:prob_fin_stoch}. As
in \sref{subsec:prob_fin_stoch} the system dynamics includes the stochastic variable $\vw_n$ as
\begin{equation}
    \vx_{n+1}=\vf(\vx_n,\vu_n)+\vw_n.
    \label{eq:prob_inf_stoch}
\end{equation}
Since the stochastic cost cannot be minimized directly, we wish to minimize the
expectation of the cost in  \sref{subsec:prob_infin_det} denoted as
\begin{equation}
    J=E \left[\sum_{k=0}^{\infty} \alpha^k L(\vx_k,\vu_k) \right],   \hspace{10mm} \alpha \in [0,1)
    \label{eq:cost_infin_sto}
\end{equation}
again with a discount factor $\alpha<1$, time invariant systems dynamics
$\vf(\cdot)$ and instantaneous costs $L(\cdot)$ as in the deterministic case
described in \sref{subsec:prob_infin_det}.

\paragraph*{Optimal value function} \label{subsec:val_infin_stoch} 
The optimal value function resembles \eref{eq:v_n}, but is extended by the expectation $E$
\begin{equation}
V^*(\vx)=\min_\mu E \left[ \sum\limits_{n=0}^{\infty} \alpha^n L(\vx_n,\vu_n) \right] 
\label{eq:infinite_horizon_stochastic}
\end{equation}

\paragraph*{Bellman equation} \label{subsec:bell_infin_stoch} 
The Bellman equation for the infinite time horizon stochastic system also uses
the expectation and gives
\begin{equation}
V^*(\vx)=\min\limits_{\vu_n}\{L(\vx,\vu)+\alpha E[V^*(\vx')] \}
\label{eq:infinite_horizon_stochastic_opt}
\end{equation}

\section{Hamilton-Jacobi-Bellman Equation}

We now wish to extend the results obtained in the previous section to \emph{continuous time} systems. As you will see shortly, in the continuous time setting, optimal solutions become less straight forward to derive analytically. We will first derive the solution to the optimal control problem for deterministic systems operating over a finite time horizon. This solution is famously known as the Hamilton-Jacobi-Bellman (HJB) equation. We will then extend the results to systems with an infinite time horizon, then to stochastic finite time horizon systems, and finally to stochastic infinite time horizon systems.

\subsection{Finite time horizon, deterministic system}
\label{subsec:HJP_fin_det}
\paragraph*{Problem Definition}
In this section we will consider a continuous time, non-linear deterministic system of the form,
\begin{equation}
\dot{\vx}(t)=\vf_t(\vx(t),\vu(t))
\label{eq:continuous_time_deterministic system equation}
\end{equation}
Its corresponding cost function over a finite time interval, $t \in  [t_0,t_f]$, starting from initial condition $\vx_0$ at $t_0$, is
\begin{equation}
J = e^{-\beta (t_f-t_0)} \Phi(\vx(t_f))+ \int_{t_0}^{t_f} { e^{-\beta (t-t_0)} L(\vx(t),\vu(t))dt},
\label{eq:continuous_time_deterministic cost function}
\end{equation}
where $L(\vx(t),\vu(t))$ and $\Phi(\vx(t_f))$ are the intermediate cost and the
terminal cost respectively. $\beta$ is a parameter $0 \leq \beta$ called the
decay or discount rate, that continuously reduces the effect of costs further
away in time (in the finite time horizon case this discount factor is usually set
to 0). Our goal is to find a control policy which minimizes this cost.
\paragraph*{HJB equation}
In order to obtain an expression for the optimal cost-to-go starting from some
$x$ at some time $t \in [t_0,t_f]$, we need to evaluate the cost function over
the optimal trajectory, $x^*(t)$, using the optimal control policy, $u^*(t)$,
during the remaining time interval $[t,t_f]$.
\begin{equation}
V^*(t,\vx)= e^{-\beta (t_f-t)} \Phi(\vx^*(t_f)) +  \int_{t}^{t_f} {e^{-\beta (t'-t)} L (\vx^*(t'),\vu^*(t'))dt'},
\label{eq:optimal_cost_deterministc_cont_time}
\end{equation}
We can informally derive the HJB equation from the Bellman equation by discretizing 
the system into $N$ time steps as follows
\begin{align*}
& \delta t =\frac{t_f-t_0}{N} \\
& \alpha = e^{-\beta \delta t} \approxeq 1 - \beta \delta t \\
& t_{n} = t_0+n\delta t \\
& \vx_{k+1} = \vx_k+\vf(\vx_k,\vu_k)\cdot \delta t
\label{eq:discrete appproximation system equation}
\end{align*}

The value function can now be approximated as a sum of instantaneous costs at each time step.
\begin{equation}
\tilde{V}(t_n,\vx) = \alpha^{N-n} \Phi(\vx_{N})+ \sum_{k=n}^{N-1}{\alpha^{k-n} L(\vx_k,\vu_k)\delta t},
\label{eq:continuous_time_discrete_approx cost function}
\end{equation}
where $n \in \{0,1,...,N\}$, and $\vx$ is the state at time $t_n$.

This is similar to the value function of a discrete time system
(equation~\eqref{eq:discrete_time_deterministic_system_cost_function}). The discrete approximation of the optimal value function is
\begin{equation}
\tilde{V}^{*}(t_n,\vx)=\min\limits_{\vu \in \vU} \{ L(\vx,\vu)\delta t + \alpha \tilde{V}^*\left(t_{n+1},\vx_{n+1} \right) \}.
\label{eq:descretized_principle_of_optimality}
\end{equation}

For small $\delta t$, we can use the Taylor Series Expansion of $\tilde{V}^{*}$ to expand the term on the right of the equation above.
\begin{align}
	\tilde{V}^{*}(t_{n+1},\vx_{n+1}) &= \tilde{V}^{*}(t_{n}+\delta t,\vx + \vf(\vx,u) \delta t) \notag \\
	&= \tilde{V}^{*}(t_n,\vx) + \Delta \tilde{V}^{*}(t_n,\vx) \notag \\
	&=\tilde{V}^{*}(t_n, \vx)+\frac{\partial \tilde{V}^{*}(t_n, \vx)}{\partial t} \delta t + \left(\frac{\partial \tilde{V}^{*}(t_n, \vx)}{\partial \vx}\right)^T \vf(\vx,\vu) \delta t
\label{eq:taylor_series_V}
\end{align}
Higher order terms are omitted because they contain $\delta t$ to the second power or higher, making their impact negligible as $\delta t$ approaches $0$.

Plugging \eqref{eq:taylor_series_V} into \eqref{eq:descretized_principle_of_optimality}, subtracting $\tilde{V}^{*}(t_n,\vx)$ from both sides, and rearranging terms we get
\begin{equation}
-\alpha \frac{\partial \tilde{V}^{*}(t_n, \vx)}{\partial t} \delta t + \beta \tilde{V}^{*}(t_n, \vx) \delta t =\min\limits_{\vu \in \vU} \Big\{ L(\vx,\vu)\delta t + \alpha \left(\frac{\partial \tilde{V}^{*}(t_n, \vx)}{\partial \vx}\right)^T \vf(\vx,\vu) \delta t \Big\}
\label{eq:substitute_in_taylor_exp_hjb}
\end{equation}
Here we assumed that $\delta t$
is small enough to allow us to approximate $\alpha$ with $ (1 - \beta \delta t)$. Letting $t=t_n$, using $\delta t \ne 0$ to remove it from both
sides of the equation, and assuming $\tilde{V}^{*}\rightarrow V^{*}$ as $\delta t
\rightarrow 0$, we get the HJB equation (notice that $\lim_{\delta t \to 0} \alpha = 1$)
\begin{equation}
\beta V^{*} -\frac{\partial V^{*}}{\partial t}=\min\limits_{\vu \in \vU} \Big\{ L(\vx,\vu) + \left(\frac{\partial V^{*}}{\partial \vx}\right)^T \vf(\vx,\vu) \Big\}
\label{eq:hjb_cont_time_determininstic}
\end{equation}

\subsection{Infinite time horizon, deterministic system}
\paragraph*{Problem Definition}
We will now consider the case where the cost function includes an infinite time
horizon. The cost function takes the form:
\begin{equation}
J=\int_{t_0}^{\infty} {e^{-\beta (t-t_0)} L(\vx(t),\vu(t))dt} 
\label{eq:continuous_time_deterministic_inf_horizon_cost function}
\end{equation}
where the terminal cost (formerly $\Phi(\vx(t_f))$) has dropped out. Furthermore,
like the discrete case, the system dynamics and the intermediate cost are
time-invariant. $\beta$ is a parameter $0 \leq \beta$ called the decay or
discount rate, that in the infinite time horizon case is usually set to greater
than 0.

\paragraph*{HJB equation}
As it was the case for discrete-time systems, if we consider an infinite time horizon problem for a continuous-time system, $V^{*}$ is not a function of time. This means that $\frac{\partial V^{*}}{\partial t}=0$ and the HJB equation simply becomes:
\begin{equation}
\beta V^* = \min\limits_{\vu \in \vU} \{L(\vx,\vu)+\left(\frac{\partial V^{*}}{\partial \vx}\right)^T \vf(\vx,\vu)\}
\label{eq:HJB_infinite_horizon}
\end{equation}

\subsection{Finite time horizon, stochastic system}
\label{subsec:HJP_fin_sto}
\paragraph*{Problem Definition}
We will now consider a continuous time, stochastic system of the form
\begin{equation}
\dot{\mathbf \vx}(t) = \vf_t(\vx(t),\vu(t)) + \mathbf B (t)\vw(t) \condition{$\vx(0)=\vx_0$}
\label{eq:stochastic_cont_time_system_eq}
\end{equation}
We assume that the stochasticity of the system can be expressed as additive white noise with mean and covariance given by:
\begin{align}
& E[\mathbf w(t)] = \bar{\mathbf w} = 0 \\
& E[\mathbf w(t) \mathbf w(\tau)^T] = \mathbf W(t) \delta(t-\tau)
\label{eq:Noise model}
\end{align}

The Dirac delta in the covariance definition signifies that the noise in the
system is uncorrelated over time. Unless $t=\tau$, the covariance will equal zero
($E[\mathbf w(t) \mathbf w(\tau)^T]=0$).

Since the system is not deterministic, the cost function is defined as the
expected value of the cost used in (\ref{eq:continuous_time_deterministic cost
function}). Once again, $\beta \in [0,+\infty)$ is the decay or discount factor.
\begin{equation}
J=E \left\{e^{-\beta (t_f-t_0)} \Phi(\vx(t_f))+ \int_{t_0}^{t_f} {e^{-\beta (t'-t_0)} L(\vx(t'),\vu(t'))dt'} \right\}
\label{eq:stochastic_continuous_time_deterministic cost function}
\end{equation}

\paragraph*{HJB equation\footnote{The following derivation can be found in Section 5.1
of \cite{Stengel1986}}} In order to obtain an
expression for the optimal cost-to-go starting from some $\vx$ at some time $t \in
[t_0,t_f]$, we need to evaluate the cost function over the optimal trajectory,
$\vx^*(t)$, using the optimal control policy, $\vu^*(t)$, during the remaining time
interval
$[t,t_f]$.
\begin{equation}
V^*(t,\vx)= E \left\{e^{-\beta (t_f-t)} \Phi(\vx^*(t_f)) +  \int_{t}^{t_f} {e^{-\beta (t'-t)} L (\vx^*(t'),\vu^*(t'))dt'} \right\},
\label{eq:optimal_cost_stochastic_cont_time}
\end{equation}
Taking the total time derivative of $V^*$ with respect to $t$ gives the following
\begin{align}
\label{eq:time_deriv_optimal_stochastic_cost}
\frac{dV^*(t,x)}{dt} & = E \left\{ \beta e^{-\beta (t_f-t)} \Phi(\vx^*(t_f)) + \beta \int_{t}^{t_f} {e^{-\beta (t'-t)} L (\vx^*(t'),\vu^*(t'))dt'} - L (\vx,\vu^*(t)) \right\} \notag \\
& = \beta V^*(t,\vx) - E \left\{L (\vx,\vu^*(t)) \right\}
\end{align}
note that $\Phi(\vx^*(t_f))$ is independent of $t$, and $t$ only occurs in the
lower limit of the integral in (\ref{eq:optimal_cost_stochastic_cont_time}).
Since $L (\vx^*(t),\vu^*(t))$ is only a function of the optimal trajectory and
control input at initial time $t$, it is known with certainty (i.e. the system noise
has no impact on its value at any point in time). This means that we can remove
the expectation to give the following, which will be used again at the end of the derivation.
\begin{equation}
\frac{dV^*(t,\vx)}{dt} = \beta V^*(t,\vx) -L (\vx,\vu^*(t))
\label{eq:time_deriv_optimal_stochastic_cost_2}
\end{equation}
We can express the incremental change of $V^*$ over time by its Taylor series
expansion, considering that $V^*$ depends on $t$ both directly, and indirectly
through its dependence on $\vx(t)$.
\begin{align}
\Delta V^*(t,\vx) &\approx \frac{dV^*(t,\vx)}{dt} \Delta t \notag \\
&= E \left\{     
\frac{\partial V^*(t,\vx)}{\partial t} \Delta t  
+ \left( \frac{\partial V^*(t,\vx)}{\partial \vx} \right)^T \dot{\vx} \Delta t
+ \frac{1}{2} \dot{\vx}^T \frac{\partial^2 V^*(t,\vx)}{\partial \vx^2} \dot{\vx} \Delta t^2 \right\}.
\label{eq:taylor_exp_stochastic_opt_cost_1}
\end{align}
Note that the expectation must be taken here because of the appearance of
$\dot{\vx}$, which depends on the system noise. 
All second-order and higher terms can be dropped from the expansion since the
value of $\Delta t^2$ is negligible as $\Delta t \to 0$. The last term on the
right is kept, though, because as you will see shortly, the second partial
derivative includes the covariance of our system noise, which has been modeled as a
Dirac delta function (equation~\ref{eq:Noise model}).

We can now plug in the system equation~(\ref{eq:stochastic_cont_time_system_eq}),
using the simplified notation $\mathbf{f} \vcentcolon = \mathbf{f}_t(\vx(t),\vu(t))$.
In addition, $V_\vz^*\vcentcolon = \frac{\partial V^*}{\partial \vz}$ is used to
simplify notation of the partial derivatives.
\begin{equation}
\frac{dV^*}{dt} \Delta t = E [ V_t^* \Delta t + V_\vx^{*T} (\vf + \mathbf{Bw}) \Delta t + \frac{1}{2} (\mathbf{f +Bw})^T V_{\mathbf {xx}}^*(\mathbf{f + Bw}) \Delta t^2]
\end{equation}

We can pull the first term out of the expectation since $V_t^*$ only depends on
$\vx$, which is known with certainty. We can also pull the second term out of
the expectation, and use $E[\mathbf w (t)]=0$ to simplify it. The third term is
replaced by its trace, since the trace of a scalar is equal to itself. This will
be useful in the next step. Dividing both sides by $\Delta t$, the time
derivative becomes:
\begin{equation}
\frac{dV^*}{dt}= V_t^* + V_\vx^{*T} \mathbf f + \frac {1}{2} \mathtt{Tr} \{E[(\mathbf{f+Bw})^T V_{\vx\vx}^*(\mathbf{f+Bw})] \Delta t \}.
\end{equation}

Using the matrix identity $\mathtt{Tr}[\vA\vB]=\mathtt{Tr}[\vB\vA]$, we rearrange
the terms inside of the expectation. Then, since $V_{xx}^*$ only depends on $x$,
which is known without uncertainty, we can remove it from the expectation.
\begin{equation}
\frac{dV^*}{dt}= V_t^* + V_\vx^{*T} \mathbf f + \frac {1}{2} \mathtt{Tr} \left[V_{\vx\vx}^* E[(\mathbf{f+Bw}) (\mathbf{f+Bw})^T] \Delta t \right].
\end{equation}
Expanding the terms inside of the expectation, and removing all terms known with certainty from the expectation gives:
\begin{equation}
\frac{dV^*}{dt}= V_t^* + V_\vx^{*T} \mathbf f + \frac {1}{2} \mathtt{Tr} \left[V_{\vx\vx}^* \bigg(  \vf\vf^T \Delta t + 2\mathbf{f}E(\mathbf{w}^T)\mathbf{B}^T \Delta t + \mathbf B E(\mathbf{ww}^T) \mathbf B^T \Delta t \bigg) \right].
\end{equation}
After plugging in our noise model (equation \ref{eq:Noise model}), the second
term in $\mathtt{Tr()}$ drops out and the last term simplifies to give
\begin{equation}
\frac{dV^*}{dt}= V_t^* + V_\vx^{*T} \mathbf f + \frac {1}{2} \mathtt{Tr} \left[V_{\vx\vx}^* \bigg( \vf\vf^T\Delta t + \mathbf{BW} \mathbf B^T \delta(t) \Delta t \bigg) \right].
\end{equation}
Assuming that $\lim\limits_{\Delta t \to 0} \delta (t) \Delta t = 1$, and taking
the limit as  $\Delta t \to 0$,
\begin{equation}
\frac{dV^*}{dt}= V_t^* + V_\vx^{*T} \mathbf f + \frac {1}{2} \mathtt{Tr} \left[V_{\vx\vx}^* \mathbf{BWB}^T \right].
\end{equation}
Plugging equation~\eqref{eq:time_deriv_optimal_stochastic_cost_2} into the left
side, and indicating that the optimal cost requires minimization over $\vu(t)$
gives the stochastic principle of optimality
\begin{equation}
\beta V^*(t,\vx) - V_t^*(t,\vx) = \min_{\vu(t)} \left\{L(\vx, \vu(t)) + V_\vx^{*T} \vf_t (\vx,\vu(t)) + \frac{1}{2} \mathtt{Tr} [V_{\vx\vx}^* \vB(t) \vW(t) \vB^T (t)]  \right\}
\label{eq:stchastic_principle_of_optimality}
\end{equation}
with the terminal condition,
\begin{equation}
V^*(t_f,\vx) = \Phi(\vx).
\end{equation}

\subsection{Infinite time horizon, stochastic system}
\paragraph*{Problem Definition}
Finally, consider a stochastic time-invariant system similar to the form in
equation~\eqref{eq:stochastic_cont_time_system_eq} over an infinite time horizon.
The cost function is similar to the cost function in
equation~\eqref{eq:continuous_time_deterministic_inf_horizon_cost function}, 
but we now must consider the expected value of the cost due to the stochasticity in the system.
\begin{equation}
J = E \left\{  \int_{t_0}^{\infty} {e^{-\beta (t-t_0)} L(\vx(t),\vu(t))dt }\right\} ,
\label{eq:continuous_time_stochastic_inf_horizon_cost function}
\end{equation}
\paragraph*{HJB equation}
Once again, over an infinite time horizon, the value function is no longer a
function of time. As a result, the second term in the left hand side of equation
\ref{eq:stchastic_principle_of_optimality} equals zero, and the HJB equation
becomes
\begin{equation}
\beta V^*(\vx)= \min_{\vu(t)} \left\{L(\vx, \vu(t)) + V_\vx^*(x) \vf_t (\vx,\vu(t)) + \frac{1}{2} \mathtt{Tr} [V_{\vx\vx}^* \vB(t) \vW(t) \vB^T (t)]  \right\}
\label{eq:stchastic_inf_horizon_principle_of_optimality}
\end{equation}

\newpage

\section{Summary of Results}
In the following table, a summary of the results in the two preceding sections is presented.
\begin{table}[H]
\centering
\begin{tabular}{|c|l|l|r|}
\hline
& \multicolumn{1}{c|}{\textbf{Discrete Time}} & \multicolumn{1}{c|}{\textbf{Continuous Time}} \\
\hline
\parbox[t]{2mm}{\multirow{19}{*}{\rotatebox[origin=c]{90}{\textbf{Stochastic System}}}} &  &\\
& Optimization Problem: & Optimization Problem: \\
&$\vx_{n+1}=\vf(\vx_n,\vu_n)+\vw_n$& $d\vx=\vf(\vx_t,\vu_t)dt + \vB(\vx_t,\vu_t)d\vw_t\hspace{23mm}$\\

& $\vw_n \sim P_{\vw}(\cdot \mid \vx_n,\vu_n)$ & $\vw_t \sim \cN(0,\Sigma)$\\
& $\min\limits_{\vu_{0 \to N-1}} E \{\alpha^N \Phi(N)+\sum\limits_{k=0}^{N -1} \alpha^k L(\vx_k,\vu_k)  \} \hspace{4mm} \alpha \in [0,1]$ & $\min\limits_{\vu_0 \to t_f} E\{e^{-\beta t_f} \Phi(t_f)+ \int_0^{t_f} {e^{-\beta t} L(\vx_t,\vu_t)dt}\}$\\
& Stochastic Bellman equation: & Stochastic HJB equation:\\
& $\begin{aligned}
V^*(n,\vx)= &\min\limits_{\vu_n} \{L(x,u_n)\\&+ \alpha E[V^*(n+1,\vx_{n+1})] \}
\end{aligned}$ 
& $\begin{aligned} 
\beta V^*(t,\vx) & -V_t^*(t,\vx) = \min\limits_{\vu_t}\{ L(\vx,\vu_t) + \\
& V_{x}^{*T}(t,\vx) \vf(\vx,\vu_t) + \frac{1}{2} Tr [V_{\vx\vx}^*(t,\vx)\vB \Sigma \vB^T] \} 
\end{aligned}$\\

&\tikzmark[xshift=18.5em]{i}&\tikzmark[xshift=1em]{j}\\
& Infinite horizon: \hspace{5 mm}  $\alpha \in [0,1)$ &   Infinite horizon:\\
& $\Phi(N)=0 \hspace{10mm} V^*$ is not function of time.\tikzmark[xshift=0.5em]{a}& $\Phi(t_f)=0 \hspace{10mm} V^*$ is not function of time.\tikzmark[xshift=0.6em]{c}\\
&&\\
\hline
\parbox[t]{2mm}{\multirow{16}{*}{\rotatebox[origin=c]{90}{\textbf{Deterministic System}}}} &  &\\
& Optimization Problem: & Optimization Problem:\\
&  $\vx_{n+1}=\vf(\vx_n,\vu_n)$ \tikzmark[xshift=10em]{b}& $d\vx=\vf(\vx_{(t)},\vu_{(t)})dt $\tikzmark[xshift=10.4em]{d}\\

& $\min\limits_{\vu_{0 \to N-1}} \{\alpha^N \Phi(N)+\sum\limits_{k=0}^{N -1} \alpha^k L(\vx_k,\vu_k)  \} \hspace{5mm} \alpha \in [0,1]$ 
& $\min\limits_{\vu_{0 \to t_f}} \{e^{-\beta t_f} \Phi(t_f)+ \int_0^{t_f} {e^{-\beta t} L(\vx_t,\vu_t)dt}\}$\\
&  Bellman equation: & HJB equation:\\
& $\begin{aligned}V^*(n,\vx)= &\min\limits_{u_n} \{L(\vx,\vu_n)\\&+\alpha V^*(n+1,\vx_{n+1}) \}\end{aligned}$ 
&$ \begin{aligned} \beta V^*(t,\vx) - V_t^*(t,\vx)=&\min\limits_{\vu_{(t)}}\{L(\vx_t,\vu_t)\\&+ V_\vx^{*T}(t,\vx) \vf(\vx,\vu)\} \end{aligned}$\\

&\tikzmark[xshift=18.5em]{e}&\tikzmark[xshift=1em]{f}\\
&Infinite time horizon: \hspace{5 mm}  $\alpha \in [0,1)$& Infinite time horizon:\\
& $\Phi(N)=0\hspace{10mm} V^*$ is not function of time.& $\Phi(t_f)=0 \hspace{10mm} V^*$ is not function of time.\\
&&\\
\hline
\end{tabular}
\drawcurvedarrow[bend left=30,-stealth, red, double distance=1pt]{a}{b}
\drawcurvedarrow[bend left=30,-stealth, red, double distance=1pt]{c}{d}
\drawcurvedarrow[bend left=60,-stealth, red, double distance=1pt]{e}{f}
\drawcurvedarrow[bend left=60,-stealth, red, double distance=1pt]{i}{j}
\annote[above left]{arrow-1}{{$\vw_n \sim P_{\vw}(\cdot \mid \vx_n,\vu_n)=\delta (\vw)$}}
\annote[above left]{arrow-2}{{$\vw_t \sim \cN(0,\mathbf{0})$ i.e. $\Sigma = \mathbf{0}$}}
\annote[above]{arrow-3}{{\sref{subsec:HJP_fin_det}}}
\annote[above]{arrow-4}{{\sref{subsec:HJP_fin_sto}}}
\end{table}

\newpage

\section {Iterative Algorithms}
In general, an optimal control problem with a nonlinear cost function and
dynamics does not have an analytical solution. Therefore numerical methods are
required to solve it. However, the computational complexity of the optimal
control problem scales exponentially with the dimension of the state space. Even
though various algorithms have been proposed in the literature to solve the
optimal control problem, most of them do not scale to the high dimension problems. In
this Section, we introduce a family of methods which approximate this complex
problem with a set of tractable sub-problems before solving it.

This section starts by introducing a method for solving optimization problems
with \emph{static constraints}. This is a simpler problem to solve than our optimal control problem, since system dynamics appear there as \emph{dynamic constraints}. We will then show the extension of this method to an optimal control problem where the
constraints are dynamic. Finally we will conclude the section by introducing an
algorithm which implements this idea.

\subsection{Sequential Quadratic Programming: SQP}
Assuming that $f(x)$ is a general nonlinear function, a problem of the form 
\begin{equation}
\begin{split}
&\min\limits_{x} f(x)    \hspace{2cm} x \in \mathbb{R}^n \\
& s.t. \hspace{1cm}   f_j(x) \leq 0,   \hspace{2cm}      j = 1,\dots,N \\
&      \hspace{1.5cm} h_j(x) =    0,   \hspace{2cm}      j = 1,\dots,N 
\end{split}
\label{eq:conv_opt}
\end{equation}
is called a \emph{nonlinear programming problem} with inequality and equality constraints. Setting the gradient of the Lagrangian function equal to zero and solving for $x$ will usually not return a closed form solution (e.g for $\nabla (x\sin x) = \sin x + x\cos x = 0$). However, there will always exist a closed form solution for the two following special cases:
\begin{itemize}
\item{Linear Programming (LP):} 
In LP, the function $f$ and all the constraints are linear w.r.t. $x$. An example algorithm for solving an LP is ``Dantzig's Simplex Algorithm''.
\item{Quadratic Programming (QP):}
In QP, the function $f$ is quadratic, but all the constraints are linear w.r.t.
$x$. Example algorithms for solving a QP are the ``Broyden–Fletcher–Goldfarb–Shanno'' (BFGS) algorithm and the ``Newton-Raphson Method''.
\end{itemize} 

One way to approach a general nonlinear problem of the form \eqref{eq:conv_opt}, is to iteratively approximate it by a QP. We start by guessing a solution $\tilde{x}_0$. Then we approximate $f(x)$ by its first and second order Taylor expansion around this initial point. We also approximate all the constraints by their first order Taylor expansion around this point. 
\begin{equation}
\newcommand{\fo}{f(\tilde{x}_0)}
\newcommand{\fj}{f_j(\tilde{x}_0)}
\newcommand{\hj}{h_j(\tilde{x}_0)}
\newcommand{\xmx}{(x - \tilde{x}_0)}
\begin{split}
&f(x) \approx \fo + \xmx^T \nabla\fo + \frac{1}{2} \xmx^T \nabla^2\fo \xmx       \\
&f_j(x) \approx \fj + \xmx^T \nabla\fj                                     \\
&h_j(x) \approx \hj + \xmx^T \nabla\hj                                     \\ 
\end{split}
\end{equation}

This problem can now be solved by one of the QP solvers mentioned previously to
obtain a new solution $\tilde{x}_1$. We then iteratively approximate the
objective function and the constraints around the new solution and solve the QP
problem. It can be shown that for a convex problem this algorithm converges to
the optimal solution as $\lim_{i\to \infty} \tilde{x}_i = x^*$.

\paragraph*{Example of a SQP solver - the Newton-Raphson Method:}
\label{subsec:newton} One method for iteratively solving a SQP, e.g. finding the
zeros of the function $f'(x)$ if no closed form solution exists\footnote{for
the sake of simplicity we have omitted all the constraints in this example.}, is
the \emph{Newton-Raphson Method}. Consider the following update law: %
\begin{equation} \label{eq:newton_method} 
\tilde{x}_1 = \tilde{x}_0 - \frac{f'(\tilde{x}_0)}{f''(\tilde{x}_0)}
\end{equation} 
Starting from an initial guess $\tilde{x}_0$, and knowing the first derivative
$f'(\tilde{x}_0)$ and second derivative $f''(\tilde{x}_0)$ at that specific
point $\tilde{x}_0$, a better approximation of our optimization problem is given
by $\tilde{x}_1$. This procedure is repeated using the new approximation
$\tilde{x}_1$ until convergence. %
\begin{figure}[htb]
  \centering  
\subfigure{\resizebox{0.5\columnwidth}{!}{
\includegraphics{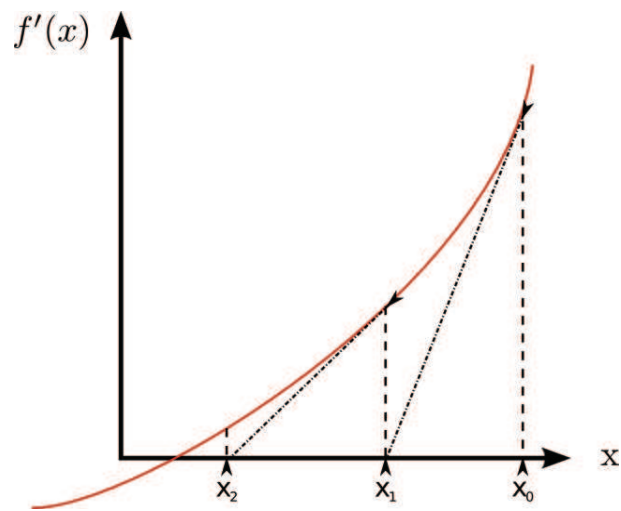}}}
  \caption{Newton-Raphson Method for finding the zeros (x-axis intersection) 
   of a function $f'(x)$ shown in red, which is equivalent to finding the 
   extreme points of $f(x)$.} 
  \label{fig:newton_method}
\end{figure}  

Note that for finding the minimum of a quadratic function $f(x) = 0.5ax^2+bx + c$, the Newton-Raphson
Method will find the minimizer $x^*$ of the function in only one iteration
step, independent of the initial guess $x_0$:
\begin{equation} \label{eq:newton_method_lin} 
\newcommand{\xt}{\tilde{x}_0}
x^* = \xt - \frac{f'(\xt)}{f''(\xt)} = \xt - \frac{a\xt+b}{a} = - \frac{b}{a}
\end{equation} 

\subsection{Sequential Linear Quadratic Programming: SLQ}
\label{Sequential Linear Quadratic Programming: SLQ}
Sequential Linear Quadratic Programming (SLQ) is a family of algorithms for solving optimal control problem involving a non-linear cost function and non-linear system dynamics. In discrete
time this problem is defined as follows
\begin{align}
\label{eq:slqp_cost_function} 
&\min\limits_{\mu} \left[ \Phi(\vx(N))+\sum\limits_{n=0}^{N-1} L_n(\vx(n),\vu(n)) \right] \\
\label{eq:slqp_dynamics} 
& s.t. \hspace{1cm} \vx(n+1) = \vf(\vx(n),\vu(n)) \hspace{1.5cm} \vx(0) = \vx_0\\
&\hspace{1.5cm} \vu(n,\vx) = \mu(n,\vx) \nonumber
\end{align}
SLQ methods are a class of algorithms which are based on the idea of fitting
simplified subproblems over the original problem. Assuming that solving the
optimality conditions is easier for these subproblems, the solution can be
improved iteratively by optimizing over these subproblems. 

Since linear-quadratic problems are almost the most difficult problems
which have a closed form solution, the subproblems are chosen to be
linear-quadratic. One can see the similarity between SLQ and SQP through the way
that the original problem is decomposed, and the way subproblems are chosen.
In both algorithms the solution of the primary problem is derived by iteratively
approximating a linear-quadratic subproblem around the latest update of the
solution and optimizing over this subproblem respectively. A general overview of 
the SLQ algorithm is as follows:
\begin{enumerate} 
\item
Guess an initial (stabilizing) control control policy $\mu^0(n,x)$. 
\item \label{it:roll_out} 
``Roll out'': Apply the control policy to the non-linear system
\eqref{eq:slqp_dynamics} (forward integration), which yields the state trajectory
$\mathbf{X}_k = \{\vx(0), \vx(1), \dots, \vx(N)\}$ and input trajectory
$\mathbf{U}_k=\{\vu(0), \vu(1), \dots, \vu(N-1)\}$.
\item \label{it:appr_val} 
Starting with $n=N-1$, approximate the value function as a quadratic function
around the pair $(\vx(N-1), \vu(N-1))$.
\item \label{it:solve_lqr} 
Having a quadratic value function, the Bellman equation can be solved efficiently.
The output of this step is a control policy at time $N-1$ which minimizes the
quadratic value function.
\item \label{it:backward_pass} 
``Backward pass'': Repeat steps \ref{it:appr_val} and \ref{it:solve_lqr} for
every state-input pair along the trajectory yielding $\delta \mu^k=\{\bar{\vu}(0,\vx)
\dots, \bar{\vu}(N-1,\vx)\}$. The updated optimized control inputs are then
calculated with an appropriate step-size $\alpha_k$ from
\begin{equation}
\mu^{k+1} = \mu^k + \alpha_k \cdot \delta \mu^k
\end{equation} 
\item \label{it:iteration} Iterate through steps \ref{it:roll_out} $\to$
\ref{it:backward_pass} using the updated control policy $\mu^{k+1}$ until
a termination condition is satisfied, e.g. no more cost improvement or no 
more control vector changes.
\end{enumerate}

Notice that if our system dynamics is already linear, and the cost function
quadratic, then only one iteration step is necessary to find the globally optimal
solution, similar to \eqref{eq:newton_method_lin}. In this case the SLQ
controller reduces to a LQR controller. In the next section, a member of the SLQ
algorithm family, called the Iterative Linear Quadratic Controller
(ILQC), will be introduced.

\section{Iterative Linear Quadratic Controller: ILQC}
\label{ILQC}
The ILQC is an iterative Linear-Quadratic method for locally-optimal feedback
control of nonlinear, deterministic discrete-time systems. Given an initial,
feasible sequence of control inputs, we iteratively obtain a local
linear approximation of the system dynamics and a quadratic approximation of the
cost function, and then an incremental improvement to the control law, until we reach convergence at a local minimum of our cost function.

\subsection{ILQC Problem statement}
Similar to Section~\ref{subsec:prob_fin_det}, we consider the discrete-time, finite-horizon, nonlinear dynamic system
\begin{equation}
\mathbf x_{n+1}=\mathbf f_n(\mathbf x_n,\mathbf u_n),\hspace{10mm} \mathbf x(0)=\mathbf x_0, \hspace{10mm} n \in \{0,1,...,N-1\}
\label{eq:nonlinear_dt_system}
\end{equation}
with state-vector $\mathbf{x}_n$ and control input vector $\mathbf{u}_n$. Let the cost function be
\begin{equation}
J=\Phi(\mathbf x_{N})+\sum_{n=0}^{N-1} L_n (\mathbf x_n,\mathbf u_n) \ \text{,}
\end{equation}
which corresponds to
Equation~\eqref{eq:discrete_time_deterministic_system_cost_function} with $\alpha
= 1$. We denote $L_n(\mathbf{x}_n,\mathbf{u}_n)$ the intermediate, non-negative
cost rate at time-step $n$ and $\Phi(\mathbf{x}_N)$ the terminal, non-negative
cost at time-step $N$. Therefore, the cost function $J$ gives the undiscounted
cost associated with a particular trajectory starting from state $\mathbf x_{0}$
at time step $n=0$ up to the final state $\mathbf x_N$ at time step $n=N$ under a
deterministic control policy
\begin{equation}
\vect \mu = \{ \mathbf u_0,\mathbf u_1,...,\mathbf u_{N-1}
\} \ \text{.} 
\end{equation}

The general goal of optimal control is to find an optimal policy $\vmu^*$
that minimizes the total cost $J$. Finding the optimal controls in a general,
nonlinear setting by solving the Bellman Equation~\eqref{eq:bellman_det_opt} is
typically infeasible, because there is no analytic solution for the Bellman
Equation if the system is nonlinear and the cost-function is non-quadratic.
Instead, we aim at finding a locally optimal control law which approximates
$\vmu^*$ in the neighborhood of a local minimum.

\subsection{Local Linear-Quadratic Approximation}
\label{sec:local_linear_quadratic_approx}
The locally-optimal control law is constructed in an iterative way. In each
iteration, we begin with a stable control policy $\vmu(n,\vx)$. Starting at
initial condition $\mathbf{\bar{x}}(0)=\mathbf{x}_0$, the corresponding nominal
state-trajectory $\{\bar{\vx}_n\}$ and control input trajectory $\{\bar{\vu}_n\}$
for the nonlinear system can be obtained by forward integration of
Equation~\eqref{eq:nonlinear_dt_system} using the policy $\vmu$. 

Now, we linearize the system dynamics and quadratize the cost function around
every pair ($\bar{\mathbf x}_n, \bar{\mathbf u}_n$). To do so, we introduce
the state and control input increments as follows
\begin{align}
\label{eq:ILQC_state_control}
\delta \mathbf x_n &\triangleq \mathbf x_n - \bar{\mathbf x}_n \notag \\
\delta \mathbf u_n &\triangleq \mathbf u_n - \bar{\mathbf u}_n  
\end{align}
Since $\{\bar{\mathbf x}_n\}$ and $\{\bar{\mathbf u}_n\}$ are satisfying the system dynamics in equation~\eqref{eq:nonlinear_dt_system}, we will have $\delta\vx(0) = 0$. 

For linearizing the system dynamics we substitute $\vx_n$ and $\vu_n$ by the definitions in
equation~\eqref{eq:ILQC_state_control} and then approximate $\vf$ by its first order Taylor expansion
\begin{align}
\bar{\mathbf x}_{n+1} + \delta \mathbf x_{n+1} &= \mathbf f_n (\bar{\mathbf x}_n + \delta \mathbf x_n, \bar{\mathbf u}_n+ \delta \mathbf u_n) \notag \\
&\approx \vf_n(\bar{\vx}_n,\bar{\vu}_n) + \frac{\partial \vf(\bar{\vx}_n,\bar{\vu}_n)}{\partial\vx} \delta\vx_n+ \frac{\partial \vf(\bar{\vx}_n,\bar{\vu}_n)}{\partial\vu} \delta\vu_n
\end{align}
Using $\bar{\vx}_{n+1}=\vf_n(\bar{\vx}_n,\bar{\vu}_n)$ to simplify the approximation, we obtain
\begin{align}
\label{eq:lin_dynamics}
\delta \vx_{n+1} &\approx \vA_n \delta\vx_n+ \vB_n \delta\vu_n \\
\vA_n &= \frac{\partial\vf(\bar{\vx}_n,\bar{\vu}_n)}{\partial\vx} \notag \\
\vB_n &= \frac{\partial\vf(\bar{\vx}_n,\bar{\vu}_n)}{\partial\vu}
\end{align}
where $\vA_n$ and $\vB_n$ are independent of $\delta\vx_n$ and $\delta\vu_n$. Notice that as long as the nominal trajectories $\bar{\vx}_n$ and $\bar{\vu}_n$ are time dependent, $\vA_n$ and $\vB_n$ are time varying. Therefore, the linear approximation transforms a \textbf{nonlinear} (either time-variant or time-invariant) system into a \textbf{linear time-variant} system.

We wish also to quadratize the cost function with respect to the nominal state and control trajectories.
\begin{align}
\label{eq:quad_cost}
J \approx &q_N +\vdx_N^T\vq_N+\frac{1}{2}\vdx_N^T\vQ_N\vdx_N \notag \\
	&+\sum_{n=0}^{N-1}{\{q_n+ \vdx_n^T\vq_n + \vdu_n^T \vr_n+ 	\frac{1}{2}\vdx_n^T\vQ_n\vdx_n+\frac{1}{2}\vdu_n^T\vR_n\vdu_n + \vdu_n^T\vP_n\vdx_n\}}
\end{align} 
where the cost function elements are defined as
\begin{align}
\label{eq:quad_cost_elements}
& \forall n \in \{0,\cdots,N-1\}: \notag \\
& q_n = L_n(\bar{\vx}_n,\bar{\vu}_n), \hspace{10mm} 
\vq_n = \frac{\partial L(\bar{\vx}_n,\bar{\vu}_n)}{\partial\vx}, \hspace{10mm}
\vQ_n = \frac{\partial^2 L(\bar{\vx}_n,\bar{\vu}_n)}{\partial\vx^2} \notag \\
& \vP_n = \frac{\partial^2 L(\bar{\vx}_n,\bar{\vu}_n)}{\partial\vu \partial\vx}, \hspace{6mm} 
\vr_n = \frac{\partial L(\bar{\vx}_n,\bar{\vu}_n)}{\partial\vu}, \hspace{10mm}
\vR_n = \frac{\partial^2 L(\bar{\vx}_n,\bar{\vu}_n)}{\partial\vu^2} \\
& n = N: \notag \\
& q_N = \Phi(\bar{\vx}_N), \hspace{17mm} 
\vq_N = \frac{\partial \Phi(\bar{\vx}_N)}{\partial\vx}, \hspace{15mm}
\vQ_N = \frac{\partial^2 \Phi(\bar{\vx}_N)}{\partial\vx^2} \notag 
\end{align}
Note that all derivatives w.r.t. $\vu$ are zero for the terminal time-step
$N$. Using the above linear-quadratic approximation to the original problem, we
can derive an \emph{approximately optimal} control law.

\subsection{Computing the Value Function and the Optimal Control Law}
In this section we will show that, if the value function (cost-to-go function) is quadratic in $\delta \mathbf x_{n+1}$ for a certain time-step $n+1$, it will stay in quadratic form during back-propagation in time, given the linear-quadratic approximation presented in Equations~\eqref{eq:lin_dynamics} to~\eqref{eq:quad_cost_elements}.

Now, suppose that for time-step $n+1$, we have the value function of the state deviation $\delta \mathbf x_{n+1}$ given as a quadratic function of the form 
\begin{equation}
\label{eq:cost_to_go}
V^*(n+1,\delta \mathbf x_{n+1}) = s_{n+1} + {\delta \mathbf x_{n+1}^T \mathbf s_{n+1}} + \frac{1}{2} \delta \mathbf x_{n+1}^T \mathbf S_{n+1} \delta \mathbf x_{n+1} \ \text{.}
\end{equation}
We can write down the Bellman Equation for the value function at the previous time-step $n$ as
\begin{equation*}
V^*(n,\delta \mathbf{x}_n) = \min_{\vu_n} \left[ L_n(\mathbf{x}_n,\mathbf{u}_n)+V^*(n+1,\delta \mathbf{x}_{n+1}) \right]
\end{equation*}
Assuming that a control input $\delta \mathbf u_n $ is given from a policy $\vect \mu$ and plugging in Equation~\eqref{eq:quad_cost} leads to 
\begin{align*}
V^*(n,\delta \mathbf{x}_n) &= \min_{\vu_n} \bigg[ q_n + \delta \mathbf x_n^T (\mathbf q_n + \frac{1}{2} \mathbf Q_n \delta \mathbf x_n) + \delta \mathbf u_n^T( \mathbf r_n + \frac{1}{2} \mathbf R_n \delta \mathbf u_n) + \delta \mathbf u_n^T \mathbf P_n \delta \mathbf x_n \\
&\quad + V^*(n+1, \mathbf A_n \delta \mathbf x_n + \mathbf B_n \delta \mathbf u_n ) \bigg]
\end{align*}
Using equation~\eqref{eq:cost_to_go} for  $V(n+1, \mathbf A_n \delta \mathbf x_n + \mathbf B_n \delta \mathbf u_n )$ and re-grouping terms results in
\begin{align}
\label{eq:ILQC_V_star_intermediate}
V^*(n,\delta \mathbf{x}_n) = \min_{\vu_n} \bigg[ & q_n + s_{n+1} + \delta \mathbf x_n^T (\mathbf q_n + \mathbf A_n^T \mathbf s_{n+1}) \\
& + \frac{1}{2} \delta \mathbf x_n^T (\mathbf Q_n +\mathbf A_n^T \mathbf S_{n+1} \mathbf A_n) \delta \mathbf x_n 
+ \highlight{\delta \mathbf u_n^T (\mathbf g_n + \mathbf G_n \delta \mathbf x_n) + \frac{1}{2} \delta \mathbf u_n^T \mathbf H_n \delta \mathbf u_n} \bigg] \notag
\end{align}
where we have defined the shortcuts
\begin{align}
\label{eq:g_G_H}
\mathbf g_n & \triangleq \mathbf r_n + \mathbf B_n^T \mathbf s_{n+1} \notag\\
\mathbf G_n & \triangleq \mathbf P_n + \mathbf B_n^T \mathbf S_{n+1} \mathbf A_n \notag\\
\mathbf H_n & \triangleq \mathbf R_n + \mathbf B_n^T \mathbf S_{n+1} \mathbf B_n
\end{align}
for each time-step $n$. At this point we notice that the expression for
$V^*(n,\delta \mathbf{x}_n)$ is a quadratic function of $\mathbf u_n$ (see
highlighted box in equation~\ref{eq:ILQC_V_star_intermediate}). In order to
minimize $V^*(n,\delta \mathbf{x}_n)$, we set $\mathbf u_n$ to the value which
makes its gradient w.r.t. $\mathbf u_n$ vanish. Therefore, we obtain the
optimal control law for this update step as
\begin{equation}
\label{eq:ILQC_optimal_control_input}
\delta \mathbf u_n = -\mathbf H_n^{-1} \mathbf g_n - \mathbf H_n^{-1} \mathbf G_n \delta \mathbf x_n
\end{equation}
It can be seen that the optimal control input consists of a feed-forward term
$\delta \vu_n^{ff} = -\mathbf H_n^{-1} \mathbf g_n$ and a feedback term 
$\mathbf K_n \delta \vx_n$ with feedback gain matrix 
$\mathbf K_n:=-\mathbf H_n^{-1} \mathbf G_n$. Replacing 
$\delta \mathbf u_n$ by 
$\delta \mathbf u_n= \delta \vu_n^{ff} + \mathbf K_n \delta \vx_n$, 
the expression which is highlighted in equation~\eqref{eq:ILQC_V_star_intermediate} becomes
\begin{align}
\label{eq:ILQC_quadr_in_x}
 &\delta {\mathbf u^{ff}}^T \mathbf g + \frac{1}{2} \delta {\mathbf u^{ff}}^T \mathbf H \delta \mathbf u^{ff} + \delta \mathbf x^T (\mathbf G^T \delta \mathbf u^{ff} + \mathbf K^T \mathbf g + \mathbf K^T \mathbf H \delta \mathbf u^{ff}) \\
 &+ \frac{1}{2} \delta \mathbf x^T ( \mathbf K^T \mathbf H\mathbf K + \mathbf K^T\mathbf G + \mathbf G^T \mathbf K) \delta \mathbf x \notag
\end{align}
where all time-step indices have been omitted for readability. Obviously,
Equation~\eqref{eq:ILQC_quadr_in_x} is quadratic in $\delta \mathbf x$ and
therefore the whole value-function remains quadratic in $\delta \mathbf x$
throughout the backwards step in time. Thus, assuming the terminal cost is also
approximated quadratically w.r.t $\delta \mathbf x$, the value-function remains
quadratic for all $n$.

By plugging \eqref{eq:ILQC_quadr_in_x} back into
equation~\eqref{eq:ILQC_V_star_intermediate}  and using the quadratic assumption for value
function introduced in \eqref{eq:cost_to_go} to replace the left hand
side of \eqref{eq:ILQC_V_star_intermediate}, we will obtain a quadratic
functional equation of $\delta \vx$. This functional should always remain
identical to zero for any arbitrary $\delta \vx$. Thus we can conclude, all the
coefficients of this quadratic functional should be identical to zero which will
lead to the following recursive equation for $\vS_n$, $\vs_n$, and $s_n$.
\begin{align}
\label{eq:ilqc_sm}
\vS_n & = \mathbf Q_n + \mathbf A_n^T \mathbf S_{n+1} \mathbf A_n + \mathbf K_n^T \mathbf H_n \mathbf K_n + \mathbf K_n^T \mathbf G_n + \mathbf G_n^T \mathbf K_n  \\
\label{eq:ilqc_sv}
\vs_n & = \mathbf q_n + \mathbf A_n^T \mathbf s_{n+1} + \mathbf K_n^T \mathbf H_n \delta \mathbf u_n^{ff} + \mathbf K_n^T \mathbf g_n + \mathbf G_n^T \delta \mathbf u_n^{ff} \\
\label{eq:ilqc_s}
s_n & = q_n + s_{n+1} + \frac{1}{2}\delta \mathbf {u_n^{ff}}^T \mathbf H_n \delta \mathbf u_n^{ff} + \delta {\mathbf u_n^{ff}}^T \mathbf g_n
\end{align}
which is valid for all $n \in \{0,\cdots,N-1\}$. For the final time-step $N$ we have the following terminal conditions:
\begin{equation}
\label{eq:ilqc_terminal_condition}
\vS_N = \vQ_N, \hspace{10mm} \vs_N = \vq_N, \hspace{10mm}  s_N =  q_N
\end{equation}
Equation~\eqref{eq:ilqc_sm} to \eqref{eq:ilqc_s} should be solved backward in
time starting from the final time-step $N$ with the terminal conditions
~\eqref{eq:ilqc_terminal_condition}. As we are propagating backward in time, the
optimal policy will also be calculated through
equation~\eqref{eq:ILQC_optimal_control_input}. However we should notice that
this policy is in fact the incremental policy. Therefore for applying this policy
to the system we first should add the nominal control trajectory to this policy.
\begin{equation}
\label{eq:ilqc_optimal_policy}
\vu(n,x)= \bar{\vu}_n + \delta \vu_n^{ff} + \vK_n (\vx_n - \bar{\vx}_n)
\end{equation}

\subsection{The ILQC Main Iteration}
In this section, we summarize the steps of the ILQC algorithm:
\begin{itemize}
\item[0.]
\emph{Initialization:} we assume that an initial, feasible policy $\vect \mu$ and initial state $\mathbf x_0$ is given.
Then, for every iteration $(i)$: 
\item[1.]
\emph{Roll-Out:} perform a forward-integration of the nonlinear system dynamics~\eqref{eq:nonlinear_dt_system} subject to initial condition $\mathbf x_0$ and the current policy $\vect \mu$. Thus, obtain the nominal state- and control input trajectories $\bar{\mathbf u}_n^{(i)}, \bar{\mathbf x}_n^{(i)}$ for $n=0,1,\ldots,N$.
\item[2.]
\emph{Linear-Quadratic Approximation:}
build a local, linear-quadratic approximation around every state-input pair 
$(\bar{\mathbf u}_n^{(i)}, \mathbf{\bar x}_n^{(i)})$ as described in
Equations~\eqref{eq:lin_dynamics} to~\eqref{eq:quad_cost_elements}.
\item[3.]
\emph{Compute the Control Law:}
solve equations~\eqref{eq:ilqc_sm} to \eqref{eq:ilqc_s} backward in time and design the affine control policy through equation~\eqref{eq:ilqc_optimal_policy}.
\item[4.]
Go back to 1. and repeat until the sequences  $\bar{\mathbf u}^{(i+1)}$ and $\bar{\mathbf u}^{(i)}$ are sufficiently close.
\end{itemize}

\section {Linear Quadratic Regulator: LQR}
\label{sec: lqr_intro} In this section, we will study the familiar LQR for both
discrete and continuous time systems. LQR stands for \textbf{L}inear (as system
dynamics are assumed linear) \textbf{Q}uadratic (as the cost function is purely
quadratic w.r.t states and inputs) \textbf{R}egulator (since the states are
regulated to zero). As the naming implies, it is an optimal control problem for a
linear system with a pure quadratic cost function to regulate the system's state
to zero. The LQR problem is defined for both discrete and continuous time systems
in a deterministic setup.

This section starts by deriving the LQR solution for the discrete time case. To
do so, we will use the results obtained in \sref{ILQC} for the ILQC controller. The
similarity between these two problems readily becomes clear by comparing their
assumptions for the underlying problem. In \sref{ILQC}, we have seen that ILQC
finds the optimal controller by iteratively approximating the nonlinear problem
with a subproblem which has linear dynamics and quadratic cost function.
Therefore if the original problem has itself linear dynamics and quadratic cost,
the algorithm converges to the optimal solution at the fist iteration i.e.
the solution to the first subproblem is the solution to the original problem.
Furthermore since the LQR problem even has more restricted assumptions, the
algorithm can even be simplified more. In \sref{sec:
lqr_infinite_time_discrete_time}, the infinite time horizon LQR solution is
obtained by using the result we derive in finite time horizon case.

For the continuous time case, we go back to the original deterministic HJB
equation. Using the linear, quadratic nature of the LQR problem, we derive
an efficient algorithm to find the global optimal solution of the problem.
Finally we extend these results to the infinite time horizon
problem. Note that although we have chosen different approach for deriving
the discrete-time LQR solution, starting
from the original Bellman equation would give an identical solution.

\subsection{LQR: Finite time horizon, discrete time}
\label{sec: lqr_finite_time_discrete_time}
From the results derived in the previous section, we can easily derive the
familiar discrete-time Linear Quadratic Regulator (LQR). The LQR setting
considers very similar assumptions to those used in ILQC, but used in a much
stricter sense. It assumes that the system dynamics are linear. Furthermore it
assumes that the cost function only consists of pure quadratic terms (no bias and
linear terms). As discussed in \sref{Sequential Linear Quadratic Programming:
SLQ} for the case of linear system with quadratic cost function the SLQ algorithm
converges to the global optimal solution in the first iteration. Therefore in
order to derive the LQR controller, we just need to solve the first
iteration of the iLQC algorithm.

Here we consider a discrete time system with linear (or linearized) dynamics of
the form:
\begin{equation}
\mathbf \vx_{n+1} = \vA_n \mathbf \vx_n + \vB_n \mathbf \vu_n
\label{eq:LQR_system_equation}
\end{equation}
This is very similar to the linearized dynamics from
equation~\eqref{eq:lin_dynamics}, but note that here $\vA$ and $\vB$ are not seen as
local approximations around some nominal trajectories, as they were in
\sref{sec:local_linear_quadratic_approx}. We are also no longer representing the
state and control input as deviations from nominal trajectories, as in
~\eqref{eq:ILQC_state_control}. This is because in the LQR setting, we assume
that we are regulating the system to zero states, which also implies zero control
input since the system is linear and has a unique equilibrium point at the
origin, Therefore we have
\begin{align*}
\delta \mathbf x_{n} = \mathbf{x}_n \\
\delta \mathbf u_{n} = \mathbf{u}_n
\end{align*}
In addition, we assume that we are trying to determine a control policy which 
minimizes a pure quadratic cost function of the form
\begin{equation}
J = \frac{1}{2} \vx_N^T \vQ_N \vx_N + \sum_{n=0}^{N-1} {\frac{1}{2} \vx_n^T \vQ_n \vx_n + \frac{1}{2} \vu_n^T \vR_n \vu_n + \vu_n^T \vP_n \vx_n}.
\label{eq:LQR_cost_function}
\end{equation}
Therefore the first derivative of the cost function with respect to both $\vx$
and $\vu$ will be zero. From equations~\eqref{eq:quad_cost_elements}, $q_n$, $\vq_n$ 
and $\vr_n$ must equal $0$ for all $n$. This then implies that in
equations~\eqref{eq:g_G_H}, \eqref{eq:ilqc_sv}, and \eqref{eq:ilqc_s}, $\vg$,
$\vs_n$, and $s_n$ must always equal $0$.
\begin{align*}
& q_n = L_n(\bar{\vx}_n,\bar{\vu}_n) = 0 \hspace{10mm} 
\vq_n = \frac{\partial L(\bar{\vx}_n,\bar{\vu}_n)}{\partial\vx} = \mathbf{0} \hspace{10mm} 
\vr_n = \frac{\partial L(\bar{\vx}_n,\bar{\vu}_n)}{\partial\vu} = \mathbf{0} \\
& \vg_n  = \vr_n + \vB_n^T \vs_{n+1} = \mathbf{0} \hspace{6mm}
\delta \vu_n^{ff} = -\vH_n^{-1} \vg_n = \mathbf{0} \\
& \vs_n = \vq_n + \vA_n^T \vs_{n+1} + \mathbf K_n^T \mathbf H_n \delta \mathbf u_n^{ff} + \mathbf K_n^T \mathbf g_n + \mathbf G_n^T \delta \mathbf u_n^{ff} = \mathbf{0}\\
& s_n = q_n + s_{n+1} + \frac{1}{2} {\delta \vu_n^{ff}}^T \vH_n \delta \vu_n^{ff} +  {\delta \vu_n^{ff}}^T \vg_n = 0
\end{align*}
Substituting these results in the equation \eqref{eq:cost_to_go}, will result a value function which has a pure quadratic form with respect to $\vx$.
\begin{equation}
\label{eq:LQR_finite_time_value_function}
V^*(n,\vx) = \frac{1}{2} \vx^T \vS_{n} \vx
\end{equation}
where $\vS_n$ is calculated from the following final-value recursive equation
(derived from equation~\eqref{eq:ilqc_sm})
\begin{align}
\label{eq:LQR_finite_time_reccati_equation}
\vS_n &= \vQ_n + \vA_n^T \vS_{n+1} \vA_n + \vK_n^T \vH_n \vK_n + \vK_n^T \vG_n + \vG_n^T \vK_n \notag \\
&= \vQ_n + \vA_n^T \vS_{n+1} \vA_n - \vG_n^T \vH_n^{-1} \vG_n \\
&= \vQ_n + \vA_n^T \vS_{n+1} \vA_n - (\vP_n^T + \vA_n^T \vS_{n+1} \vB_n)  (\vR_n + \vB_n^T \vS_{n+1} \vB_n)^{-1} (\vP_n + \vB_n^T \vS_{n+1} \vA_n) \notag
\end{align}
This equation is known as the discrete-time Riccati equation.
The optimal control policy can be also derived form equation~\eqref{eq:ILQC_optimal_control_input} as follows
\begin{align}
\label{eq:LQR_discrete_time_policy}
\vmu^*(n,\vx) &= -\vH_n^{-1} \vG_n \vx \notag \\
&= - (\vR_n + \vB_n^T \vS_{n+1} \vB_n)^{-1} (\vP_n + \vB_n^T \vS_{n+1} \vA_n) \vx
\end{align}
We can derive the LQR controller by starting from the terminal condition, $\vS_N = \vQ_N$, and then solving for $\vS_n$ and $\vmu^*(n,\vx)$ iteratively, backwards in time.

\subsection{LQR: Infinite time horizon, discrete time}
\label{sec: lqr_infinite_time_discrete_time}
In this section we will derive the discrete-time LQR controller for the case
that the cost function is calculated over an infinite time horizon. Furthermore
we have assumed that the system dynamics are time invariant (e.g. a LTI system)
\begin{equation}
J = \sum_{n=0}^{\infty} {\frac{1}{2} \vx_n^T \vQ \vx_n + \frac{1}{2} \vu_n^T \vR \vu_n + \vu_n^T \vP \vx_n}.
\end{equation}
Note that the coefficients of the cost function are also assumed to be time
invariant. Furthermore the decay factor in the cost function is assumed to be 1.
Basically, in order to keep the cost function bounded in an infinite horizon
problem, the decay factor should be smaller than one. However for infinite
horizon LQR problem, it can be shown that under mild conditions the optimal
quadratic cost function is bounded.

As discussed in \sref{subsec:prob_infin_det}, the value function in the infinite
time horizon problem is not a function of time. This implies that the value
function in equation~\eqref{eq:LQR_finite_time_value_function} changes as follows
\begin{equation}
\label{eq:LQR_infinite_time_value_function}
V^*(\vx) = \frac{1}{2} \vx^T \vS \vx
\end{equation}
Since $\vS$ in not a function of time, the
equation~\eqref{eq:LQR_finite_time_reccati_equation} and the optimal policy are
reduced to the following forms
\begin{align}
& \vS = \vQ + \vA^T \vS \vA - (\vP^T + \vA^T \vS \vB)  (\vR + \vB^T \vS \vB)^{-1} (\vP + \vB^T \vS \vA) \label{eq:LQR_infinite_time_reccati_equation} \\
& \vmu^*(\vx) = - (\vR + \vB^T \vS \vB)^{-1} (\vP + \vB^T \vS \vA) \vx
\end{align}
Equation~\eqref{eq:LQR_infinite_time_reccati_equation} is known as the
discrete-time algebraic Riccati equation.

\subsection{LQR: Finite time horizon, continuous time}
\label{sec: lqr_finite_time_continuous_time}

In this section, we derive the optimal LQR controller for a \emph{continuous}
time system. This time we will start from the basic HJB equation which we derived
for deterministic systems in \sref{subsec:HJP_fin_det}. The system dynamics are
now given as a set of linear time varying differential equations.
\begin{equation}
\dot{\vx}(t) = \vA(t) \vx(t) + \vB(t) \vu(t).
\label{eq:lqr_fin_cont_sys_dyn}
\end{equation}
and the quadratic cost function the controller tries to optimize is given by
\begin{equation}
\newcommand{\half}{\frac{1}{2}}
J = \half \vx(T)^T \vQ_T \vx(T) 
  + \int_{0}^{T} {\left( \half \vx(t)^T \vQ(t) \vx(t) + \half \vu(t)^T \vR(t) \vu(t) + \vu(t)^T \vP(t) \vx(t) \right)dt},
\label{eq:LQR_cont_cost_function}
\end{equation}
where the matrices $\vQ_T$ and $\vQ$ are symmetric positive semidefinite, and $\vR$ is symmetric positive definite. 
Starting from the HJB equation \eref{eq:hjb_cont_time_determininstic} and
substituting in \eref{eq:lqr_fin_cont_sys_dyn} and \eref{eq:LQR_cont_cost_function} 
we obtain
\begin{align*}
-\frac{\partial V^{*}}{\partial t} 
 &=\min\limits_{u \in U} \{ 
  L(x,u) 
  +\left(\frac{\partial V^{*}}{\partial x}\right)^T f(x,u) \} \\
 &=\min\limits_{u \in U} \{ 
  {\frac{1}{2}} \vx^T \vQ \vx  +  {\frac{1}{2}} \vu^T \vR \vu  +  \vu^T \vP \vx   
  +\left(\frac{\partial V^{*}}{\partial x}\right)^T  (\vA \vx(t) + \vB \vu(t))\}, 
  \numberthis \label{eq:hjb_subst}\\ %
  V(T,\vx)&= {\frac{1}{2}} \vx^T \vQ_T \vx. 
  \numberthis \label{eq:hjb_boundary_cond} %
\end{align*}
This partial differential equation must hold for the value function to be
optimal. Considering the spacial form of this equation and also the quadratic
final cost, one sophisticated guess for the solution is a quadratic function as
follows
\begin{equation}
V^*(t,\vx) = \frac{1}{2} \vx^T \vS(t) \vx
\label{eq:hjb_val_fun}
\end{equation} 
where the Matrix $\vS$ is symmetric. The time and state derivatives of value function are 
\begin{align}
\label{eq:hjb_val_fun_dev_time}
\frac{\partial V^{*}(t,\vx)}{\partial t}   &= \frac{1}{2} \vx^T \dot{\vS}(t) \vx \\
\label{eq:hjb_val_fun_dev_state}
\frac{\partial V^{*}(t,\vx)}{\partial \vx} &= \vS(t) \vx
\end{align} 
By substituting equations~\eqref{eq:hjb_val_fun_dev_time} and 
\eqref{eq:hjb_val_fun_dev_state} in equation~\eqref{eq:hjb_subst} we obtain
\begin{equation}
-\vx^T \dot{\vS}(t) \vx  
 =\min\limits_{u \in U} \{ 
 \vx^T \vQ \vx  +  \vu^T \vR \vu + 2 \vu^T \vP \vx   
 + 2\vx^T \vS(t) \vA \vx + 2\vx^T \vS(t) \vB\vu\}. 
 \label{eq:hjb_subst_val_fun}
\end{equation}
In order to minimize the right hand side of the equation, we put its 
derivative with respect to $\vu$ equal to zero. 
\begin{align}
& 2\vR\vu + 2\vP\vx + 2\vB^T\vS(t)\vx = 0 \notag \\
\label{eq:lqr_opt_u}
& \vu^*(t,\vx) = -\vR^{-1} \left( \vP + \vB^T \vS(t) \right) \vx  
\end{align}
By inserting $\vu(t,\vx)$ into equation~\eqref{eq:hjb_subst_val_fun} and few more
simplification steps we obtain
\begin{align*}
\vx^T \Big[ \vS(t)\vA(t) + \vA^T(t)\vS(t) - \left( \vP(t) + \vB^T(t)\vS(t) \right)^T \vR^{-1} & \left(\vP(t) + \vB^T(t)\vS(t) \right)  \\
&+ \vQ(t) + \dot{\vS}(t) \Big] \vx = 0
\end{align*}
Since this equation should hold for all the states, the value inside the brackets
should be equal to zero. Therefore we get
\begin{equation}
\dot{\vS} = -\vS\vA - \vA^T\vS + \left( \vP + \vB^T\vS \right)^T \vR^{-1} \left( \vP + \vB^T\vS \right) - \vQ,
\quad\quad     \text{with } \vS(T) = \vQ_T.
\label{eq:cont_time_ricc}
\end{equation}
This equation is known as the continuous-time Riccati equation. If $\vS(t)$ satisfies
\eqref{eq:cont_time_ricc} then we found the optimal value function through
\eqref{eq:hjb_val_fun}. Furthermore the optimal control input can then be
computed using \eqref{eq:lqr_opt_u}

\subsection{LQR: Infinite time horizon, continuous time}
\label{sec: lqr_infinite_time_continuous_time}
In this subsection the optimal control input $\vu(t)^*$ is derived over an infinitely long time horizon. In contrast to
the finite horizon case \eqref{eq:LQR_cont_cost_function}, the cost function does
not include a terminal cost, since the evaluation is never terminated. The
decay factor is also chosen 1. As in the discrete-time case, it can be shown that
under mild conditions, the infinite-time LQR cost function will be bounded.
\begin{equation}
\newcommand{\half}{\frac{1}{2}}
J = \int_{0}^{\infty} {\Big[ \half \vx(t)^T \vQ \vx(t) + \half \vu(t)^T \vR \vu(t) + \vu(t)^T \vP \vx(t) \Big] dt}.
\label{eq:LQR_cont_cost_function_inf_time}
\end{equation}
Since the value function in an infinite horizon problem is not a function of
time, the time dependency in equation~\eqref{eq:hjb_val_fun} is dropped,
resulting in
\begin{equation}
V(\vx) = \vx^T \vS \vx , \qquad\qquad \vS: \text{n x n symmetric}.
\label{eq:hjb_val_fun_inf_time}
\end{equation} 
The evaluation is performed just as in the previous section, by deriving
\eqref{eq:hjb_val_fun_inf_time} w.r.t state $\vx$ and time $t$ and substituting
the values into the HJB equation \eqref{eq:hjb_cont_time_determininstic}. The
solution is equal to the continuous time Riccati equation
\eqref{eq:cont_time_ricc}, apart from the derivative on the left hand side
$\dot{\vS}(t)$. Due to the time independence this derivative is zero resulting in
the continuous-time algebraic Riccati equation.
\begin{equation}
\vS\vA + \vA^T\vS - \left( \vP + \vB^T\vS \right)^T \vR^{-1} \left( \vP + \vB^T\vS \right) + \vQ = 0,
\label{eq:cont_time_ricc_inf}
\end{equation} 
By solving this equation once for $\vS$, the optimal control input at every 
state $\vx$ is given by
\begin{equation}
\vu^*(\vx) = -\vR^{-1} ( \vP + \vB^T \vS) \vx 
\end{equation}

\section{Linear Quadratic Gaussian Regulator: LQG(R)}
In this section, we will study the LQG regulator. LQG stands for
\textbf{L}inear \textbf{Q}uadratic \textbf{G}aussian regulator. The assumption
behind the LQG problem is very similar to the one in LQR problem except one main
point. In contrast to the LQR problem, in LQG
the system dynamics is corrupted with a Gaussian noise. The introduction of noise
in the system dynamics causes the system to demonstrate stochastic behavior
e.g. different runs of the system under a similar control policy (or control
input trajectory) will generate different state trajectories. This fact implies
that a cost function defined for the LQR problem should be stochastic as well. Therefore in
the LQG controller, the cost function  is defined as an expectation of the LQR
cost function.

One important observation about the LQG problem is that the noise is assumed to have a Gaussian
distribution, not any arbitrary distribution. Beyond the discussion that the
Gaussian noise has some physical interpretations, the main reason for this choice
is the nice analytical feature that Gaussian noise is closed under the
linear transformation. In other words, if a Gaussian random variable is linearly
transformed, it is still a Gaussian random variable.

This section is organized as follows. In the first section we will derive the
LQG controller for discrete-time systems with finite-time horizon cost
function. We then extend this result to the infinite-time horizon problem.
We then derive the LQG controller for continuous-time systems both for the
finite-time and infinite-time horizon cost functions.

\subsection{LQG: Finite time horizon, discrete time}
The LQG controller can be seen as a LQR controller with additive Gaussian noise.
We can find it by applying the Stochastic Principle of Optimality to the
Linear-Quadratic problem. The problem formulation for the discrete time case are
as follows:

Quadratic discrete cost:
\begin{align}
J = \frac{1}{2} E \left\{ \alpha^N \vx_N^T \vQ_N \vx_N + \sum\limits_{n=0}^{N-1} \alpha^n \begin{bmatrix} \vx_n^T & \vu_n^T \end{bmatrix} 
\begin{bmatrix}
\vQ_n &  \vP_n^T \\
\vP_n & \vR_n
\end{bmatrix}
\begin{bmatrix} \vx_n \\  \vu_n \end{bmatrix}
\right\}
\end{align}
This is identical to the cost function of the discrete time LQR, except for the expectation and the decay factor. As you will see later, the decay factor for the finite horizon
problem is usually chosen to be 1. However for the infinite horizon problem, it is
absolutely necessary that $\alpha$ is smaller than 1. Again $\vQ$ is a positive
semidefinite matrix and $\vR$ is a positive definite matrix (which is basically
invertible).

Linear system dynamics (or linearized system dynamics):
\begin{align}\label{eq:linearized_system_dynamics}
\vx_{n+1} = \vA_n \vx_n + \vB_n \vu_n + \vC_n \vw_n
\hspace{10mm}
\mathbf \vx(0) = \vx_0 \text{ given}
\end{align}
Matrices $\vQ$, $\vP$, $\vR$, $\vA$, $\vB$ and $\vC$ can be functions of time. $\vw_n$ is a uncorrelated, zero-mean Gaussian process.
\begin{align}
\label{eq:lqr_gaussian_noise}
& E[\mathbf w_n] = 0 \notag \\ 
& E[\vw_n \vw_m^T] = \vI \delta(n-m)
\end{align}
In order to solve this optimal control problem, we use the Bellman equation given
by equation~\eqref{eq:Bellman_stoch}. Like in the LQR setting, we assume a quadratic value
function as Ansatz but with an additional term to account for the stochasticity.
We will call it the Quadratic Gaussian Value Function:
\begin{equation}
\label{eq:LQG_descrete_time_ansatz}
V^*(n,\vx)=\frac{1}{2} \vx^T \vS_n \vx + \upsilon_n
\end{equation}
Plugging the cost function and that Ansatz into \eqref{eq:Bellman_stoch} yields the following equation for the optimal value function.
\begin{align*} 
V^*(n,\vx)& = \min_{\vu_n} \frac{1}{2} E \left[ \vx_n^{T} \vQ_n \vx_n + 2 \vu_n^{T} \vP_n \vx_n + \vu_n^T \vR_n \vu_n + \alpha \vx_{n+1}^T \vS_{n+1} \vx_{n+1} + 2 \alpha \upsilon_{n+1} \right]
\end{align*}
By the use of the system equation (\ref{eq:linearized_system_dynamics}) and
considering $E\{w\}=0$ we can write
\begin{align*}
V^*(n,\vx) = \min_{\vu_n} \frac{1}{2} E \big[ & \vx_n^T \vQ_n \vx_n + 2 \vu_n^T \vP_n \vx_n + \vu_n^T \vR_n \vu_n + \alpha (\vA_n \vx_n + \vB_n \vu_n)^T\vS_{n+1} (\vA_n \vx_n + \vB_n \vu_n) \\
 & + \alpha \vw_n^T \vC_n^T \vS_{n+1} \vC_n \vw_n + 2 \alpha \upsilon_{n+1}  \big]
\end{align*}
Using $Tr(\vA \vB) = Tr(\vB \vA)$ and considering $E[\vw_n \vw_n^T] = \vI$, we get
\begin{align*}
V^*(n,\vx) = \min_{\vu_n} \frac{1}{2} & \{ \vx_n^T \vQ_n \vx_n + 2 \vu_n^T \vP_n \vx_n + \vu_n^T \vR_n \vu_n + \alpha (\vA_n \vx_n + \vB_n \vu_n)^T\vS_{n+1} (\vA_n \vx_n + \vB_n \vu_n) \\
 & + \alpha Tr (\vS_{n+1} \vC_n \vC_n^T) + 2 \alpha \upsilon_{n+1} \}
\end{align*}
In order to minimize the right-hand-side of the equation with respect to $\vu$,
we set the gradient with respect to $\vu$ equal to zero, and then substitute the result back into the previous equation.
\begin{equation}
\label{eq:LQG_discrete_time_policy_decayfactor}
\vu^*(n,\vx) = - (\vR_n + \alpha \vB_n^T \vS_{n+1} \vB_n)^{-1} (\vP_n + \alpha \vB_n^T \vS_{n+1} \vA_n) \vx
\end{equation} 
\begin{align*}
V^*(n,\vx) = \frac{1}{2} \vx_n^T \Big[ & \vQ_n + \alpha \vA_n^T \vS_{n+1} \vA_n 
- (\vP_n + \alpha \vB_n^T \vS_{n+1} \vA_n)^T  (\vR_n + \alpha \vB_n^T \vS_{n+1} \vB_n)^{-1} \\
& (\vP_n + \alpha \vB_n^T \vS_{n+1} \vA_n)  \Big] \vx_n + \alpha \Big[ \frac{1}{2} Tr (\vS_{n+1} \vC_n \vC_n^T) + \upsilon_{n+1} \Big]
\end{align*}
Substituting $V^*(n,\vx)$ with the Ansatz and bringing all the terms to one side, gives
\begin{align*}
\frac{1}{2} \vx_n^T \Big[ & \vQ_n + \alpha \vA_n^T \vS_{n+1} \vA_n 
- (\vP_n + \alpha \vB_n^T \vS_{n+1} \vA_n)^T  (\vR_n + \alpha \vB_n^T \vS_{n+1} \vB_n)^{-1} \\ & 
(\vP_n + \alpha \vB_n^T \vS_{n+1} \vA_n) - \vS_n \Big] \vx_n + \Big[\frac{1}{2} \alpha Tr (\vS_{n+1} \vC_n \vC_n^T) + \alpha \upsilon_{n+1} - \upsilon_{n} \Big] = 0.
\end{align*}
In order to satisfy this equality for all $\vx$, the terms inside the brackets should equal zero.
\begin{align}
\label{eq:LQG_finite_time_reccati_equation_decayfactor}
\vS_n &= \vQ_n + \alpha \vA_n^T \vS_{n+1} \vA_n - (\vP_n + \alpha \vB_n^T \vS_{n+1} \vA_n)^T  (\vR_n + \alpha \vB_n^T \vS_{n+1} \vB_n)^{-1} (\vP_n + \alpha \vB_n^T \vS_{n+1} \vA_n) \notag \\
\upsilon_{n} &= \frac{1}{2} \alpha Tr (\vS_{n+1} \vC_n \vC_n^T) + \alpha \upsilon_{n+1} , \hspace{10mm} \vS_N = \vQ_N \hspace{5mm} \upsilon_{N} = 0
\end{align}
For finite horizon problem, we normally choose $\alpha$ equal to 1. Therefore
the updating equation for $\vS$ will reduce to the well known Riccati equation
\begin{align}
\label{eq:LQG_finite_time_reccati_equation}
\vS_n &= \vQ_n + \vA_n^T \vS_{n+1} \vA_n - (\vP_n^T + \vA_n^T \vS_{n+1} \vB_n)  (\vR_n + \vB_n^T \vS_{n+1} \vB_n)^{-1} (\vP_n + \vB_n^T \vS_{n+1} \vA_n) \notag \\
\upsilon_{n} &= \frac{1}{2} Tr (\vS_{n+1} \vC_n \vC_n^T) + \upsilon_{n+1} , \hspace{10mm} \vS_N = \vQ_N \hspace{5mm} \upsilon_{N} = 0
\end{align}
and the optimal control policy
\begin{equation}
\label{eq:LQG_discrete_time_policy}
\vu_n^* = - (\vR_n + \vB_n^T \vS_{n+1} \vB_n)^{-1} (\vP_n + \vB_n^T \vS_{n+1} \vA_n) \vx_n
\end{equation} 
By comparing the optimal control policies for the LQG problem in
equation~\eqref{eq:LQG_discrete_time_policy} and the LQR problem in
equation~\eqref{eq:LQR_discrete_time_policy}, we can see that they have the
same dependency on $\vS$. Furthermore by comparing the recursive formulas for
calculating $\vS$ in equations~\eqref{eq:LQG_finite_time_reccati_equation} and
\eqref{eq:LQR_finite_time_reccati_equation}, we see that they are basically the
same. In fact, the optimal control policy and the Riccati equation in both
LQR and LQG problem are identical. 

The only difference between the LQG and LQR problems are their corresponding value functions. It can be shown that the value function in the LQG case is always greater than that in the
LQR case. In order to prove this, we need just to prove that $\upsilon_{n}$ is
always nonnegative. We will prove it by induction. First we show that the base
case ($n = N$) is correct. Then we show that if $\upsilon_{n+1}$ is
nonnegative, $\upsilon_{n}$ should also be nonnegative.

\textbf{Base case:} It is obvious because $\upsilon_{N}$ is equal to zero.

\textbf{Induction:} From equation~\eqref{eq:LQG_finite_time_reccati_equation}, 
we realize that if $\upsilon_{n+1}$ is nonnegative, $\upsilon_{n}$ will be
nonnegative if and only if $Tr (\vS_{n+1} \vC_n \vC_n^T) \geq 0$. Now we will
show this.
\begin{align*}
Tr (\vS_{n+1} \vC_n \vC_n^T) &= Tr (\vC_n^T \vS_{n+1} \vC_n) \\
&= \sum_{i}{{\mathbf{C}^i}_n^T \vS_{n+1} {\mathbf{C}^i}_n}
\end{align*}
where ${\mathbf{C}^i}_n$ is a the \textit{i}th column of $\vC_n$. Since $S_n$ is
always positive semidefinite, ${\mathbf{C}^i}_n^T \vS_{n+1} {\mathbf{C}^i}_n \geq
0$ holds for all \textit{i}. Therefore $Tr (\vS_{n+1} \vC_n \vC_n^T) \geq 0$, and $\upsilon_{n}$ is always nonnegative.

\subsection{LQG: Infinite time horizon, discrete time}
In this section the LQG optimal control problem with an infinite-time horizon cost
function is introduced. The quadratic cost function is defined as follows
\begin{align}
\label{eq:LQG_disc_cost_function_inf_time}
J = \frac{1}{2} E \left\{ \sum\limits_{n=0}^{\infty} \alpha^n \begin{bmatrix} \mathbf x_n^T & \mathbf u_n^T \end{bmatrix} 
\begin{bmatrix}
\mathbf \vQ &  \mathbf \vP^T \\
\mathbf \vP & \mathbf \vR
\end{bmatrix}
\begin{bmatrix}
\mathbf x_n \\
 \mathbf u_n
\end{bmatrix}
\right\}
\end{align}
and the system dynamics
\begin{align}
\vx_{n+1} = \vA \vx_n + \vB \vu_n + \vC \vw_n
\hspace{10mm}
\mathbf \vx(0) = \vx_0 \text{ given}
\end{align}
where $\vw_n$ is a Gaussian process with the characteristic described at
\eqref{eq:lqr_gaussian_noise}. All of the matrices are time
independent. An interesting difference between the infinite-time LQR problem and
the LQG problem is that the decay factor must be always smaller than 1 ($0 <
\alpha < 1$), otherwise the quadratic cost function will not be bounded. As you
will see shortly, although $\alpha$ can approach in limit to 1, it is
absolutely necessary that $\alpha \neq 1$.

In order to solve this problem we will use the results we have
obtained from the previous section. The only difference between these two problems is
that in the infinite time case the value function is only a function of state,
not time. Therefore the value function can be expressed as
\begin{equation}
V^*(\vx)=\frac{1}{2} \vx^T \vS \vx + \upsilon
\end{equation}
Using equation~\eqref{eq:LQG_finite_time_reccati_equation_decayfactor}, we get
\begin{align}
\label{ilqg_infinite_discrete_general_riccati}
\vS &= \vQ + \alpha \vA^T \vS \vA - (\vP + \alpha \vB^T \vS \vA)^T  (\vR + \alpha \vB^T \vS \vB)^{-1} (\vP + \alpha \vB^T \vS \vA) \\
\label{ilqg_infinite_discrete_general_riccati_increment}
\upsilon &= \frac{\alpha}{2(1-\alpha)}  Tr (\vS \vC \vC^T)
\end{align}  
and the optimal control policy
\begin{equation}
\label{ilqg_infinite_discrete_general_policy}
\vu^*(\vx) = - (\vR + \alpha \vB^T \vS \vB)^{-1} (\vP + \alpha \vB^T \vS \vA) \vx
\end{equation} 
As equation~\eqref{ilqg_infinite_discrete_general_riccati_increment} illustrates,
if $\alpha$ approaches 1, $\upsilon$ grows to infinity. Therefore $\alpha$ should
always be smaller than 1. However it could approach in limit to 1, in which case
equations~\eqref{ilqg_infinite_discrete_general_riccati} and
\eqref{ilqg_infinite_discrete_general_policy} simplify as
\begin{align}
\label{ilqg_infinite_discrete_riccati}
&\vS = \vQ + \vA^T \vS \vA - (\vP + \vB^T \vS \vA)^T  (\vR + \vB^T \vS \vB)^{-1} (\vP + \vB^T \vS \vA) \\
\label{ilqg_infinite_discrete_policy}
&\vu^*(\vx) = - (\vR + \vB^T \vS \vB)^{-1} (\vP + \vB^T \vS \vA) \vx
\end{align} 
Equation~\eqref{ilqg_infinite_discrete_riccati} is the discrete-time
algebraic Riccati equation similar to the LQR one in
equation~\eqref{eq:LQR_infinite_time_reccati_equation}.

\subsection{LQG: Finite time horizon, continuous time}
In this section we solve the LQG problem for continuous-time systems. The problem
formulation is as follows:

Quadratic cost function: 
\begin{align}
J =
\frac{1}{2} E \left\{e^{-\beta T} \mathbf x^T (T) \mathbf Q_T \mathbf x(T) + \int_{0}^{T} e^{-\beta t} \begin{bmatrix} \mathbf x^T(t) & \mathbf u^T(t)\end{bmatrix}
\begin{bmatrix}
\mathbf \vQ(t)   & \mathbf \vP^T(t)\\
\mathbf \vP(t) & \mathbf \vR(t)
\end{bmatrix}
\begin{bmatrix}
 \mathbf x (t)\\
  \mathbf u (t)
\end{bmatrix} dt \right\}
\end{align}
$\vQ$ is a positive semidefinite matrix and $\vR$ is a positive definite matrix
(which is basically invertible). As you will see later, the decay factor
($\beta$) for the finite horizon problem is usually chosen 0. However for the
infinite horizon problem, it is absolutely necessary that $\beta$ be greater than
0.

Stochastic linear system dynamics:
\begin{align}\label{eq:LQGcontSystem}
\dot{\mathbf x} (t) = \vA(t) \vx(t) + \vB(t) \vu(t) + \vC(t) \vw(t) \condition{$\mathbf x(0) = x_0$}
\end{align}
Matrices $\vQ$, $\vP$, $\vR$, $\vA$, $\vB$ and $\vC$ can be functions of the
time. $\vw(t)$ is a uncorrelated, zero-mean Gaussian process.
\begin{align}
& E[\mathbf w(t)] = \mathbf 0 \notag \\ 
& E[\vw(t) \vw(\tau)^T] = \vI \delta(t - \tau)
\label{eq:lqg_gaussian_noise}
\end{align}
The solution procedure is like in the deterministic case, except that now we are
accounting for the effect of noise. We make an Ansatz like in
equation~\eqref{eq:LQG_descrete_time_ansatz}: A quadratic value function with
stochastic value function increment.
\begin{equation}
\label{eq:LQG_continuous_time_ansatz}
V^*(t,\vx)= \frac{1}{2} \mathbf x^T (t) \mathbf S (t) \mathbf x(t) + \upsilon(t)
\end{equation}
where $\mathbf{S}(t)$ is a symmetric matrix and $\upsilon(t)$ is the stochastic
value function increment. 

From the Ansatz \eqref{eq:LQG_continuous_time_ansatz} we can derive the first
partial derivative with respect to $t$ and the first and second partial
derivatives with respect to $x$ as:
\begin{align}\label{eq:partial_deriv_LQG}
V^*_{t}(t,\vx) & = \frac{1}{2} \mathbf x^T (t) \dot{\mathbf S} (t) \mathbf x(t) + \dot{\upsilon}(t) \notag \\
V^*_{\mathbf x}(t,\vx) & = \mathbf S(t) \mathbf x(t) \notag \\
V^*_{\mathbf {xx}}(t,\vx) & = \mathbf S (t)
\end{align}

Plugging the Ansatz and its derivatives into the Stochastic HJB equation and following a similar simplification strategy as in the discrete case, we get:
\begin{equation*}
\beta V^* - V_t^*= \min_{\vu} \frac{1}{2} \{ \vx^T \vQ \vx + 2 \vu^T \vP \vx + \vu^T \vR \vu + 2\vx^T \vS (\vA \vx + \vB \vu) + Tr(\vS \vC \vC^T) \}
\end{equation*}
The optimal control is found following the same procedure as in the LQR case,
namely differentiating the right side of the equation with respect to $\vu$ and
setting the result to $0$:
\begin{equation}
\label{eq:LQG_continuous_time_policy}
\vu^*(t,\vx) = -{\vR(t)}^{-1} \left( \vP(t) + \vB^T(t) \vS(t) \right) \vx  
\end{equation}
In order to find the unknown $S(t)$, we substitute $\mathbf u$ and $V_t^*$ into the previous equation 
\begin{align*}
\frac{1}{2} \vx^T \Big[ \vS(t)\vA(t) + \vA^T(t)\vS(t) &- \left( \vP(t) + \vB^T(t)\vS(t) \right)^T \vR^{-1} \left(\vP(t) + \vB^T(t)\vS(t) \right)  \\
&+ \vQ(t) + \dot{\vS}(t) -\beta \vS \Big] \vx + \Big[ \dot{\upsilon}(t) - \beta \upsilon(t) + \frac{1}{2} Tr(\vS \vC \vC^T) \Big]= 0
\end{align*}
In order to satisfy this equality for all $\vx$, the terms inside the brackets should equal zero. Therefore we have
\begin{align}
\label{eq:cont_time_ricc_lqg_general}
& \dot{\vS} = \beta \vS -\vS\vA - \vA^T\vS + \left( \vP + \vB^T\vS \right)^T \vR^{-1} \left( \vP + \vB^T\vS \right) - \vQ,
\quad\quad     \text{with } \vS(T) = \vQ_T. \\
\label{eq:cont_time_ricc_lqg_inc_general}
& \dot{\upsilon} = \beta \upsilon(t) - \frac{1}{2} Tr(\vS \vC \vC^T)
\hspace{65mm}     \text{with } \upsilon(T) = 0.
\end{align}
For finite horizon problems, we normally choose $\beta$ equal to 0. Therefore
the updating equation for $\vS$ will reduce to the following:
\begin{align}
\label{eq:cont_time_ricc_lqg}
& \dot{\vS} = -\vS\vA - \vA^T\vS + \left( \vP + \vB^T\vS \right)^T \vR^{-1} \left( \vP + \vB^T\vS \right) - \vQ,
\quad\quad     \text{with } \vS(T) = \vQ_T. \\
\label{eq:cont_time_ricc_lqg_inc}
& \dot{\upsilon} = - \frac{1}{2} Tr(\vS \vC \vC^T)
\hspace{69mm}     \text{with } \upsilon(T) = 0.
\end{align}
Equation~\eqref{eq:cont_time_ricc_lqg} is known as the continuous time Riccati equation. By comparing this equation and the one in \eqref{eq:cont_time_ricc}, we realize that the Riccati equation is the same for both LQR and LQG problems. Furthermore the control policies are identical in both cases. However the value functions are different. One can easily prove that the value function in the LQG problem is grater and equal to the LQR value function.

\subsection{LQG: Infinite time horizon, continuous time}
In this section, the continuous time LQG problem with an infinite time horizon cost function is studied. The quadratic cost function is defined as follows
\begin{align}
J =
\frac{1}{2} E \left\{\int_{0}^{\infty} e^{-\beta t} \begin{bmatrix} \mathbf x^T(t) & \mathbf u^T(t)\end{bmatrix}
\begin{bmatrix}
\mathbf \vQ   & \mathbf \vP^T\\
\mathbf \vP & \mathbf \vR
\end{bmatrix}
\begin{bmatrix}
 \mathbf x (t)\\
  \mathbf u (t)
\end{bmatrix} dt \right\}
\end{align}
and the system dynamics
\begin{align}
\dot{\vx} (t) = \vA \vx(t) + \vB \vu(t) + \vC \vw(t)
\hspace{10mm}
\mathbf \vx(0) = \vx_0 \text{ given}
\end{align}
Matrices $\vQ$, $\vP$, $\vR$, $\vA$, $\vB$ and $\vC$ are all time invariant.
$\vw(t)$ is an uncorrelated, zero-mean Gaussian process with the characteristic
described at \eqref{eq:lqg_gaussian_noise}. The defined cost function is
comparable to the one introduced in \eqref{eq:LQR_cont_cost_function_inf_time}
except the expectation and the decay factor. The expectation is introduced
because of the stochasticity of the problem. The decay rate is chosen to be non-zero
since the pure quadratic cost function will never be bounded in the LQG case though
it is bounded for LQR. As you will see shortly, although $\beta$ can
approach in limit to 0, it is absolutely necessary that $\beta > 0$.

In order to solve this problem we will use the results we have
obtained from the previous section. The only difference between these two problems is
that in the infinite time case the value function is only a function of state,
not time. Therefore the value function can be expressed as
\begin{equation}
V^*(\vx)=\frac{1}{2} \vx^T \vS \vx + \upsilon
\end{equation}
Using equations~\eqref{eq:cont_time_ricc_lqg_general} and \eqref{eq:cont_time_ricc_lqg_inc_general}, we get
\begin{align}
\label{eq:cont_time_ricc_lqg_infinite_decayfactor}
& -\beta \vS +\vS\vA + \vA^T\vS - \left( \vP + \vB^T\vS \right)^T \vR^{-1} \left( \vP + \vB^T\vS \right) + \vQ = 0\\
\label{eq:cont_time_ricc_lqg_inc_infinite_decayfactor}
& \upsilon(t) = \frac{1}{2 \beta} Tr(\vS \vC \vC^T)
\end{align}
and the optimal control policy at every state $\vx$ is given by
\begin{equation}
\vu^*(\vx) = -\vR^{-1} ( \vP + \vB^T \vS) \vx 
\end{equation}
As equation~\eqref{eq:cont_time_ricc_lqg_inc_infinite_decayfactor} illustrates,
if $\beta$ approaches 0, $\upsilon$ grows to infinity. Therefore $\beta$ should
always be greater than 0. However it could be arbitrary small. Thus if $\beta$
approaches to 0, equation~\eqref{eq:cont_time_ricc_lqg_infinite_decayfactor}
simplifies to
\begin{align}
\label{eq:cont_time_ricc_lqg_infinite}
& \vS\vA + \vA^T\vS - \left( \vP + \vB^T\vS \right)^T \vR^{-1} \left( \vP + \vB^T\vS \right) + \vQ = 0\\
\label{eq:cont_time_ricc_lqg_policy_infinite}
& \vu^*(\vx) = -\vR^{-1} ( \vP + \vB^T \vS) \vx 
\end{align}
Equation~\eqref{eq:cont_time_ricc_lqg_infinite} is the continuous-time
algebraic Riccati equation similar to the LQR one in
equation~\eqref{eq:cont_time_ricc_inf}.

\thispagestyle{empty}\cleardoublepage
\chapter{Classical Reinforcement Learning}  \label{ch:classical_RL}
In the field of Reinforcement Learning (RL) different notations are used even
though the problem setting is the same as in the optimal control case. In order
to keep consistency with the previous chapters of the script we will use the
notation of optimal control also in this chapter. The following table summarizes
the relationship between important terms used in RL and optimal control. It
should help the reader link the material here to the external literature.
\begin{center}
  \begin{tabular}{ | p{5cm} | p{5cm} |}
    \hline
        \begin{center}\textbf{RL}\end{center} & \begin{center}\textbf{Optimal Control}\end{center}  \\ \hline\hline
        environment & system \\[0.1cm] \hline
     	agent & controller \\ [0.1cm] \hline
     	state: $s$ & state: $x$ \\ [0.1cm] \hline
     	control action: $a$ & control input: $u$ \\ [0.1cm] \hline
     	reward: $r$ & intermediate cost: $L$ \\ [0.1cm] \hline
     	discount factor: $\gamma$ & discount factor $\alpha$ \\ [0.1cm] \hline
     	stochastic policy: $\pi$ & deterministic policy $\mu$ \\ [0.1cm] \hline         
  \end{tabular}
\end{center}
One notation that we will adopt from RL is the reward. In oder to evaluate a
specific control policy RL does not punish elements that lead to unwanted
behavior but it rewards behavior that is desired. Therefore instead of defining a
cost function like in optimal control, RL computes a reward function, which has
to be maximized. Conceptually, these concepts are trivially equivalent
though, since $cost = - reward$.

\begin{minipage}{\textwidth}
\vspace{1cm}
\emph{Note that this chapter follows in parts very closely Sutton and Barto's\footnote{Richard S. Sutton and Andrew G. Barto. Introduction to Reinforcement Learning. MIT Press,
Cambridge, MA, USA, 1st edition, 1998}
  classical and excellent RL text book. Following the aim of the
  course to give a unified view on the two fields of RL and Optimal
  Control, we have reproduced here some
  figures and developments from the book using the unified
  notation.} 
\end{minipage}

\section{Markov Decision Process} \label{sec:MDP}
A stochastic process satisfies the Markov property if the following conditional probability distribution holds.
\begin{equation*}
Pr \{ x_{n+1} \mid x_n, x_{n-1}, \dots, x_0 \} = Pr \{ x_{n+1} \mid x_n \}
\end{equation*}
This means that the behavior of the process (system) at each time step is based only on information from the previous time step. In this sense, there is no memory in the system and its behavior in each state does not depend on how it got there, but only on the action taken at that point in time. According to this definition, all of the dynamical systems we have defined so far
in Chapter~\ref{ch:optimal_control} are Markov processes.

In RL, a system that satisfies the Markov property is called \textbf{M}arkov
\textbf{D}ecision \textbf{P}rocess (MDP). It consists of a finite set of states
and actions (control inputs) and a reward function. Loosely speaking, states are
defined based on information we gather from the environment and the actions
(e.g. a motor command of a robot) are the way we affect the surrounding
environment. The reward function is a memoryless stochastic process that assesses
the value of the current state and the subsequent action of the agent. The term ``agent" in RL refers to the entity which interacts with the environment in order to
optimize some criteria of its performance.

In most cases, states and control inputs are regarded as finite sets
\begin{equation}
x \in \vX = \{x_1, \dots , x_{\norm{X}}\} {,} \quad u \in \vU = \{u_1, \dots , u_{\norm{U}}\}
\end{equation}
where $\norm{X}$ and $\norm{U}$ are the number of possible states and actions at each time step
respectively (for simplicity we have assumed that the number of the possible
actions in each state is the same). The process's dynamics are defined by what is called the transition probability distribution.  This defines the probability of moving from state $x_n=x$ to state $x_{n+1}=x'$ after applying the control input $u_n=u$:
\begin{equation}
\mathcal P^u_{xx'} = Pr \{ x_{n+1} = x'\mid x_n=x,u_n=u \}
\label{eq:transition_probability}
\end{equation}
The process's rewards are also stochastic. The expected intermediate reward received at time $n$ by performing
action $u$ in state $x$ and transiting to state $x'$ is defined as:
\begin{equation}
\mathcal R^u_{xx'} = E\{ r_{n} \mid x_{n+1} = x', u_n = u, x_n = x \}
\label{eq:expected_reward}
\end{equation}
Three points worth noticing here are: 
\begin{itemize}
\item The reward function is stochastic in contrast to the optimal control
cost function which has always been defined as a deterministic function.
\item We have only defined the first moment of the reward process. As
we will see later, this is the only piece of information needed to find the optimal policy.
\item Last but not the least, in contrast to the cost function that is defined
based on the current state and current control input, the reward function in this
chapter is defined based on the triple of the current state, the current control input,
and the next state. However, one can derive a reward function only based on the
current state and control input from this reward function as follows
\begin{equation}
\mathcal R^u_{x} = \sum_{x'}{\mathcal P^u_{xx'} \mathcal R^u_{xx'}}
\end{equation}
Actually, as we will see later, whenever $\mathcal R^u_{xx'}$ appears, it is always
marginalized with respect to $x'$. Therefore, the actual probability distribution
that we are interested in is $\mathcal R^u_{x}$. Using $\mathcal R^u_{xx'}$ is a
design choice which sometimes makes the definition of the reward function more
intuitive.
\end{itemize}

In the problems of previous chapters we have mostly considered deterministic policies
that we denoted by $\mu(x)$. On the contrary, MDPs assume more general policies, 
which include the stochastic policy class. In the following sections, stochastic policies will be expressed as $\pi(x,u)$, or more precisely
$\pi(u|x)$. $\pi$ is a probability distribution over the action (control input)
set and conditioned over the current state.

Except for the restricting assumption that the state and action are discrete, the
modeling assumptions behind the MDP problem are more general than in the optimal
control problem of the previous chapter. We should note, however, that the
discrete state and action assumptions are often impractical when working on real-world 
robotic platforms.

\section{The RL Problem} \label{sec:the_RL_problem}
In the optimal control problem, our objective was to find a policy (controller)
that minimizes the expected total cost. Similarly, in the RL problem we want to
find a policy, $\pi$, which maximizes the expected accumulated reward. The
optimal control counterpart of the RL problem is the infinite-horizon discrete
time optimal control problem. However in RL the state and the actions are
discrete (or discretized). Furthermore the reward function is not a deterministic
function.

\paragraph{Problem Statement:} Assuming an MDP problem, we seek a stochastic
policy that maximizes the expected accumulated reward. The accumulated reward is
defined as
\begin{equation} \label{eq:accumulated_reward_0}
R_0 = r_{0} + \alpha \, r_{1} + \alpha^2 \, r_{2} + \dots + \alpha^n r_{n} + \dots = \sum_{k = 0}^{\infty}{\alpha^k \, r_{k}}
\end{equation}
and the optimal policy as
\begin{equation} \label{eq:RL_optimal_policy}
\pi^* = \operatorname{arg\,max}_{\pi} E[R_0]
\end{equation}
In equation~\eqref{eq:accumulated_reward_0}, $\alpha \in [0,1]$ is the decay or
discount factor. In an episodic task (i.e a task that has some terminal states or
finishes after a finite number of time steps) $\alpha$ can be 1. Otherwise it should be
chosen in such a way to ensure that the infinite summation of intermediate rewards exists. Furthermore, we should notice
that $R_0$ is a stochastic variable. Its stochasticity originates from three
sources. First, the reward function is a stochastic process. Therefore
$R_0$ as a summation of random variables is also a random variable. Second, the
state transition is governed by a stochastic process. Therefore, the state sequence,
which the reward function depends on, is stochastic. Finally, the policy can also
be stochastic, which affects both state and control sequences.

In equation~\eqref{eq:RL_optimal_policy}, the expectation is taken with respect
to all sources of stochasticity in the accumulated reward. As you can see, since the rewards at
each time step are summed (which is a linear operation), we can
easily pull the expectation inside of the sum, resulting in a sum of expected intermediate rewards. This means that, in the interest of finding the optimal policy, the only relevant characteristic of
the reward function's distribution is its first moment.

\begin{figure}[htb]
\centering
  \includegraphics[width=0.7\linewidth]{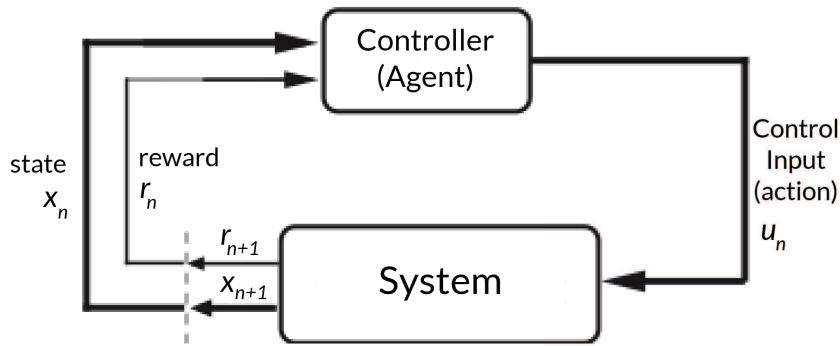}
  \caption{Flow chart of state, control input and reward in the RL
    notation. Reproduced from \cite{Sutton1998}.}
  \label{fig:RL_problem}
\end {figure}

For the sake of the notion simplicity, we introduce a more general definition for
the accumulated reward. In this definition, the accumulated reward at time step
$n$ is defined as
\begin{equation} \label{eq:accumulated_reward}
R_n = \sum\limits_{k=0}^{\infty} \alpha^k \, r_{n+k}
\end{equation}
One can verify that $R_n$ at time zero is the originally defined
accumulated reward. In Figure~\ref{fig:RL_problem}, a flow chart of state, control input
and reward in RL is given.

\section{State Value Function and the Bellman Equation}  \label{sec:state_value_function}
As was the case with the optimal control problem, in order to find a policy which maximizes a process's expected reward, we first need to define a value function which quantifies how effective a given policy is. In this chapter, in order to clearly distinguish
between the value function and the action value function (which will be introduced in
the next section), we will refer to the value function as the \textit{state value function}.
Similar to the optimal control setting, the state value function $V^{\pi}(x)$ captures the accumulated reward expected when starting at $x$ and following the given policy $\pi$.
\begin{equation} \label{eq:State_value_fun}
V^{\pi}(x) = E \{ R_n \mid x_n = x \} = E \left \{ \sum\limits_{k=0}^{\infty} \alpha^k \, r_{n+k} \mid x_n = x \right \}
\end{equation}
From equation~\eqref{eq:State_value_fun} the Bellman equation can be derived by
relating the state value function of state $x$ with the state value function of
its successor states $x'$. Here we are using the fact that in the infinite-horizon
problem the value function is not a function of time, but just a function of
states. For a more detailed discussion, refer to \sref{subsec:val_infin_det}.

Extracting the first reward term from the summation in
equation~\eqref{eq:State_value_fun} and then factoring out an $\alpha$ gives
\begin{equation*}  
V^\pi (x) = E \left \{ r_{n} + \alpha \sum_{k=0}^\infty \alpha^k \, r_{n+k+1} \mid x_n =x \right\} 
\end{equation*}
Since the expectation is a linear operator, we can change its order with the 
first summation
\begin{equation*}
   V^\pi (x) = E \left \{ r_{n} \mid x_n =x \right\} + E \left \{ \alpha \sum_{k=0}^\infty \alpha^k \, r_{n+k+1} \mid x_n =x \right\}
\end{equation*}
Using the law of total expectation ( $E \left\{\dots \mid y \right\} = E \left\{ E \left\{ \dots \mid z, y \right\} \mid y \right\}$), we can additionally condition the term on the right by the control input applied and the subsequent value of the state, adding another expectation over these additional conditional arguments.
\begin{align*}
   & V^\pi (x) = E \left\{ r_{n} \mid x_n =x \right\} \\
   &+ E_{u_n} \left\{ E_{x_{n+1}} \left \{\highlight{E_{\pi} \left \{ \alpha \sum_{k=0}^\infty \alpha^k \, r_{n+k+1} \mid x_{n+1} =x', u_n =u, x_n =x \right\}} \mid u_n =u, x_n =x \right\} \mid x_n =x  \right\}
\end{align*}
Here we explicitly include the variables over which the expectation is taken for clarity.

By means of the problem's Markov property (cf. the definition of reward in equation
\ref{eq:expected_reward}), we can verify that for all \textit{k}s greater
and equal to zero
\begin{equation*}
Pr \left\{ r_{n+k+1} \mid x_{n+1} = x' \right\} = Pr \left\{ r_{n+k+1} \mid x_{n+1} = x', u_{n} = u, x_{n} = x \right\}, \qquad \forall k \geq 0.
\end{equation*}
Therefore the inner expectation can be simplified to the following
\begin{align*}
   V^\pi (x) &= E \left \{ r_{n} \mid x_n =x \right\} \\ 
   &+ E_{u_n} \left\{ E_{x_{n+1}} \left \{\highlight{E_{\pi} \left \{ \alpha \sum_{k=0}^\infty \alpha^k \, r_{n+k+1} \mid x_{n+1} =x' \right\}} \mid u_n =u, x_n =x \right\} \mid x_n =x  \right\} \\
   &= E \left \{ r_{n} \mid x_n =x \right\} + E_{u_n} \left\{ E_{x_{n+1}} \left \{\highlight{\alpha V^\pi(x')} \mid u_n =u, x_n =x \right\} \mid x_n =x  \right\} \\
\end{align*}
In the last equality, we have substituted the most inner expectation by the value 
function at state $x'$. By rolling back the expanded expectations, we obtain  
\begin{equation*}
   V^\pi (x) = E \left \{ r_{n} \mid x_n =x \right\} + E \left \{ \alpha V^\pi(x') \mid x_n =x \right\}
\end{equation*}
And then pulling the expectation outside of the sum:
\begin{equation} \label{eq:Bellman_RL}
  V^\pi (x) = E_{u_n,r_n,x_{n+1}} \left \{ r_{n} + \alpha V^\pi(x') \mid x_n =x \right\}
\end{equation}
Equation~\eqref{eq:Bellman_RL} is the Bellman equation for a given policy $\pi$.
This equation is essentially the same as the Bellman equation for the optimal
control problem in stochastic systems
(Equation~\ref{eq:infinite_horizon_stochastic_opt}). The only difference is that
Equation~\eqref{eq:Bellman_RL} evaluates the expectation with respect to the
reward function $r_n$ and policy, as well as the state transition, since the reward function and policy may now be stochastic as well. 

We can solve the Bellman equation in equation \eqref{eq:Bellman_RL} using 
the state transition probability, the policy, and the reward function.
\begin{align*}
   V^\pi (x) &=  E_{u_n,r_n,x_{n+1}} \left \{ r_{n} + \alpha V^\pi(x') \mid x_n =x \right\} \\
   &= E_{u_n} \left \{ E_{r_n,x_{n+1}} \left \{ r_{n} + \alpha V^\pi(x') \mid u_n = u, x_n =x \right\} \mid x_n =x \right\}
\end{align*}
Substituting the appropriate probability distributions for each of these expectations gives
\begin{equation} \label{eq:Bellman_RL_solved}
   V^\pi (x) = \sum\limits_u \pi (x,u) \sum_{x'} \mathcal P_{xx'}^u [\mathcal R_{xx'}^u + \alpha V^\pi (x') ]
\end{equation}
Equation~\eqref{eq:Bellman_RL_solved} has a few interesting features. First, it
shows that the value of the state value function at state $x$ is related to the
value of other states' value function. Therefore a changes in one state's value
function will affect the value of the other states' value functions as well. Second it shows that the
relationship between the value function of different states is linear. Therefore
if we collect the value functions associating to all the states in a vector of the size
$\norm{X}$ (number of states in MDP) and call it $\vV$, we will have
\begin{align} \label{eq:linear_equation_for evaluating_value_function}
 \vV &= \vA \vV + \vB \notag \\
 [\vA_{i,j}] &= \alpha \sum\limits_u {\pi (x_i,u) \mathcal P_{x_i x_j}^u}  \\
 [\vB_i] &= \sum\limits_u {\pi (x_i,u) \sum_{x'} {\mathcal P_{x_i x'}^u \mathcal R_{x_i x'}^u} } 
 \notag
\end{align}
\section{Action Value Function and the Bellman Equation}
We introduce the action value function as the adversary to the state value
function: It is the expected accumulated reward starting at $x$ and choosing
control input $u$, then following the policy $\pi$. It is defined as follows. Notice the additional conditioning with respect to $u$ inside the brackets of the expectation.
\begin{align}
Q^\pi(x,u) = E_{\pi} \{ R_n \mid x_n =x, u_n = u \} = E \left \{ \sum\limits_{k=0}^\infty {\alpha^k \, r_{n+k} \mid x_n = x, u_n = u} \right \} \notag
\end{align}

The difference between the
action value function and the state value function is that in the state value function
the control input sequence is totally generated according to the given policy,
$\pi$, while in the action value function, the \emph{first} control input is fixed and
is not extracted form the policy $\pi$ (note that the rest of the control input
sequence will be extracted form $\pi$). Based on this definition, we can write
the relationship between state value function and action value function as
follows
\begin{equation} \label{eq:V_Q_relationship}
V^\pi (x) = \sum\limits_u {\pi(x,u) Q^\pi (x,u)}
\end{equation}

In some literature, the action value function is also referred as state-value
function. It is also common to call it as Q-table. The reason for calling
it a table is that we can represent a set of action value functions as a table, associating rows to the states and columns to the control inputs. If the
number of the states is $\norm{X}$ and the the number of control inputs is $\norm{U}$, then the system's complete
Q-table will be of the size $\norm{X}$-by-$\norm{U}$.

Similar to equation~\eqref{eq:Bellman_RL}, we can derive the Bellman equation for
the action value function. Due to the similarity between this derivation and the
one for the value function, we will skip some steps. 
\begin{align*}
    Q^\pi(x,u) &= E \{ R_n \mid x_n = x, u_n=u \} \\
    &= E \left\{ r_{n} + \alpha \sum_{k=0}^\infty \alpha^k \, r_{n+k+1} \mid x_n =x, u_n=u \right\} \\
    &= E_{r_n,x_{n+1}} \left\{ r_{n} + \alpha E \left\{ \sum_{k=0}^\infty \alpha^k \, r_{n+k+1} \mid x_{n+1} =x' \right\} \mid x_n =x, u_n=u \right\} \\ 
    &= E_{r_n,x_{n+1}} \left\{ r_{n} + \alpha V^\pi(x') \mid x_n =x, u_n=u \right\}
\end{align*}
By plugging in equation~\eqref{eq:V_Q_relationship}
\begin{equation} \label{eq:Bellman_Q_RL}
    Q^\pi(x,u) = E_{r_n,x_{n+1}} \left\{ r_{n} + \alpha \sum\limits_{u'} {\pi(x',u') Q^\pi (x',u')} \mid x_n =x, u_n=u \right\}
\end{equation}
Equation~\eqref{eq:Bellman_Q_RL} is the Bellman equation for the action value function. Again we can solve this equation using the state transition probability, the policy and the reward function.
\begin{equation} \label{eq:Bellman_Q_RL_solved}
    Q^\pi(x,u) = \sum_{x'} \mathcal P_{xx'}^u \left[\mathcal R_{xx'}^u + \alpha \sum\limits_{u'} {\pi(x',u') Q^\pi (x',u')} \right]
\end{equation}
Since the control input is fixed in the action value function, the outer summation here is over the finite set $\vX$ of future states. All the statements we made in the previous section for
equation~\eqref{eq:Bellman_RL_solved} hold for this equation as well.

\section{Optimal Policy}
Solving the RL problem means finding a policy that achieves the
maximum accumulated reward over an infinite-time horizon. This means we want to
find a policy whose state value function satisfies the following for all $x
\in \vX$ and for any possible policy $\pi$.
\begin{equation}
V^*(x) \geq V^{\pi} (x)  
\end{equation}
In other words, we can write
\begin{equation}
V^*(x) = \max_{\pi} V^{\pi} (x)
\end{equation}
All the optimal policies which result in $V^*$ are denoted as $\pi^*$. Equivalently,
the optimal action value function is defined for all $x \in \vX$
and $u \in \vU$ as,
\begin{equation}
Q^* (x,u) = \max_{\pi} Q^{\pi} (x,u)
\end{equation}
The relationship between $V$ and $Q$ is given by
equation~\eqref{eq:V_Q_relationship}, therefore we can write the same for the 
optimal policy
\begin{equation}
V^*(x) = \sum\limits_a {\pi^*(x,u) Q^*(x,u)}
\end{equation}
Since $\pi^*(x,u)$ is always between 0 an 1 for all the control inputs, 
the following inequality holds
\begin{equation} \label{eq:V_Q_optimal_relationship_1}
V^*(x)= \sum\limits_u {\pi^*(x,u) Q^{*}(x,u)} \leq \max\limits_u Q^{*}(x,u)
\end{equation}
Therefore, if we have a policy which always chooses a control input that has the
maximum action value function, its value function will be higher or equal to the
optimal value function. However, this can only be correct if the policy is
actually the optimal policy. Therefore in the previous equation, the inequality should be replaced by equality.
\begin{equation} \label{eq:V_Q_optimal_relationship_2}
V^*(x) =  \max\limits_u Q^{*}(x,u)
\end{equation}
Equation~\eqref{eq:V_Q_optimal_relationship_2} is the relationship between the
optimal state value function and the optimal action value function.

Applying the Bellman equation~\eqref{eq:Bellman_Q_RL} for $Q^*$ leads to
\begin{equation*}
    Q^*(x,u) = E \left\{ r_{n} + \alpha \sum\limits_{u'} {\pi^*(x',u') Q^* (x',u')} \mid x_n =x, u_n=u \right\}
\end{equation*}
By using the results from equations~\eqref{eq:V_Q_optimal_relationship_1} and \eqref{eq:V_Q_optimal_relationship_2}, we will have
\begin{equation} \label{eq:optimal_Bellman_Q_RL}
    Q^*(x,u) = E \left\{ r_{n} + \alpha \max\limits_{u'} Q^{*}(x',u') \mid x_n =x, u_n=u \right\}
\end{equation}
This is the optimal Bellamn equation for the action value function. 
We can again solve this equation to give,
\begin{equation} \label{eq:optimal_Bellman_Q_RL_solved}
Q^*(x,u) = \sum_{x'} \mathcal P_{x x'}^u \left [ \mathcal R_{x x'}^u + \gamma \max\limits_{u'} Q^* (x', u')\right ].
\end{equation}
In order to derive the Bellman equation for the optimal state value function, 
we use the equation~\eqref{eq:Bellman_RL} for the optimal policy
\begin{equation} \label{eq:optimal_Bellman_RL}
   V^* (x) = \max_{u}{E \left \{ r_{n} + \alpha V^*(x') \mid x_n =x \right\}}
\end{equation}
This is the optimal Bellman equation for the optimal value function. We can also solve this equation to give,
\begin{equation} \label{eq:optimal_Bellman_RL_solved}
V^*(x) = \max\limits_{u \in \vU} \sum \limits_{x'} \mathcal P_{x x'}^{u} [ \mathcal R_{x x'}^{u} + \alpha V^* (x') ].
\end{equation}

\section{Policy Evaluation}  \label{sec:policy_evaluation}
We would like to be able to compute the optimal policy, given our state
transition probability distribution (equation~\ref{eq:transition_probability})
and the expected reward (equation~\ref{eq:expected_reward}). In order to do this,
we first need to be able to calculate the value function $V^\pi$, given some
arbitrary policy, $\pi$. This is referred to as Policy Evaluation.

From equation~\eqref{eq:linear_equation_for evaluating_value_function}, we can see that performing
Policy Evaluation involves solving a system of $\norm{X}$ (the number of possible
states) linear equations with $\norm{X}$ unknowns (each $V^\pi (x)$). While finding a
closed-form solution to this system of equations is often not feasible due to the
size of the state space, a solution can be found iteratively. If we start with
some arbitrary approximation of the value function, $V_0$, we can eventually
obtain the true value function by iteratively pushing our value function towards the true value function.

Iterative improvement of the approximated value functions requires an update rule
which guarantees that the subsequent value function is always a better
approximation. Here, we will use the Bellman equation which was derived in the previous
section (equation~\ref{eq:Bellman_RL_solved}).
\begin{equation}
V_{k+1}(x) = \sum\limits_u \pi(x,u) \sum\limits_{x'} \mathcal P_{xx'}^u [ \mathcal R_{xx'}^u + \alpha V_k(x') ]
\end{equation}
You can see that $V_k = V^\pi$ is a fixed point of this equation, since substituting it in gives the form of the Bellman equation. Though it will not be proven here, this method is guaranteed to converge to the true value function as long as either $\alpha < 1$ or "eventual termination is guaranteed from all states under policy $\pi$" (Sutton p.88). Furthermore since it is not feasible in practice to iterate infinitely, evaluation is typically terminated once $\max\limits_{x \in \vX}|V_{k+1}(x)-V_k(x)|$ is sufficiently small.
\begin{algorithm}[H] \caption{Iterative Policy Evaluation Algorithm}
\begin{algorithmic}
\Require $\pi$, the policy to be evaluated
\State Initialize $V(x) = 0$, for all $x \in \mathcal X^+$
\Repeat
    \State $\Delta \gets 0$
    \State {{\textbf for each }$x \in \mathcal X $}
      \State $v \gets V(x)$
       \State $V(x) \gets \sum_u \pi(x,u) \sum_{x'} \mathcal P_{xx'}^u [ \mathcal R_{xx'}^u + \gamma V(x') ]$
       \State $\Delta \gets \max(\Delta, |v- V(x)|)$
\Until {$\Delta < \theta \text{ (a small positive number)}$}
\Ensure $V \approx V^\pi$
\end{algorithmic}
\end{algorithm}
This algorithm can similarly be used to find the action Value Function. Once again, we simply set some initial guess $Q_0(x,u)$ for $\forall x \in \vX,\forall u \in \vU$, and use the Bellman equation as the update rule, and iterate until convergence.
\begin{equation}
Q_{k+1}(x,u) = \sum_{x'} \mathcal P_{xx'}^u \left [ \mathcal R_{xx'}^u + \alpha \sum\limits_{u'} {\pi^*(x',u')  Q_k(x', u')} \right ]
\end{equation}

\section{Policy Improvement}  \label{sec:policy_improvement}
Now that we can evaluate a policy, the next step is to determine how to improve a
policy. We will first define what makes one policy ``better'' than another. A policy $\pi'$ is better than
policy $\pi$ if the following two conditions hold:
\begin{align}
	&\forall \ x \in \vX: \quad \quad V^{\pi'}(x) \geq V^\pi(x) \notag \\
	&\exists \ x \in \vX: \quad \quad V^{\pi'}(x) > V^\pi(x)
	\label{eq:value_function_better_conditions}
\end{align}
This means that a superior policy must perform better than an inferior policy in
at least one state, and cannot perform worse in any state.

Practically speaking, what we would like to be able to do is the following: Given
some arbitrary policy, $\pi$, whose corresponding value function $V^\pi$ has
already been found using the Iterative Policy Evaluation Algorithm, we want to be
able to decide if modifying this policy at some state yields a better policy. We
can evaluate this decision by considering how our reward would change if we took
action $u$, which doesn't follow out existing policy, at $x$, then continued
following the existing policy $\pi$ afterwards. Evaluating this decision requires
comparing the original policy's value function $V^\pi(x)$ to the action value
function of the altered policy, which can conveniently be expressed as $Q^\pi(x,u)$. In order to compare these quantities, we will use
the Policy Improvement Theorem, which is explained below.

To better illustrate the Policy Improvement Theorem, we will temporarily consider
deterministic policies, even though the results derived also apply to stochastic 
policies. To simplify the notation, we define the function $\mu(x)$ to be:
\begin{equation} \label{eq:det_policy}
\mu(x) = a,   \qquad \text{where: } \pi(u=a \mid x) = 1.
\end{equation}
Assume that we have two policies, $\pi$ and $\pi'$. These policies are identical
at all states except $x$ (i.e. $\pi'(x) \neq \pi(x)$). If we assume that the value
of taking the action of policy $\pi'$ at $x$, and then following policy $\pi$
afterwards will yield more reward than always following policy $\pi$, we can say,
\begin{equation}
Q^\pi(x,\mu'(x)) \geq V^\pi(x)
\label{eq:policy_improvement_theorem_1}
\end{equation}
where $\mu'(x)$ is defined as equation~\ref{eq:det_policy} for the deterministic 
policy $\pi'$.

We can not directly, however, perform this action update at every state and expect this result to hold. It would be impossible to follow the old policy after the first time step because new policy would completely override the old policy. Because of this, the question becomes, if we greedily choose the action with higher action value function in
each state, will the new policy $\pi'$ always be a better policy than $\pi$.

In order to show this, we recursively expand 
equation~\eqref{eq:policy_improvement_theorem_1}, and show that choosing a greedy
action (control input) in each step will indeed increase the value function
\begin{align*}
	V^\pi(x) &\leq \highlight{Q^\pi(x,\mu'(x))} \\
	&\leq \highlight{E_{\pi} \left\{ r_{n} + \alpha V^{\pi}(x_{n+1}) | u_n= \mu'(x), x_n=x \right\}} 
\end{align*}	
note that the highlighted terms are equivalent.

By continuing to choose a control input which has a higher action value function 
(being greedy) on the next time step $n+1$, we should substitute $V^{\pi}(x_{n+1})$
by $Q^{\pi}(x_{n+1},\mu'(x_{n+1}))$.
\begin{align*}
	V^\pi(x) &\leq E_{\pi} \left\{ r_{n} + \alpha \highlight{Q^{\pi}(x_{n+1},\mu'(x_{n+1}))} \mid u_n= \mu'(x), x_n=x \right\} \\
	V^\pi(x) &\leq E_{\substack{u_{n+1} \\ x_{n+1}}} \left\{ r_{n} + \alpha \highlight{E_{\pi} \left\{ r_{n+1} + \alpha V^{\pi}(x_{n+2}) \mid u_{n+1}= \mu'(x'), x_{n+1}=x'\right\}} | u_n= \mu'(x), x_n=x \right\} \\
	V^\pi(x) &\leq E_{\substack{u_{n+1} \\ x_{n+1}}} \left\{ E_{\pi} \left\{ r_{n} + \alpha r_{n+1} + \alpha^2 V^{\pi}(x_{n+2}) | u_{n+1}= \mu'(x'), x_{n+1}=x', u_n= \mu'(x), x_n=x \right\} \right\}
\end{align*}
where the last equation indicates that we choose the first two control inputs
according to the policy $\pi'$, then we follow the policy $\pi$ afterwards. To
simplify the notation, we show this equation as
\begin{equation*}
V^\pi(x) \leq E_{ \substack{u_{[n,n+1]} \sim \pi' \\ u_{[n+2,\dots]} \sim \pi} } \left\{ r_{n} + \alpha r_{n+1} + \alpha^2 V^{\pi}(x_{n+2}) | x_n=x \right\}
\end{equation*}
Following the same procedure, we can expand the inequality to the end.
\begin{equation*}
V^\pi(x) \leq E_{\substack{u_{[n,n+1,n+2,\dots]} \sim \pi' \\ u_{\infty} \sim \pi}} \left\{ r_{n} + \alpha r_{n+1} + \alpha^2 r_{n+2} + \alpha^3 r_{n+3} + ... \mid  x_n=x \right\}
\end{equation*}
Since we are following the policy $\pi'$ from time $n$ to the end, we can
simply omit the subscription $\pi$ from the expectation.
\begin{align}
	V^\pi(x) &\leq E_{\pi'} \left\{ r_{n} + \alpha r_{n+1} + \alpha^2 r_{n+2} + \alpha^3 r_{n+3} + ... \mid  x_n=x \right\} = V^{\pi'}(x) \notag \\
	V^\pi(x)  &\leq V^{\pi'}(x)
	\label{eq:policy_improvement_theorem_2}
\end{align}
Here we have shown that the altered policy, $\pi'$ is indeed better than the
original policy. Let us now consider greedy improvement of our policy. Consider a
policy update rule of the form
\begin{align}
	\pi'(x) &= \arg\!\max\limits_u Q^\pi(x,u) \notag \\
	&= \arg\!\max\limits_u E \left\{ r_{n} + \alpha V^\pi(x_{n+1}) | x_n=x,u_n=u \right\}
	\label{eq:policy_improvement}
\end{align}
The original policy is modified to take the action which maximizes the reward in
the current time step, according to $Q^\pi$. From the \textit{Policy
Improvement Theorem}, we know that this greedy policy update will \textbf{always}
yield a policy which is greater than or equal to the original one. This greedy
approach is called \textit{Policy Improvement}.

Now, suppose that the new greedy policy is strictly equal to the original policy ($V^{\pi'} = V^\pi$). From equation~\eqref{eq:policy_improvement},
\begin{align}
	V^{\pi'}(x) &= \max\limits_u E \left\{ r_{n} + \alpha V^{\pi'}(x_{n+1}) | x_n=x,u_n=u \right\} \notag \\
	&= \max\limits_u \sum_{x'} \mathcal{P}_{xx'}^u [ \mathcal{R}_{xx'}^u + \alpha V^{\pi'}(x') ]    
\end{align}
which is the same as the optimal Bellman equation
(equation~\ref{eq:optimal_Bellman_RL_solved})! This means that \textit{Policy
Improvement} will always give us a better policy, unless the policy is already
optimal. Once again, this section only considered deterministic policies, even
though the methods and results described here do in general extend to stochastic
policies, as we will show later for a class of stochastic policies known as
$\varepsilon$-greedy policies (See \cite{Sutton1998}, p. 94 for more details).

\section{Model-Based RL: Generalized Policy Iteration} \label{sec:model_based_Rl}
\begin{figure} [h]
\centering
  \includegraphics[width=0.3\linewidth]{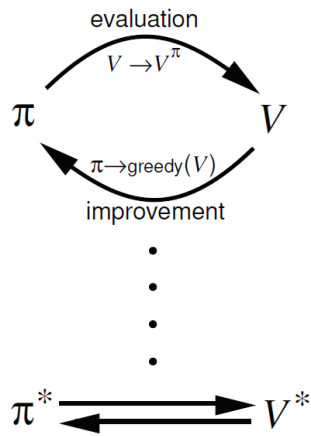}
  \caption{Generalized Policy Iteration. Reproduced from \cite{Sutton1998}.}
  \label{fig:GPI}
\end{figure}

In this section, we will introduce three algorithms for solving the RL
optimization problem. These algorithms are based on the idea of iteratively
evaluating a policy and then improving it (Figure \ref{fig:GPI}). The granularity
of the interaction between the Policy Evaluation and the Policy Improvement can
be at any level from completing one exhaustively before starting the other to alternating between
them after every single step of both processes.

This section starts with introducing the Policy Iteration algorithm, in which the
Policy Evaluation and the Policy Improvement process are completely separated.
Then the Value Iteration algorithm is introduced. The Value Iteration
algorithm has a finer level of interaction between Policy Evaluation and the
Policy Improvement. Finally we introduce a more general algorithm referred
to as Generalized Policy Improvement (GPI) which essentially allows to use any level of
granularity between the Policy Evaluation and the Policy Improvement.

\subsection{Policy Iteration} \label{sec:policy_iteration}
Now armed with the techniques of \textit{Policy Evaluation} and \textit{Policy
Improvement}, finding an optimal policy is quite straight forward. Starting from
some arbitrary policy, $\pi_0$, we can iteratively find the corresponding value
function, $V^{\pi_0}$. We can then greedily improve our policy with respect the
that value function to obtain a new policy $\pi_1$. We then use the previous
value function as the initialization for the following \textit{Policy
Evaluation}. We can repeat this process until our policy converges. Each
subsequent policy is guaranteed to be a strict improvement to the previous one
unless it is already optimal. Also, since a finite MDP has only a finite number
of possible actions, and therefore a finite number of possible policies, we know
our solution will converge in a finite number of iterations. This process is
called \textit{Policy Iteration}. The complete algorithm is given in
Algorithm~\ref{alg:policy_iteration}.

\begin{algorithm} [H] 
\caption{Policy Iteration}
\label{alg:policy_iteration}
\begin{algorithmic}
	\State \textbf{1. Initialization}
	\State select $V(x) \in \Re$\, and \,$\pi(x) \in \vU$ \,arbitrarily for all $x \in \vX$
	\State \textbf{2. Policy evaluation}
		\Repeat
			\State$\Delta \gets 0$
				\hspace{4mm}\ForEach {$x \in \mathcal X $}
				\State $v \gets V(x)$
				\State $V(x) \gets \sum_{u} {\pi(x,u) \sum_{x'} \mathcal P_{xx'}^{u} [ \mathcal R_{xx'}^u + \alpha V(x')]}$
				\State $\Delta \gets \max(\Delta, |v- V(x)|)$
		\Until {$\Delta < \theta \text{ (a small positive number)}$}
	\State \textbf{3. Policy Improvement}
		\State $policyIsStable \gets true$
		\For {$x \in \mathcal X$}
			\State $b \gets \pi(x)$
			\State $\pi(x) \gets \arg\!\max_u \sum_{x'} \mathcal P_{xx'}^u [\mathcal R_{xx'}^u + \alpha V(x')]$
			\If {$b \neq \pi(x)$}
				\State $policyIsStable \gets false$
			\EndIf
		\EndFor
	\If {$policyIsStable$} 
		\State stop 
	\Else 
		\State go to 2 
	\EndIf
	\Ensure a policy, $\pi$, such that: $\pi(x)= \operatorname{arg} \max_u \sum_{x'} \mathcal P_{x x'}^u [\mathcal R_{x x'}^u + \alpha V(x')]$	
\end{algorithmic}
\end{algorithm}

\subsection{Value Iteration} \label{sec:value_iteration}
One of the drawbacks of the Policy Iteration is that each iteration of the
algorithm involves a Policy Evaluation, which may require several sweeps of the
whole state space. Since Policy Evaluation typically converges to the optimal policy in the
limit, we normally need to exit after a finite number of sweeps. This means
that we need to truncate the Policy Evaluation part after some iterations.
One extreme of this approach would be to exit after the first sweep of the Policy
Evaluation and then perform a Policy Improvement process. In this case the policy
Evaluation and Policy Improvement can be merged in a single update rule as
\begin{equation}
V_{k+1}(x) = \max\limits_u \sum_{x'} \mathcal P_{x x'}^u [\mathcal R_{x x'}^u + \alpha V_k(x')]
\end{equation}
The algorithm which is based on this update rule is called \textit{Value Iteration}. It
is proven that Value Iteration converges to the optimal state value function in
limit. Also, this update rule has another interesting interpretation. By comparing it
to the optimal Bellman equation for value function
(\eref{eq:optimal_Bellman_RL_solved}), you can see that
it is an iterative backup rule for solving this equation. The complete
algorithm is given as Algorithm~\ref{alg:value_iteration}.
\begin{algorithm}[H]
\caption{Value Iteration}
\label{alg:value_iteration}
\begin{algorithmic}
\State \textbf{Initialization:} $V(x) \in \Re$\, and \,$\pi(x) \in \vU$ \,arbitrarily for all $x \in \vX$
\Repeat
\State$\Delta \gets 0$
\For {$x \in \mathcal \vX $}
	\State $v \gets V(x)$
	\State $V(x) \gets \max_u \sum_{x'} \mathcal P_{x x'}^u [ \mathcal R_{x x'}^u + \alpha V(x')]$
	\State $\Delta \gets \max(\Delta, |v- V(x)|)$
\EndFor
\Until {$\Delta < \theta \text{ (a small positive number)}$}
\Ensure a policy, $\pi$, such that: $\pi(x)= \operatorname{arg} \max_u \sum_{x'} \mathcal P_{x x'}^u [\mathcal R_{x x'}^u + \alpha V(x')]$
\end{algorithmic}
\end{algorithm}

\subsection{Generalized Policy Iteration}  \label{sec:generalized_policy_iteration}
Generalized Policy Iteration (GPI) refers to the class of algorithms which
are based on the general idea of synchronously performing Policy Evaluation and Policy
Improvement, independent of the granularity of the two processes. For
example in the Policy Iteration algorithm these two process are performed one after the other, each
completing before the other starts. In value Iteration algorithm, only after a
single sweep of the Policy Evaluation, the Policy Improvement is performed. In
general, there two processes can be interleaved in any fine level. Almost all of the
algorithms in this chapter conform to this GPI idea. The convergence
to the optimal value is typically guaranteed as long as both processes continue
to update one after the other.
\newpage

\section{Sample-Based RL: Monte Carlo Method}
Monte Carlo methods propose a sample-based approach to solve the RL problem,
previously introduced in \sref{sec:the_RL_problem}. Unlike the model-based
approaches, they are based on \emph{samples} generated from the agent's interaction
with its environment  instead of requiring the reward and the transition
probability distributions. Since the information for the Monte Carlo Algorithm is
provided through samples, there should be a practical way to extract data and
save it. This simply implies that tasks with infinite execution time will not
be considered in these approaches. Therefore in this section, we restrict our 
discussion to tasks with limited durations which we refer to as episodic tasks.

The main idea of the Monte Carlo approach carries over from the GPI algorithm
(section~\ref{sec:generalized_policy_iteration}). However, looking at a policy
iteration algorithm (see e.g. Algorithm~\ref{alg:policy_iteration}), we realize
that model-knowledge is required in the form of the transition probabilities and
the reward function, in both the policy evaluation and policy improvement steps.
Furthermore, in order to obtain the optimal policy from the calculated optimal
value function, we need to use the following equation, which is based on the
state transition model
\begin{align} 
   \pi^* (x) &= \arg\!\max_{\pi}{E \left \{ r_{n} + \alpha V^*(x') \mid x_n =x \right\}} \notag \\
   &= \arg\!\max\limits_{\pi} \sum \limits_{x'} \mathcal P_{x x'}^{u} [ \mathcal R_{x x'}^{u} + \alpha V^* (x') ] \notag 
\end{align}
However, if the action value function is calculated directly, there is no need for
a transition model, since
\begin{equation*} 
   \pi^* (x) = \arg\!\max_{\pi} Q^*(x,u)
\end{equation*}
As we shall see, if a model is not available, it is particularly useful to
\emph{estimate action value functions} $Q^\pi(x,u)$ \emph{instead of state value
functions} $V^\pi(x)$.

\subsection{Sample-Based GPI with Action Value Function}  \label{sec:sample_based_GPI} 
Algorithm~\ref{alg:general_policy_iteration_with_Q} states a GPI that utilizes
action value functions instead of state value functions (note that the policy
evaluation stopping criterion is left open as design choice). In the following
sections, we will examine how the fundamental elements of GPI (namely Policy
Evaluation and Policy Improvement) can be carried over to the sample-based case.
\begin{algorithm}[t]
\caption{Generalized Policy Iteration (GPI) using the action value function $Q^\pi(x,u)$}
\label{alg:general_policy_iteration_with_Q}
\begin{algorithmic}
	\State \textbf{1. Initialization}
		\State $Q^\pi (x,u) \in \mathbb{R} $
	\State \textbf{2. Policy Evaluation (PE)}
		\Repeat
			\For {select a pair $(x,u) \ \text{with} \ x \in \mathcal X, \ u \in \mathcal U $}
				\State $v \gets Q^\pi(x,u)$
				\State $Q^\pi(x,u) \gets \sum_{x'}\mathcal P^u_{xx'} \left[ \mathcal R_{xx'}^u + \alpha \sum_{u} \pi\left(x',u \right) Q^\pi \left( x',u \right) \right]$
			\EndFor
		\Until \text{"individual PE criterion satisfied"}
	\State \textbf{3. Policy Improvement}
		\State policy-stable $\gets$ true
		\For {$x \in \mathcal X$}
			\State $b \gets \pi(x)$
			\State $\pi(x) \gets \arg\!\max\limits_{\pi} \sum \limits_{x'} \mathcal P_{x x'}^{u} [ \mathcal R_{x x'}^{u} + \alpha V(x') ]$
			\If {$b \neq \pi(x)$}
				\State policy-stable$ \gets $false
			\EndIf
		\EndFor
	\If {(policy-stable $==$ true)} 
		\State stop; 
	\Else 
		\State go to 2. 
	\EndIf
\end{algorithmic}
\end{algorithm}
\paragraph{Model-free Policy Improvement}
First, we consider the Policy Improvement step of the GPI as given in
Algorithm~\ref{alg:general_policy_iteration_with_Q}. Given an action value
function $Q^\pi(x,u)$, policy improvement is done by making the policy greedy
w.r.t. the current action value function. In other words, for every state $x$ we
evaluate all possible control inputs $u$ and choose the one which leads to the
the best possible combination of reward and action-value of the successor state.
This can be equivalently written as
\begin{equation*}
\pi(x) = \arg \max \limits_{u}   Q^\pi(x,u)
\end{equation*}
which does not require a model to construct and is compliant with the Policy
Improvement Theorem~\eqref{eq:policy_improvement_theorem_2}.

\paragraph{Model-free Policy Evaluation}  \label{sec:model_free_policy_evaluation}
Consider the Policy Evaluation step of GPI as given in
Algorithm~\ref{alg:general_policy_iteration_with_Q}. Recall that the action value
is the expected (discounted) accumulative reward starting from a certain state $x$
by choosing control $u$ and then following a policy $\pi$, thus
\begin{align}
Q^\pi(x,u) = E_{\pi} \{ R_n \mid x_n =x, u_n = u \} \quad \text{.} \notag
\end{align}
Taking an episodic point of view, an obvious way to estimate $Q^\pi(x,u)$ is
simply by averaging the returns observed after visits to that state-action pair.
After a total of $N$ visits (each observation is indexed with $i$), we write the
averaged discounted return $\widetilde{Q}_N^\pi(x,u)$ as
\begin{align}
\label{eq:MonteCarlo_Q_N_est}
\widetilde{Q}_N^\pi(x,u) \approx \frac{1}{N}\sum\limits_{i=1}^N R_n^i (x,u) = \frac{1}{N} \sum\limits_{i=1}^N \left( r_{n}^i + \alpha \, r_{n+1}^i + \alpha^2 \, r_{n+2}^i + \dots \right) \quad \text{.} 
\end{align}
Equation~\eqref{eq:MonteCarlo_Q_N_est} implies that for calculating the action
value function for each state-control pair, we need to store all the retrieved
samples as they are received at runtime and then calculate the average for each. This
approach is usually called batch computation. However as you may guess, it is an
inefficient approach both in terms of memory usage and computation requirements. On the other hand, we can perform this averaging in a recursive
way. To do so, assume that a new sample indexed as $N+1$ is acquired. We
decompose $\widetilde{Q}_{N+1}^\pi(x,u)$ as
\begin{align}
\label{eq:MonteCarlo_Q_N_est_recursive}
\widetilde{Q}_{N+1}^\pi(x,u) &= \frac{1}{N+1}\sum\limits_{i=1}^{N+1} R_n^i (x,u) \notag \\ 
&= \frac{1}{N+1}\left(\sum_{i=1}^N R_n^i(x,u) + R_n^{N+1}(x,u) \right) \notag \\
&= \frac{N}{N+1} \cdot \widetilde{Q}_N^\pi(x,u) + \frac{1}{N+1} \cdot R_n^{N+1}(x,u) \qquad \text{by Equation \eqref{eq:MonteCarlo_Q_N_est}} \notag \\
&=  \widetilde{Q}_N^\pi(x,u) + \underbrace{\frac{1}{N+1}}_{=:\omega_{N+1}} \left( R_n^{N+1} - \widetilde{Q}_N^\pi(x,u) \right)  
\end{align}
which yields a recursive update equation for the estimated action-value function
with ``learning rate" $\omega_{N+1} = 1/(N+1)$. In the stationary case, by taking
the limit $N \rightarrow \infty $,
expression~\eqref{eq:MonteCarlo_Q_N_est_recursive} converges towards the true
value of the action value function
\begin{equation*}
\lim \limits_{N \rightarrow \infty} \widetilde{Q}_N^\pi(x,u) = Q^\pi(x,u) \quad \textbf{.}
\end{equation*}
However it can be shown, for any arbitrary learning rate sequence $\{\omega_N\}$
that meets the \emph{Robbins-Monro} conditions
(equation~\eqref{eq:Robbins_Monro_1} to \eqref{eq:Robbins_Monro_3}) the recursive
update equation converges to the true value in the stationary case (i.e. the case
that the stochastic property of the problem does not change over time).
\begin{align}
\lim \limits_{N \rightarrow \infty} \omega_N = 0 \label{eq:Robbins_Monro_1}\\
\sum \limits_{N = 1}^{\infty} \omega_N = \infty   \label{eq:Robbins_Monro_2}\\
\sum \limits_{N = 1}^{\infty} \omega_N^2 < \infty  \label{eq:Robbins_Monro_3}
\end{align}
The first condition~\eqref{eq:Robbins_Monro_1} ensures that the successive
corrections decrease in magnitude so that the process can converge to a limit.
The second condition~\eqref{eq:Robbins_Monro_2} is required to ensure that
the algorithm does not converge prematurely and the third
condition~\eqref{eq:Robbins_Monro_3} is needed in order to ensure that the
accumulated noise has finite variance and and hence does not spoil convergence.
However, the above conditions are tailored to the \emph{stationary case}. Note
that in sample-based methods, the involved distributions are typically
\emph{non-stationary}. This is because the return depends on the policy, which is
changing and improving over time. It can be shown that, in order to reach a good
approximation of the true expected value by sampling in the non-stationary case,
conditions~\eqref{eq:Robbins_Monro_1} and~\eqref{eq:Robbins_Monro_3} are not
required to hold. Therefore we normally choose a constant learning rate. 

To summarize, we define the update-rule for the $(N+1)$-th action value after
receiving a new sample reward $R_{n+1}$ for a particular action-value pair ($x$,
$u$) as
\begin{equation}
\widetilde{Q}_{N+1}^\pi(x,u) = \widetilde{Q}_N^\pi(x,u) + \omega_{N+1} \cdot \left( R_{N+1} - \widetilde{Q}_{N}^\pi(x,u) \right)
\end{equation}
which can be initialized with any arbitrary values (e.g. zero or a random number). 

\subsection{Naive Monte Carlo Control with ``Exploring Starts''}
\label{sec:mc_exploring_starts}
As shown in \sref{sec:sample_based_GPI}, the GPI algorithm can be used in a
purely model-free way. The key elements for moving from a model-based to a
sample-based algorithm were: 1) estimating the \textit{action} value function instead of the \textit{state} value
function, 2) replacing the Bellman updating rule in the Policy Evaluation by a
numerical averaging. 3) using the estimated action value function in the Policy
Improvement process. This implementation of the GPI algorithm which is based on the
raw samples rather than the model is referred to as the Monte Carlo method.

Equipped with the Monte Carlo method, we can solve the RL problem by only
using samples. A sample in this context is defined as a sequence of
state-action-rewards which is acquired by initializing the agent in an arbitrary
state-action pair and then interacting with the environment according to the policy
at hand. As indicated by Algorithm~\ref{alg:general_policy_iteration_with_Q}, we
need to estimate the action value function for every possible state-action pair.
This is ensured by initializing the agent in every one of these pairs.

Furthermore, in order to increase the sampling efficiency, we can reuse the
already extracted samples by considering each of the sub-sequences as a new sample
which has a different starting state-action pair. Although this reusing scheme
helps to acquire samples from state-action pairs that the agent is not
initialized from, it cannot guarantee a perfect coverage of all the state-action
pairs. This is the general problem of \emph{maintaining exploration}. Since the
current version of the algorithm uses a greedy policy (greedy with respect to
latest approximation of the action value function), in order to maintain the
exploration, we should start from each possible state-input pair.
\begin{align*}
\underbrace{x_{n}, u_{n}}, r_{n}, x_{n+1}, u_{n+1}, r_{n+1}, x_{n+2}, u_{n+2}, r_{n+2}, \dots, x_{T-1}, u_{T-1}, r_{T-1}, x_{T}, u_{T}, r_{T}  &\\
\underbrace{x_{n+1}, u_{n+1}}, r_{n+1}, x_{n+2}, u_{n+2}, r_{n+2}, \dots, x_{T-1}, u_{T-1}, r_{T-1}, x_{T}, u_{T}, r_{T}  &\\
\underbrace{x_{n+2}, u_{n+2}}, r_{n+2}, \dots, x_{T-1}, u_{T-1}, r_{T-1}, x_{T}, u_{T}, r_{T}  \\
\vdots &\\
\underbrace{x_{T-1}, u_{T-1}}, r_{T-1}, x_{T}, u_{T}, r_{T} &
\end{align*}
Relying only on samples to solve the RL problem makes it possible to use the
algorithm on real world scenarios where the samples are directly drawn from the
interaction between the agent and the world. Therefore, the agent can improve its
performance while it interacts with the environment for fulfilling its task. This
process is referred to in the literature as ``Learning while Living''.

However, as noted before, in order to guarantee that the algorithm actually
converges to the optimal action value function (and the optimal corresponding
policy), we need to initialize it from every possible state-input pair. This is
contrary to the idea of ``Learning while Living'' since we need to artificially
place the agent in different initial states. This approach is called \emph{``exploring
starts"} and it actually is a naive implementation of Monte Carlo methods - hence
we also call the corresponding algorithm \emph{``Naive Monte Carlo"}. A pseudo-code of the algorithm  is shown in Algorithm ~\ref{alg:naive_monte_carlo_on_policy}.  In the next
section, we will introduce another implementation of the Monte Carlo method which
overcomes this limitation by introducing a non-greedy exploration policy. 
\begin{algorithm}[ht] 
\caption{Naive Monte Carlo Algorithm Assuming Exploring Starts}
\label{alg:naive_monte_carlo_on_policy}
\begin{algorithmic}
\State Initialize, for all $x \in \mathcal X$, $u \in \mathcal U$
\State \hspace{4mm}$Q(x,u) \gets$ arbitrary
\State \hspace{4mm}$\pi \gets$ an arbitrary deterministic policy 
\State \textbf{Repeat forever:}
\State \hspace{4mm}(a) Exploring start: select random pair ($x,u$)
\State \hspace{4mm}(b) Select a policy $\pi$ and generate an episode: $x_0, u_0, r_0,x_1, u_1, r_1, \ldots , x_{N}, u_{N}, r_N.$
\State \hspace{4mm}(c) Sample-based \textbf{Policy Evaluation}: 
\State \hspace{8mm} for each pair $x,u$ appearing in the episode:
\State \hspace{12mm} $R \gets$ return following the first occurrence of $x,u$
\State \hspace{12mm} $Q(x,u) \gets Q + \omega \cdot \left( R - Q(x,u) \right)$
\State \hspace{4mm} \normalsize (d) \textbf{Policy improvement}:
\State \hspace{8mm}$\pi(x) \gets \operatorname{arg} \max_u Q(x,u)$
\end{algorithmic}
\end{algorithm}

\subsection{On-Policy Monte Carlo Control with $\varepsilon$-soft Policy}
\label{sec:epsilon-soft-monte-carlo}

In this section we introduce an on-policy method, that is, a method that attempts
to improve the policy that is used to make the decisions. In on-policy Monte
Carlo control methods, the policies are generally \emph{soft}, meaning that
$\pi(x,u)>0$ for all control inputs in each state.

While in the previous algorithm, we should initialize the agent in each
state-action pair in order to balance exploration-exploitation, this
implementation removes this restriction by introducing a soft policy (called
$\varepsilon$-greedy policy) instead of the greedy one.

Therefore in the Policy Improvement process of the algorithm, we  use an
$\varepsilon$-greedy policy improvement instead of the greedy policy improvement.
That means, with higher probability we choose an action that has maximum
estimated action value, but with some non-zero probability $\varepsilon$, we
select other actions randomly. The $\varepsilon$-greedy policy is defined as
follows
\begin{equation}
\label{eq:greedy_non_greedy}
\pi(u \mid x) = 
\begin{cases} 
\frac{\varepsilon}{\norm{\mathcal U}} \qquad  \qquad \qquad \enspace \text{for the non-greedy action} \\
1 - \varepsilon \left( 1 - \frac{1}{\norm{\mathcal U}} \right)   \quad \text{for the greedy action} 
\end{cases}
\end{equation}
for some $\varepsilon > 0$ and $\norm{\mathcal U}$ being the number of the control
inputs.

The overall idea of on-policy Monte Carlo is still based on GPI, however, with
the assumption of the $\varepsilon$-greedy policy, we cannot improve the policy
by making it greedy w.r.t. the current action value function. In the following,
we show that the Policy Improvement theorem still holds for $\varepsilon$-greedy
policies. In particular, we show that any $\varepsilon$-greedy policy with
respect to the latest estimation of $Q^\pi$ is an improvement over the current
$\varepsilon$-greedy policy. This is similar to what we showed in
equation~\eqref{eq:policy_improvement_theorem_1} for the greedy Policy
Improvement (see \sref{sec:policy_improvement}), but now for the
$\varepsilon$-greedy case. Let $\pi'$ be the $\varepsilon$-greedy policy w.r.t.
the $Q^{\pi}$ (the action value function associated to the policy $\pi$).
The proof steps are very similar to the one for the greedy policy. First, we 
show that if at an arbitrary time step e.g. $n$, we perform w.r.t. the new policy
$\pi'$, then we follow the old policy $\pi$ afterwards, the expected accumulated
reward is greater or equal to the value function of the old policy,
$V^{\pi}$.
\begin{align} \label{eq:epsilon-soft-policy-improvement-theorem}
E_{\substack{u_{n} \sim \pi' \\ u_{[n+1,\dots]} \sim \pi}} \left\{ R_n \mid  x_n=x \right\} &= \sum \limits_{u} \pi'(x,u) Q^\pi(x,u) \notag\\
&=\frac{\varepsilon}{\norm{\mathcal U}} \sum \limits_{u}Q^\pi(x,u)+(1-\varepsilon) \max \limits_{u} Q^\pi(x,u) 
\end{align}
Now the goal is to find the relation between the left hand side of the equation
and the value function of the policy $\pi$. To do so, we use the following trick
which implies that the weighted average of a set of numbers is always less or equal
to their maximum. Therefore, we can write
\begin{equation} \label{eq:max_average_ineq}
\max \limits_{u} Q^\pi(x,u) \geq \sum \limits_{u}{\frac{\pi(x,u)-\frac{\varepsilon}{\norm{\mathcal U}}} {1-\varepsilon}Q^\pi(x,u)}.
\end{equation}
Note that $\left\{ \frac{\pi(x,u)-\frac{\varepsilon}{\norm{\mathcal U}}}
{1-\varepsilon} \right\}$ are the nonnegative averaging weights which actually
sum up to one.
\begin{align*}
\sum \limits_{u}{\frac{\pi(x,u)-\frac{\varepsilon}{\norm{\mathcal U}}}{1-\varepsilon}} &= \sum \limits_{u}{\frac{\pi(x,u)}{1-\varepsilon}} - \sum \limits_{u}{\frac{\frac{\varepsilon}{\norm{\mathcal U}}}{1-\varepsilon}} \\
&= \frac{1}{1-\varepsilon} \sum \limits_{u}{\pi(x,u)} - \frac{\frac{\varepsilon}{\norm{\mathcal U}}}{1-\varepsilon} \sum \limits_{u}{1} \\
&= \frac{1}{1-\varepsilon} - \frac{\frac{\varepsilon}{\norm{\mathcal U}}}{1-\varepsilon} \norm{\mathcal U} = 1
\end{align*}
By substituting $\max \limits_{u} Q^\pi(x,u)$ in the
equation~\eqref{eq:epsilon-soft-policy-improvement-theorem} using the inequality
\eqref{eq:max_average_ineq}, we get
\begin{equation*}
E_{\substack{u_{n} \sim \pi' \\ u_{[n+1,\dots]} \sim \pi}} \left\{ R_n \mid  x_n=x \right\} \geq \frac{\varepsilon}{|\mathcal U (x)|} \sum \limits_{u}Q^\pi(x,u)+(1-\varepsilon) \sum \limits_{u}{\frac{\pi(x,u)-\frac{\varepsilon}{|\mathcal U(x)|}}{1-\varepsilon}Q^\pi(x,u)}
\end{equation*}
The whole right-hand side expression can be rewritten as
\begin{align*}
\frac{\varepsilon}{|\mathcal U (x)|} \sum \limits_{u}Q^\pi(x,u)- 
\frac{\varepsilon}{|\mathcal U (x)|} \sum \limits_{u}Q^\pi(x,u)+
\sum \limits_{u} \pi(x,u) Q^\pi(x,u) = \sum \limits_{u} \pi(x,u) Q^\pi(x,u)
\end{align*}
Therefore we have
\begin{equation*}
E_{\substack{u_{n} \sim \pi' \\ u_{[n+1,\dots]} \sim \pi}} \left\{ R_n \mid  x_n=x \right\} \geq \sum \limits_{u} \pi(x,u) Q^\pi(x,u)
\end{equation*}
which clearly resolves to be equal to $V^\pi(x)$, hence
\begin{equation}
E_{\substack{u_{n} \sim \pi' \\ u_{[n+1,\dots]} \sim \pi}} \left\{ R_n \mid  x_n=x \right\} \geq V^\pi(x)
\end{equation}
In the second step of the proof, we show that the expected accumulated
reward in case we perform the first two steps according to the policy
$\pi'$ and then follow the policy $\pi$ is greater or equal to
$E_{\substack{u_{n} \sim \pi' \\ u_{[n+1,\dots]} \sim \pi}} \left\{ R_n \mid 
x_n=x \right\}$. Since this proof is very similar to the previous it will be skipped but the results will be used:
\begin{align*}
E_{\substack{u_{[n,n+1]} \sim \pi' \\ u_{[n+2,\dots]} \sim \pi}} \left\{ R_n \mid  x_n=x \right\} &\geq E_{\substack{u_{n} \sim \pi' \\ u_{[n+1,\dots]} \sim \pi}} \left\{ R_n \mid  x_n=x \right\} \\
&\geq V^\pi(x)
\end{align*}
Following the same procedure, we can replace the policy $\pi'$ until the end. Therefore
\begin{equation}
E_{\substack{u_{[n,n+1,\dots]} \sim \pi'}} \left\{ R_n \mid  x_n=x \right\} \geq V^\pi(x)
\end{equation}
The left-hand side of this inequality is actually the value function for the
policy $\pi'$. So we have
\begin{equation}
V^{\pi'}(x) \geq V^\pi(x)
\end{equation}
which implies that the policy $\pi'$ is better than the policy $\pi$. Therefore,
we have proven the Policy Improvement theorem for the $\varepsilon$-greedy policy case.
It can be shown that the Policy Improvement for $\varepsilon$-greedy policies
converges when the policy is optimal amongst the $\varepsilon$-greedy 
policies (for more details refer to Sutton section 5.4).

A major difference between this method and the Policy Iteration from the previous sections is that it
does not find the \emph{best overall policy} -- it just converges to the
\emph{best $\varepsilon$-greedy policy}, which contains occasional exploration
moves (which are suboptimal). But at the same time, it eliminates the need for
exploring starts. To tackle this issue, we can eventually decrease the
$\varepsilon$ to zero. Therefore, in the end the algorithm converges to the
optimal greedy policy. However it is important that the decrease has a reasonable
rate since a fast decrease can cause a premature convergence. The reason is that
by decreasing the $\varepsilon$, we favor more exploitation against exploration
which can cause some favorable states to never be evaluated. The $\varepsilon$-soft,
On-Policy Monte Carlo Control is given as
Algorithm~\ref{alg:epsilon_soft_monte_carlo}.
\begin{algorithm}[ht] 
\caption{$\varepsilon$-soft, On-Policy Monte Carlo Algorithm}
\label{alg:epsilon_soft_monte_carlo}
\begin{algorithmic}
\State \hspace{4mm} choose a constant learning rate, $\omega$
\State \hspace{4mm} choose a positive $\varepsilon \in (0,1]$
\State \hspace{4mm}$Q^\pi(x,u) \gets$ arbitrary
\State \hspace{4mm}$\pi \gets$ an arbitrary $\varepsilon$-soft policy 
\State \textbf{Repeat forever:}
\State \hspace{4mm}(a) generate an episode using $\pi$ 
\State \hspace{4mm}(b) Policy Evaluation
\State \hspace{8mm} \textbf{for each} pair ($x, u$) appearing in the episode
\State \hspace{12mm}$R \gets$ return following the first occurrence of $(x,u)$
\State \hspace{12mm}$Q^\pi(x,u) \gets Q^\pi(x,u) + \omega \left( R - Q^\pi(x,u) \right)$ 
\State \hspace{4mm}(c) Policy Improvement
\State \hspace{8mm} \textbf{for each:} $x$ in the episode:
\State \hspace{12mm}$u^* \gets \operatorname{arg} \max_u Q^\pi(x,u)$
\State \hspace{12mm} For all $a \in \mathcal U (x)$: 
\State \hspace{18mm}
	$ \pi(x,u) \gets 
	\begin{cases} 
	\frac{\varepsilon}{|\mathcal U (x)|} \qquad  \qquad \qquad \ \text{if } u \neq u^* \\
	1 - \varepsilon \left( 1 - \frac{1}{|\mathcal U (x)|} \right)   \quad \text{if } u = u^* 
	\end{cases}$
\State \hspace{4mm}(d) (\textit{optional}) decrease $\varepsilon$.
\end{algorithmic} 
\end{algorithm}

\newpage
\section{Sample-Based RL: Q-Learning}  \label{sec:q-learning}
One of the most important breakthroughs in reinforcement learning was the
development of \emph{Q-learning}. Roughly speaking, it is a combination of Monte
Carlo ideas and Dynamic Programming ideas. Like Monte Carlo methods, Q-learning
learns directly from raw experience without requiring model information. However
like Dynamic Programming, it updates its estimates based partly on the
neighboring state estimates, without waiting for the episode end.

We start the discussion be recalling the Bellman Equation for the optimal action
value functions which is given in Equation~\eqref{eq:optimal_Bellman_Q_RL} as
\begin{equation*}
    Q^*(x_n,u_n) = E \left\{ r_{n} + \alpha \max\limits_{u'} Q^{*}(x',u') \mid x_n =x, u_n=u \right\}
\end{equation*}
where we denote the control action that follows $u$ as $u'$. Since we would like
to learn the action value $Q^*(x,u)$ of a certain state-action pair from
experience without using a model, we sample and average action values over $N$
occurrences of $(x,u)$. This leads us to conventional averaging as in
equation~\eqref{eq:MonteCarlo_Q_N_est}. However, this time, we are not
considering the whole tail of episode until its termination -- instead, we only
consider the following state $x_i'$ and append the current reward ($r_n^i$) to
the best action value function that can be gained by selecting a particular
$u'$. Therefore, figuratively speaking, we average over instances of the
right-hand side of the Bellman optimality equation as we are sampling.
\begin{equation*}
    \widetilde Q(x,u) = \frac{1}{N} \sum \limits_{i=1}^N \left( r_{n}^i + \alpha \max\limits_{u} Q(x_i',u) \right)
\end{equation*}
which can also be formulated recursively similar to what we did in
Equation~\eqref{eq:MonteCarlo_Q_N_est_recursive}. That leads us to the following
update equation for a sample that appears in episode $i+1$ at time $n$:
\begin{equation}
\widetilde Q^{i+1}(x_n,u_n) = \widetilde Q^i(x_n,u_n) + \omega_{i+1} \left[r_n^{i+1} + \alpha \max \limits_{u_n'} \widetilde Q^{i}(x_n',u_n') - \widetilde Q^i(x_n, u_n)
\right] 
\end{equation}
Note that so far we have not talked about the policy $\pi$. The learned
action-value function, $\widetilde Q (x,u)$, directly approximates $ Q^*(x,u)$,
the optimal action-value function, independent of the policy being followed. Of
course, the policy has an effect in that it determines which state-action pairs
are visited and updated. However, all that is required for correct convergence is
that all pairs continue to be updated. Therefore, Q-learning is an off-policy
control method. It learns about the greedy policy, independent of what policy is
actually being followed by the agent (as long as it tries all state-action
pairs). Thus, it is learning about one policy while following another, which is
the defining characteristic of off-policy methods.

Q-learning learns its estimates in part on the basis of other estimates, i.e. it
looks ahead one timestep, pre-selects an optimal following action and action
value and appends it to the immediate reward. This is what is commonly called
``bootstrapping''. Q-learning is shown in procedural form in
Algorithm~\ref{alg:Q-learning}.

\begin{algorithm}[H] \caption{Q-Learning} \label{alg:Q-learning}
\begin{algorithmic}
\State Initialize $Q(x,u)$ arbitrarily
\State \textbf{Repeat} for each episode:
\State $\hspace{4mm}$Initialize $x$
\Repeat{ (for each step of episode):}
\State Choose $u$ from $x$ using policy derived from $Q$
\State $\hspace{4mm}$ (e.g., $\varepsilon-\text{greedy}$)
\State Take action $u$, observe $r,x'$
\State $Q(x,u) \gets Q(x,u) + \omega [ r + \gamma \max_{u'} Q(x',u') - Q (x,u)]$
\State $x \gets x'$
\Until {$x$ is terminal}
\end{algorithmic}
\end{algorithm}

\thispagestyle{empty}\cleardoublepage
\chapter{Path Integrals}

\section{The Brownian Motion}
Brownian motion, or random walk, is defined as a stochastic process of some variable, $w$, with a Gaussian probability
distribution as follows
\begin{equation} \label{eq:brownian_motion_pdf} 
\mathbb{P}_{\textbf{w}}(t,w) = \frac{1}{\sqrt{2 \pi \sigma^2 t}} \exp{\left(- \frac{(w-\mu t)^2}{2 \sigma^2 t} \right)}
\end{equation}
At each instance in time, this process has a Gaussian distribution with the following mean
and variance:
\begin{align}
&\mathbb{E} \{w(t)\} = \mu t \notag \\
&\mathbb{V}ar \{w(t)\} = \sigma^2 t 
\end{align}
To define the probability distribution of a stochastic process, we need to
define not only $P_{\textbf{w}}(t,w)$ but also the joint distribution of the
stochastic process for an arbitrary subset of time steps i.e.
$P_{\textbf{w}}(t_1,w_1,\dots,,t_N,w_N)$.

\begin{figure} [h]
\centering
\includegraphics[width=0.6\textwidth]{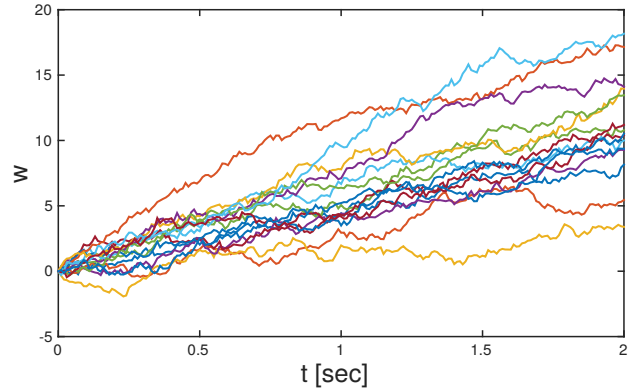}
\caption{Brownian Motion with $\mu = 5$ and $\sigma^2 = 4$}
\label{fig:brownian_motion}
\end{figure}

Instead of defining the joint probability distribution directly, we will define the
Brownian motion's increment process, $ dw(t) = \lim\limits_{\Delta t \to 0}{w(t + \Delta t) - w(t)}$ by the two
following characteristics:
\begin{enumerate}
\item The increment process, $dw(t)$, has a Gaussian distribution with the
mean and the variance, $\mu \Delta t$ and $\sigma^2 \Delta t$ respectively.
\item The increment process, $dw(t)$, is statistically independent of
$w(s)$ for all $s \leq t$.
\end{enumerate}
With these characteristics of the increment process and the probability distribution introduced
in equation~\eqref{eq:brownian_motion_pdf}, we can derive the joint distribution of any arbitrary set of time instances. For example, we can show that the joint distribution for $w(t)$ and
$w(s)$ is a Gaussian distribution with the cross covariance
\begin{equation*}
\mathbb{E}{ \left\{\left( w(t)-\mu t \right) \left( w(s) - \mu s \right) \right\} } = \sigma^2 \min(t,s)
\end{equation*}
Figure \ref{fig:brownian_motion} illustrates 15 samples extracted from a Brownian
motion with $\mu=5$ and $\sigma^2 = 4$ within a time period [0 2]. An interesting
aspect of Brownian motion is that even though it is continuous over time,
it is not differentiable.

In order to simulate the Brownian motion, either we should find the joint probability
distribution of $\{ w(0), w(\Delta t), w(2\Delta), \dots, w(N \Delta t) \}$
and then extract samples from it or we should forward integrate the following equation
(which is in fact a discretized stochastic differential equation)
\begin{equation*}
w(t+\Delta t) = w(t) + \mu \Delta t + \sqrt{\Delta t \sigma^2} \varepsilon, \qquad w(0) = 0
\end{equation*}
where $\varepsilon$ is extracted from a normal distribution with zero mean and
variance one. In this equation, the terms $\mu \Delta t + \sqrt{\Delta t \sigma^2}
\varepsilon$ are used to approximate the increment process, $dw$. In the limit
when $\Delta t$ approaches to zero, this term will be equal to the increment
process. Throughout this chapter we will mostly use the increment process instead of the Brownian motion itself.

Finally, the Brownian motion can be easily extended to n-dimension vector form as
\begin{equation*}
\mathbb{P}_{\textbf{w}}(t,\vw) = \frac{1}{(2\pi)^{n/2} (\mathbf{\Sigma} t)^{1/2}} \exp{\left(- \frac{1}{2}(\vw-\vmu t)^T (\mathbf{\Sigma} t)^{-1} (\vw-\mathbf{\vmu} t) \right)}
\end{equation*}
\section{Stochastic Differential Equations}
The following equation describes a Stochastic Differential Equation (SDE)
\begin{equation} \label{eq:sde_equation}
d\vx = \vf(t,\vx)dt + \vg(t,\vx) d\vw
\end{equation}
where $\vx(t)$ is a stochastic vector process of dimension $n$ and $\vw(t)$ is a
Brownian motion with zero mean and identity covariance matrix ($d\vw$ is the
increment process). $\vf(t,\vx)$ and  $\vg(t,\vx)$ are called respectively the
drift and the diffusion coefficients. Note that for $ \vg(t,\vx) = \mathbf{0}$
the SDE reduces to the well-known Ordinary Differential Equation (ODE).
Furthermore defining the mean and covariance of the Brownian motion to zero and
the identity matrix, respectively, is not a restrictive assumption. Since any
arbitrary Brownian motion can be put in this form by adding the mean
to the drift coefficient and multiplying the diffusion coefficient by the square
root of the covariance matrix.

One simple approach for simulating the SDE in equation~\eqref{eq:sde_equation} is to
discretize it over time by some sufficiently small time step $\Delta t$ and
replace $d\vw$ by $\sqrt{\Delta t} \varepsilon$. In doing this, we assume
that during the time step $\Delta t$ the Brownian motion has a constant increment
with mean zero and covariance $\Delta t \vI$. The discretized SDE is then expressed as follows:
\begin{equation} \label{eq:sde_equation_discretized}
\vx(t_{n+1}) = \vx(t_n) + \vf(t_n,\vx(t_n))\Delta t + \vg(t_n,\vx(t_n)) \sqrt{\Delta t}
\varepsilon,  \qquad \varepsilon \sim \mathcal{N}(\mathbf{0},\vI)
\end{equation}
Note that when equation~\eqref{eq:sde_equation} is non-linear, as is typically the case,
$\vx$ will have a non-Gaussian distribution. However the following conditioned
probability distribution for an infinitely small time step is always Gaussian.
\begin{equation} \label{sde_conditioned_pdf}
\mathbb{P}_{\textbf{x}}(t+\Delta t,\vx \mid t, \vy) = \mathcal{N} \Big( \vy + \vf(t,\vy)\Delta t, \vg(t,\vy) \vg^T(t,\vy) \Delta t\Big)
\end{equation}
\section{The Fokker Planck Equation} \label{sec:fokker_planck_equation}
Consider the SDE in equation~\eqref{eq:sde_equation}. Despite the fact that the Brownian
motion has a Gaussian distribution, generally the resulting process does not have
a Gaussian distribution (except in the linear case). However the probability
distribution associated to the SDE in equation~\eqref{eq:sde_equation} can be
derived as a solution to an initial value Partial Differential Equation (PDE)
known as Fokker Plank equation.

We define $\mathbb{P}_{\textbf{x(t)}}(t ,\vx \mid s, \vy)$ as the conditioned
probability distribution of $x(t)$ knowing that the process at time $s$ has
value $\vy$ ($\vx(s) = \vy$). The governing PDE on this probability
distribution is defined by the following Fokker Planck equation
\begin{equation} \label{eq:fokker_planck_equation}
\partial_t\mathbb{P} = -\nabla_x^T (\vf\mathbb{P}) + \cfrac{1}{2} \operatorname{Tr}{\left[\nabla_{xx} (\vg \vg^T \mathbb{P})\right]}
\end{equation}
with initial condition $\mathbb{P}_{\textbf{x(t)}}(t=s ,\vx \mid s, \vy) = \delta(\vx-\vy)$.
The operators $\nabla_x()$ and $\nabla_{xx}()$ are defined as:
\begin{equation*}
\nabla_x() = \begin{bmatrix} \frac{\partial}{\partial x_1} \\ \frac{\partial}{\partial x_2} \\ \vdots \\ \frac{\partial}{\partial x_n} \end{bmatrix}   \qquad \qquad
\nabla_{xx}() = 
\begin{bmatrix} 
\frac{\partial}{\partial x_{11}}  & \dots & \frac{\partial}{\partial x_{1n}} \\
\vdots & \ddots & \vdots \\
\frac{\partial}{\partial x_{n1}}  & \dots & \frac{\partial}{\partial x_{nn}}
\end{bmatrix}
\end{equation*}
where $x_{i}$ is the \textit{i}th element of the n-dimension stochastic process vector $\vx$.

\section{Linearly-Solvable Markov Decision Process}  \label{sec:LMDP}
The Linearly-Solvable Markov Decision Process (LMDP) is a class of optimal control
problems in which the optimality equation (characterized by the Bellman equation or
the HJB equation) can be transformed into a linear form. In this section we will
consider a special case of this class with stochastic dynamics defined as follows
\begin{equation}  \label{eq:LMDP_sde}
d\vx = \vf(t,\vx) dt + \vg(t,\vx) \left(\vu dt + d\vw\right), \qquad d\vw \sim \mathcal{N}(\mathbf{0},\mathbf{\Sigma}dt)
\end{equation}
where $\vx$ is the state vector of size $n$, $\vu$ is the control input of
size $m$, and $\vw$ is the $m$-dimension Brownian motion with zero mean and
covariance $\mathbf{\Sigma} dt$. In the control literature, the system described
by equation~\eqref{eq:LMDP_sde} is called a control affine system since the
control input appears linearly in the differential equations. Note that if we
divide the both  sides of the equation by $dt$, and then substitute
$\frac{d\vw}{dt}$ for $\vvep$ (stationary white noise), we get a system
equation in the class of systems introduced in \sref{subsec:HJP_fin_sto}
(equation~\ref{eq:stochastic_cont_time_system_eq}).
\begin{equation*}
\dot{\vx} = \vf(t,\vx) + \vg(t,\vx) \left(\vu + \vvep\right), \qquad \vvep \sim \mathcal{N}(\mathbf{0},\mathbf{\Sigma})
\end{equation*}
The cost function for this optimal control problem is defined as
\begin{equation} \label{eq:LMDP_cost}
J = E \left\{\Phi(\vx(t_f))+ \int_{t_0}^{t_f} {q(t,\vx) + \frac{1}{2} \vu^T \vR \vu \ dt} \right\}
\end{equation}
The expectation is over the trajectories extracted from the stochastic process
described by equation~\eqref{eq:LMDP_sde}. The only difference between this cost
function and the general one introduced by
equation~\eqref{eq:stochastic_continuous_time_deterministic cost function} is
that the control input cost is quadratic.

In order to make this optimal control problem a LMDP, we need to add another 
condition which is
\begin{equation} \label{eq:LMDP_condition}
\vR \mathbf{\Sigma} = \lambda \vI
\end{equation}
where $\lambda$ is an arbitrary positive real number. This condition has the following
intuitive interpretation: Let's assume for simplicity that both $\mathbf{\Sigma}$ and
$\vR$ are diagonal matrices. Then we get
\begin{equation*}
\vR_{ii} \mathbf{\Sigma}_{ii} = \lambda, \qquad \text{\textit{for all }} i \in \{1,\dots,m\}
\end{equation*}
This means that if the covariance of the noise affecting control input $i$ is
relatively high, the cost for that control input should be lower. Higher control
effort should be tolerated to counteract the noise (note that in
equation~\ref{eq:LMDP_sde} the process noise is added to the control input).

Now we will show how the optimal control problem can be transformed to
an LMDP. We will start by writing the HJB equation for this problem. We will use
the HJB equation given in equation~\eqref{eq:stchastic_principle_of_optimality}
with the following substitutions
\begin{eqnarray}	
\beta \leftarrow 0 \hspace{50mm} \notag  & & L(\vx, \vu(t)) \leftarrow q(t,\vx) + \frac{1}{2} \vu^T \vR \vu  \notag \\
\vf_t (\vx,\vu(t)) \leftarrow \vf(t,\vx) + \vg(t,\vx) \vu  \hspace{10mm} \notag & & \vB(t) \leftarrow \vg(t,\vx) \notag \\
\vW(t) \leftarrow \mathbf{\Sigma} \hspace{43mm} \notag & &
\end{eqnarray}
Then we get
\begin{equation} \label{eq:LMDP_HJB}
- \partial_t V^*(t,\vx) = \min_{\vu} \left\{q(t,\vx) + \frac{1}{2} \vu^T \vR \vu + \nabla_x^T V^*(t,\vx) (\vf(t,\vx) + \vg(t,\vx) \vu) + \frac{1}{2} \mathtt{Tr} [\nabla_{xx} V^*(t,\vx) \vg(t,\vx) \mathbf{\Sigma} \vg^T(t,\vx)]  \right\}
\end{equation}
where $V^*$ is the optimal value function. For simplicity, we will omit all the
dependencies with respect to time and state. In equation~\eqref{eq:LMDP_HJB} we
have also adopted the notation introduced in \sref{sec:fokker_planck_equation}. If we
minimize the left hand side with respect to $\vu$, we get
\begin{equation} \label{eq:LMDP_HJB_minimized}
- \partial_t V^* = q - \frac{1}{2} \nabla_x^T V^* \, \vg \vR^{-1} \! \vg^T \nabla_x V^* + \nabla_x^T V^* \, \vf + \frac{1}{2} \mathtt{Tr} [\nabla_{xx} V^* \, \vg \mathbf{\Sigma} \vg^T]
\end{equation}
and the optimal control
\begin{equation}
\vu^*(t,\vx) = - \vR^{-1} \vg^T(t,\vx) \nabla_x V(t,\vx) 
\end{equation} 
Equation~\eqref{eq:LMDP_HJB_minimized} is a nonlinear PDE, since $\nabla_x^T V^*$
appears in a quadratic form. In order to shorten the notion, we substitute $\vg \,
\vR^{-1} \! \vg^T$ by $\mathbf{\Xi}$. Using the
condition~\ref{eq:LMDP_condition}, we can also write $\vg \mathbf{\Sigma} \vg^T =
\lambda \mathbf{\Xi}$. Therefore
\begin{equation} \label{eq:LMDP_HJB_main}
- \partial_t V^* = q - \frac{1}{2} \nabla_x^T V^* \, \mathbf{\Xi} \nabla_x V^* + \nabla_x^T V^* \, \vf + \frac{\lambda}{2} \mathtt{Tr} [\nabla_{xx} V^* \, \mathbf{\Xi}]
\end{equation}
In spite of the nonlinearity of the resulting equation, it can be shown that
under a log transformation we get a linear PDE. Assume the following
transformation $\Psi(t,\vx)$ is called the desirability function
\begin{equation} \label{eq:log_transform}
V^*(t,\vx) = - \lambda \log{\Psi(t,\vx)},
\end{equation}
we can write the derivatives as 
\begin{eqnarray} \label{eq:log_transform_derivatives}
\partial_t V^*(t,\vx) & = & - \lambda \frac{\partial_t \Psi}{\Psi} \notag \\
\nabla_x V^*(t,\vx) & = & - \lambda \frac{\nabla_x \Psi}{\Psi} \notag \\
\nabla_{xx} V^*(t,\vx) & = & \frac{1}{\lambda} \nabla_x V^* \: \nabla_x^T V^* - \lambda \frac{\nabla_{xx} \Psi}{\Psi}.
\end{eqnarray}
Furthermore, for the scalar value $-\frac{1}{2} \nabla_x^T V^* \, \mathbf{\Xi}
\nabla_x V^*$ in equation~\eqref{eq:LMDP_HJB_main} we can write
\begin{equation} \label{eq:log_transform_Tr}
-\frac{1}{2} \nabla_x^T V^* \, \mathbf{\Xi} \nabla_x V^* = -\frac{1}{2} \mathtt{Tr} [\nabla_x^T V^* \, \mathbf{\Xi} \nabla_x V^*] = -\frac{1}{2} \mathtt{Tr} [\nabla_x V^* \, \nabla_x^T V^* \, \mathbf{\Xi}].
\end{equation}
By substituting equations~\eqref{eq:log_transform_derivatives} and \eqref{eq:log_transform_Tr} into equation~\eqref{eq:LMDP_HJB_main}, we get
\begin{equation*}
\lambda \frac{\partial_t \Psi}{\Psi} = q - \cancel{\frac{1}{2} \mathtt{Tr} [\nabla_x V^* \, \nabla_x^T V^* \, \mathbf{\Xi}]} - \lambda \vf^T \frac{\nabla_x \Psi}{\Psi} + \cancel{\frac{1}{2} \mathtt{Tr} [\nabla_x V^* \, \nabla_x^T V^* \, \mathbf{\Xi}]} - \frac{\lambda^2}{2} \mathtt{Tr} \left[ \frac{\nabla_{xx} \Psi}{\Psi} \mathbf{\Xi} \right].
\end{equation*}
Finally, we multiply both sides by $-\sfrac{\Psi}{\lambda}$
\begin{equation} \label{eq:LMDP_pde}
- \partial_t \! \Psi = -\frac{1}{\lambda} q \Psi +  \vf^T \nabla_x \! \Psi + \frac{\lambda}{2} \mathtt{Tr} [ \mathbf{\Xi} \nabla_{xx} \! \Psi ].
\end{equation}
Note that we used the matrix identity $\mathtt{Tr}[\vA \vB] = \mathtt{Tr}[\vB
\vA]$ to write $\mathtt{Tr} [ \nabla_{xx} \! \Psi \, \mathbf{\Xi}  ] =
\mathtt{Tr} [ \mathbf{\Xi} \nabla_{xx} \! \Psi ]$. Equation~\eqref{eq:LMDP_pde}
is linear with respect to $\Psi$ and its derivatives. This equation is normally
written in the form of the linear operator $\mathtt{H}$ as follows
\begin{equation} \label{eq:LMDP_pde_short}
- \partial_t \! \Psi = \mathtt{H} [\Psi]
\end{equation}
where $\mathtt{H}$ is defined as
\begin{align} \label{eq:LMDP_H_operator}
\mathtt{H} &= -\frac{1}{\lambda} q +  \vf^T \nabla_x + \frac{\lambda}{2} \mathtt{Tr} [ \mathbf{\Xi} \nabla_{xx}] \notag \\ 
&= -\frac{1}{\lambda} q + \sum_{i}{\vf_i \frac{\partial \;}{\partial_{x_i}}} + \frac{\lambda}{2} \sum_{i,j}{\mathbf{\Xi}_{ij} \frac{\partial^2 \quad}{\partial_{x_i} \partial_{x_j}}}
\end{align}
From the principle of optimality, we know that the original optimal control problem is a final value problem, meaning that only the terminal condition is known prior to solving. Equation~\eqref{eq:LMDP_pde} is therefore also a final value problem, which has the following
terminal value.
\begin{align} \label{eq:LMDP_pde_terminal_value}
\Psi(t_f,\vx) &= \exp \left(-\frac{1}{\lambda} V^*(t_f,\vx) \right) \notag \\
&= \exp \left(-\frac{1}{\lambda} \Phi(\vx) \right)
\end{align}
The equation~\eqref{eq:LMDP_pde_short} is consider to be an LMDP. In the next
section, we will introduce the Path Integral method to solve this PDE.

\section{Path Integral Optimal Control}
The Path Integral framework introduces a method for solving the linear backward PDE
introduced in equation~\eqref{eq:LMDP_pde_short} through a forward integration
approach. We start the proof by introducing a new function over time and state, 
$\rho(t,\vx)$. We also define the following inner product in the function space which 
is in fact a function of time.
\begin{equation}
<\rho \mid \Psi> = \int{\rho(t,\vx) \Psi(t,\vx) d\vx}
\end{equation}
We also assume that 
\begin{equation} \label{eq:rho_inifine_condition}
\lim\limits_{\norm{x} \to \infty}{\rho(t,\vx) = 0}
\end{equation}
The $\mathtt{H}^\dagger$ operator is called the Hermitian conjugate of the
operator $\mathtt{H}$ (equation_\ref{eq:LMDP_H_operator}) if it satisfies the
following equality.
\begin{equation}
<\rho \mid \mathtt{H} [\Psi]> = <\mathtt{H}^\dagger [\rho] \mid \Psi>
\end{equation}
The $\mathtt{H}^\dagger$ formulation can be derived as follows by starting from the definitions of the inner product and the $\mathtt{H}$ operator.
\begin{align}
<\rho \mid \mathtt{H} [\Psi]> &= \int{\rho(t,\vx) \mathtt{H} [\Psi(t,\vx)] d\vx}  \notag \\
&= \int{\rho(t,\vx) \, \left(-\frac{1}{\lambda} q + \sum_{i}{\vf_i \frac{\partial \;}{\partial_{x_i}}} + \frac{\lambda}{2} \sum_{i,j}{\mathbf{\Xi}_{ij} \frac{\partial^2 \quad}{\partial_{x_i} \partial_{x_j}}}\right) \Psi(t,\vx) d\vx}
\end{align}
Using integration by parts and the assumption in \ref{eq:rho_inifine_condition} 
we can easily show
\begin{align*}
<\rho \mid \mathtt{H} [\Psi]> &= \int{\Psi \left(-\frac{1}{\lambda} q \rho - \sum_{i}{ \frac{\partial \left( \vf_i \rho \right)}{\partial_{x_i}}} + \frac{\lambda}{2} \sum_{i,j}{ \frac{\partial^2 \left(\mathbf{\Xi}_{ij} \rho \right) }{\partial_{x_i} \partial_{x_j}}} \right)  d\vx} \\ 
&= \int{\Psi(t,\vx) \mathtt{H}^\dagger [\rho(t,\vx)] d\vx}
\end{align*}
Therefore we can write $\mathtt{H}^\dagger$ as
\begin{align}
\mathtt{H}^\dagger &= -\frac{1}{\lambda} q - \sum_{i}{ \frac{\partial \vf_i }{\partial_{x_i}}} + \frac{\lambda}{2} \sum_{i,j}{ \frac{\partial^2 \mathbf{\Xi}_{ij} }{\partial_{x_i} \partial_{x_j}}} \notag \\
&= -\frac{1}{\lambda} q - \nabla_x^T \vf + \frac{\lambda}{2} \mathtt{Tr} [ \nabla_{xx} \mathbf{\Xi}]
\end{align}
So far the only restriction that we have posed over $\rho(t,\vx)$ is the
condition~\ref{eq:rho_inifine_condition}. We will now pose another restriction on
$\rho(t,\vx)$ that $<\rho \mid \Psi>$ is time independent. Therefore we have
\begin{equation}
\frac{d \ }{dt} <\rho \mid \Psi> = 0
\end{equation}
Then we can write
\begin{align*}
0 &=\frac{d \ }{dt} <\rho \mid \Psi> \\
&= \int{\partial_t \Big(\rho(t,\vx) \Psi(t,\vx)\Big) d\vx} \\
&= \int{\partial_t \rho(t,\vx) \Psi(t,\vx) + \rho(t,\vx) \partial_t \Psi(t,\vx) d\vx} \\
&= <\partial_t \rho \mid \Psi> + <\rho \mid \partial_t \Psi> 
\end{align*}
By using the equation~\eqref{eq:LMDP_pde_short}, we get
\begin{equation*}
0 = <\partial_t \rho \mid \Psi> - <\rho \mid \mathtt{H} [\Psi]> 
\end{equation*}
Now we use the Hermitian conjugate operator $\mathtt{H}^\dagger$
\begin{equation*}
0 = <\partial_t \rho \mid \Psi> - <\mathtt{H}^\dagger[\rho] \mid \Psi> 
\end{equation*}
Therefore we can write
\begin{equation*}
<\partial_t \rho - \mathtt{H}^\dagger[\rho] \mid \Psi> = 0
\end{equation*}
A trivial solution to this equation is 
\begin{align} \label{eq:forward_diffusion_process_pdf}
\partial_t \rho &= \mathtt{H}^\dagger[\rho] \notag \\
&= -\frac{1}{\lambda} q \rho - \nabla_x^T (\vf \rho) + \frac{\lambda}{2} \mathtt{Tr} [ \nabla_{xx} (\mathbf{\Xi} \rho)]
\end{align}
Since equation~\eqref{eq:LMDP_pde_short} is a final value problem,
equation~\eqref{eq:forward_diffusion_process_pdf} should be an initial value
problem, otherwise the terminal value for $\rho$ cannot be defined freely, as the
inner product should be time independent. 

Furthermore it can be shown that if $\rho(t,\vx)$ satisfies the
equation~\eqref{eq:forward_diffusion_process_pdf}, it will always satisfy the
condition in \ref{eq:rho_inifine_condition}. Therefore by only determining the
initial value, we can uniquely define a $\rho(t,\vx)$ function which satisfies all
of the conditions. In the literature this equation is sometimes referred to as a
forward diffusion process. 

We will now define an appropriate
initial condition for this forward diffusion process. Equation~\eqref{eq:forward_diffusion_process_pdf} resembles the Fokker Planck
equation (equation~\ref{eq:fokker_planck_equation}). The only difference is the
term $-\frac{1}{\lambda} q \rho$. The effect of this extra term is that the probability
distribution decays over time. Based on this similarity, we can show that the
following process has the same probability distribution as the solution of
equation~\eqref{eq:forward_diffusion_process_pdf} (note that we are misusing the
term ``probability distribution''. As mentioned the term $-\frac{1}{\lambda} q
\rho$ attenuates the probability distribution over time, therefore its
integral eventually becomes less than one).
\begin{align} \label{eq:forward_diffusion_process_sde}
& d\vx(t_i) = \vf(t_i,\vx(t_i)) dt + \vg(t_i,\vx(t_i)) d\vw, \qquad \vx(t_0 = s) = \vy \notag \\
& \begin{cases}
\vx(t_{i+1}) = \vx(t_i) + d\vx(t_i) \quad \text{with probability } \exp \left(-\frac{1}{\lambda} q dt\right) \\
\vx(t_{i+1}): \text{annihilation} \hspace{9.5mm} \text{with probability } 1-\exp \left(-\frac{1}{\lambda} q dt\right)
\end{cases}
\end{align}
where $\vw$ is a Brownian motion with zero mean and the covariance
$\mathbf{\Sigma} dt$ and $t_{i+1} = t_i + dt$. Here ``annihilation'' means that
we discard the sample with some probability. Through the similarity to
the Fokker Planck equation, we also can determine an appropriate initial condition. If
the state vector at time $s$ is $\vy$ then the probability distribution of the state
at the initial time should be a delta function at $\vy$.
\begin{equation} \label{eq:forward_diffusion_process_initial}
\rho(t=s,\vx) = \delta(\vx - \vy)
\end{equation}
In order to emphasize that the $\rho(t,\vx)$ is a solution to the diffusion
process with the initial condition $\vx(s) = \vy$, sometimes the initial
condition is explicitly written in the condition i.e. we write $\rho(t,\vx\mid s, \vy)$.

Since the process described by equation~\eqref{eq:forward_diffusion_process_sde}
is a Markov Process we can write the joint probability distribution of a
trajectory $\tau = \{ \vx(t_{0}), \vx(t_{1}), \dots \vx(t_{N}) \}$, with $t_0 = s$, as
\begin{equation}
\rho(\tau \mid s,\vy) = \prod\limits_{i=0}^{N-1} {\rho(t_{i+1},\vx(t_{i+1}) \mid t_{i},\vx(t_{i}))}, \qquad \vx(t_0=s) = \vy
\end{equation}
where $N$ is the number of the time steps and the initial state at time $s$ is
$\vy$. Furthermore the conditioned probability distribution will have a Gaussian
like distribution.
\begin{equation} 
\rho(t_{i+1},\vx(t_{i+1}) \mid t_{i},\vx(t_{i})) = \operatorname{e}^{-\frac{1}{\lambda} q(t_i,\vx(t_i)) dt} \mathcal{N} \Big(\vx(t_i) + \vf(t_i,\vx(t_i)) dt, \mathbf{\Xi}(t_i,\vx(t_i)) dt \Big)
\end{equation}
Here $\mathbf{\Xi}(t_i,\vx(t_i))$ is defined as \sref{sec:LMDP}. The effect of
the annihilation process is incorporated by multiplying the distribution by
$\exp{(-\frac{1}{\lambda} q dt)}$. Therefore the probability distribution of the
trajectory $\tau$ can be written as
\begin{align} \label{eq:forward_diffusion_process_tau_pdf}
\rho(\tau \mid s,\vy) &= \prod\limits_{i=0}^{N-1} {\operatorname{e}^{-\frac{1}{\lambda} q(t_i,\vx(t_i)) dt} \mathcal{N} \Big(\vx(t_i) + \vf(t_i,\vx(t_i)) dt, \mathbf{\Xi}(t_i,\vx(t_i)) dt \Big)} \notag \\
&= \operatorname{e}^{\sum\limits_{i=0}^{N-1}{-\frac{1}{\lambda} {q(t_i,\vx(t_i)) dt}}} \ \prod\limits_{i=0}^{N-1} {\mathcal{N} \Big(\vx(t_i) + \vf(t_i,\vx(t_i)) dt, \mathbf{\Xi}(t_i,\vx(t_i)) dt \Big)}   \notag \\
=& \mathbb{P}_{uc}(\tau \mid s, \vy) \ \operatorname{e}^{\sum\limits_{i=0}^{N-1}{-\frac{1}{\lambda} {q(t_i,\vx(t_i)) dt}}}
\end{align}
where $\mathbb{P}_{uc}(\tau \mid s, \vy)$ is introduced in
equation~\eqref{eq:LMDP_tau_pdf} and it is actually the probability distribution
of the trajectory $\tau$ generated by the system in
equation~\eqref{eq:forward_diffusion_process_sde} \textbf{without the
annihilation} when initialized at $\vx(s) = \vy$. Furthermore by omitting the
annihilation term from equation~\eqref{eq:forward_diffusion_process_sde}, it will
reduce to the same system as LMDP (equation~\ref{eq:LMDP_sde}) when the $\vu$ is
set to zero. In control literature such a system is normally referred to as the
uncontrolled system (hence the subscript ``\textit{uc}'').
\begin{equation} \label{eq:LMDP_tau_pdf}
\mathbb{P}_{uc}(\tau \mid s, \vy) = \prod\limits_{i=0}^{N-1} {\mathcal{N} \Big(\vx(t_i) + \vf(t_i,\vx(t_i)) dt, \mathbf{\Xi}(t_i,\vx(t_i)) dt \Big)}
\end{equation}
Now we go back to the characteristic that the inner product of $\Psi(t,\vx)$ and
$\rho(t,\vx)$ is time independent. Since the inner product is time independent,
its value at any time will be always the same. Therefore for the initial
time and the final time we can write
\begin{align}
<\rho \mid \Psi>(t=s) &= <\rho \mid \Psi>(t=t_f) \notag \\
\int{\rho(s,\vx_s) \Psi(s,\vx_s) d\vx_s} &= \int{\rho(t_f,\vx_{t_f}) \Psi(t_f,\vx_{t_f}) d\vx_{t_f}}
\end{align}
By substituting the initial condition given in
equation~\eqref{eq:forward_diffusion_process_initial} and integrating it we will
get
\begin{align}
\int{\delta(\vx_s-\vy) \Psi(s,\vx_s) d\vx_s} &= \int{\rho(t_f,\vx_{t_f}) \Psi(t_f,\vx_{t_f}) d\vx_{t_f}} \notag \\
\Psi(s,\vy) &= \int{\rho(t_f,\vx_{t_f}) \Psi(t_f,\vx_{t_f}) d\vx_{t_f}} \notag
\end{align}
Now we will use the terminal condition of $\Psi$ from equation~\eqref{eq:LMDP_pde_terminal_value}.
\begin{equation*}
\Psi(s,\vy) = \int{\rho(t_f,\vx_{t_f}) \operatorname{e}^{-\frac{1}{\lambda} \Phi(\vx_{t_f})} d\vx_{t_f}} 
\end{equation*}
To calculate the RHS integral we need to define the probability distribution at the last time step $t_f$. We can derive $\rho(t_f,\vx_{t_f})$ by marginalizing the joint probability distribution introduced in equation~\eqref{eq:forward_diffusion_process_tau_pdf} with respect to $\{ \vx(t_{1}), \dots \vx(t_{N-1}) \}$. Note that the time index starts from $t_1$ and goes up to $t_{N-1}$. 
\begin{align*}
\rho(t_f,\vx_{t_f}) &= \int{\rho(\tau \mid s,\vy) \, d\vx(t_1) \dots \vx(t_{N-1})} \\
&= \int{ \mathbb{P}_{uc}(\tau \mid s, \vy) \ \operatorname{e}^{\sum\limits_{i=0}^{N-1}{-\frac{1}{\lambda} {q(t_i,\vx(t_i)) dt}}} \ d\vx(t_1) \dots d\vx(t_{N-1})}
\end{align*}
By substituting this in the previous equation, we get (we have changed the name of $\vx_{t_f}$ to $\vx(t_{N})$)
\begin{equation}
\Psi(s,\vy) = \int{ \mathbb{P}_{uc}(\tau \mid s, \vy) \ \operatorname{e}^{-\frac{1}{\lambda}  \Big( \Phi(x(t_{N})) + \sum\limits_{i=0}^{N-1}{{q(t_i,\vx(t_i)) dt}} \Big) } \ d\vx(t_1) \dots d\vx(t_{N-1}) d\vx(t_{N}) }
\end{equation}
This equation is equal to the following expectation
\begin{align}
\Psi(s,\vy) &= \operatorname{E}_{\tau_{uc}} \left\{ \operatorname{e}^{-\frac{1}{\lambda}  \Big( \Phi(\vx(t_N)) + \sum\limits_{i=0}^{N-1}{{q(t_i,\vx(t_i)) dt}} \Big) } \right\} \notag \\
&= \operatorname{E}_{\tau_{uc}} \left\{ \operatorname{e}^{-\frac{1}{\lambda}  \Big( \Phi(\vx(t_f))+ \int_{s}^{t_f} {q(t,\vx) \ dt} \Big) } \right\}
\end{align}
where the first equation is the discretized equivalent of the latest one. We can
estimate this expectation through a Monte Carlo method by extracting samples and averaging them. The samples of this expectation are generated through the
uncontrolled dynamics system
\begin{equation}  \label{eq:uc_sde}
d\vx = \vf(t,\vx) dt + \vg(t,\vx) d\vw, \qquad d\vw \sim \mathcal{N}(\mathbf{0},\mathbf{\Sigma}dt), \quad \vx(t=s) = \vy
\end{equation}
Therefore in order to estimate the desirability function at time $s$ and state
$\vy$, we should forward simulate the uncontrolled dynamics several times with
the initial condition $\vx(s)=\vy$. Then we can estimate the expectation by
averaging the value of $-\frac{1}{\lambda}  \Big( \Phi(\vx(t_f))+ \int_{s}^{t_f}
{q(t,\vx) \ dt} \Big) $ for each sample.

After calculating the desirability function for all times and states, we can
derive the optimal control input as
\begin{align}
\vu^*(s,\vy) &= - \vR^{-1} \vg^T(s,\vy) \nabla_y V^*(s,\vy) \notag \\
&= \lambda \vR^{-1} \vg^T(s,\vy) \frac{\nabla_y \Psi(s,\vy)}{\Psi(s,\vy)}
\end{align} 
However we can also calculate the optimal control input directly by a similar
path integral approach
\begin{equation} \label{eq:pi_original_optimal_controller}
\vu^*(s,\vy) = \lim\limits_{\Delta s \to 0}{ \frac{  \operatorname{E}_{\tau_{uc}} \left\{ \displaystyle\int_{s}^{s+\Delta s} \!\!\!\!\! d\vw \  \operatorname{e}^{-\frac{1}{\lambda}  \Big( \Phi(\vx(t_f))+ \int_{s}^{t_f} {q(t,\vx) \ dt} \Big) } \right\} }    {\Delta s \operatorname{E}_{\tau_{uc}} \left\{ \operatorname{e}^{-\frac{1}{\lambda}  \Big( \Phi(\vx(t_f))+ \int_{s}^{t_f} {q(t,\vx) \ dt} \Big) } \right\}} }
\end{equation} 
again if we use the white noise notation where $\frac{d\vw}{dt} = \varepsilon$, we can write
\begin{equation} \label{eq:pi_optimal_controller}
\vu^*(s,\vy) = \frac{  \operatorname{E}_{\tau_{uc}} \left\{ \varepsilon \  \operatorname{e}^{-\frac{1}{\lambda}  \Big( \Phi(\vx(t_f))+ \int_{s}^{t_f} {q(t,\vx) \ dt} \Big) } \right\} }    {\operatorname{E}_{\tau_{uc}} \left\{ \operatorname{e}^{-\frac{1}{\lambda}  \Big( \Phi(\vx(t_f))+ \int_{s}^{t_f} {q(t,\vx) \ dt} \Big) } \right\}}
\end{equation} 
or equivalently 
\begin{equation} \label{eq:pi_optimal_controller1}
\vu^*(s,\vy) = \operatorname{E}_{\tau_{uc}} \left\{ \varepsilon \frac{  \operatorname{e}^{-\frac{1}{\lambda}  \Big( \Phi(\vx(t_f))+ \int_{s}^{t_f} {q(t,\vx) \ dt} \Big) }  }    {\operatorname{E}_{\tau_{uc}} \left\{ \operatorname{e}^{-\frac{1}{\lambda}  \Big( \Phi(\vx(t_f))+ \int_{s}^{t_f} {q(t,\vx) \ dt} \Big) } \right\}} \right\}
\end{equation} 
The samples are generated by
\begin{equation} \label{eq:pi_uc_system}
\dot{\vx} = \vf(t,\vx) + \vg(t,\vx) \varepsilon, \qquad \varepsilon \sim \mathcal{N}(\mathbf{0},\mathbf{\Sigma}), \quad \vx(t=s)=\vy
\end{equation}
Before we move on to the next section, we would like to give few insights into
the path integral formulation for optimal control. In this discussion we will
mainly use  equations~\eqref{eq:pi_optimal_controller} and
\eqref{eq:pi_uc_system}, however the same argument holds for the formulation in
equations~\eqref{eq:pi_original_optimal_controller} and \eqref{eq:uc_sde}.

According to equation~\eqref{eq:pi_optimal_controller1} in order to derive the
optimal control at time $s$ and state $\vy$, we should repeatedly forward
simulate the noise-driven system from that moment and point until the end of the
time horizon $t_f$. Then we should weight the first noise element of each sample
by $\alpha(\tau_{uc};s,\vy)$ which is defined as
\begin{equation}  \label{eq:pi_sample_weighting}
\alpha(\tau_{uc};s,\vy) = \sfrac{  \operatorname{e}^{-\frac{1}{\lambda}  \Big( \Phi(\vx(t_f))+ \int_{t_0}^{t_f} {q(t,\vx) \ dt} \Big) }  }    {\operatorname{E}_{\tau_{uc}} \left\{ \operatorname{e}^{-\frac{1}{\lambda}  \Big( \Phi(\vx(t_f))+ \int_{t_0}^{t_f} {q(t,\vx) \ dt} \Big) } \right\}} 
\end{equation}
Since in $\alpha(\tau_{uc};s,\vy)$ the accumulated cost is in the exponential,
this wighting will be almost zero for a costly trajectory and only a very small
portion of the samples with near optimal or optimal cost will have non-negligible
weights. However, the probability of generating optimal trajectories through the
noise-driven system (a.k.a. random walk) is very low. Therefore in order to be
able to estimate the optimal input, we need abundantly many samples. This issue
becomes even more severe in high dimensional problems, quickly to the point that
calculating the above exponential becomes intractable. In the next section, an
importance sampling scheme will be introduced which can boost the sampling
efficiency of the  Path Integral.
  
\vspace{8cm}

\section{Path Integral with Importance Sampling} \label{sec:PI_SI}
In order to make path integral sampling more efficient, we will use 
importance sampling. Before introducing the method, we will
briefly describe the idea behind importance sampling.

Assume the following expectation problem where $x$ is a random variable with
probability distribution $p(x)$ and $f(x)$ is an arbitrary deterministic
function.
\begin{equation*}
\operatorname{E}_{p}{[f(x)]} = \int_{-\infty}^{\infty}{f(x) \  p(x) dx}
\end{equation*}
We also assume that we have another random variable named $y$ with the
probability distribution $q(y)$. Lets assume that calculating the expectation of
an arbitrary function for this random variable is less costly than the previous
one. Is there a way to calculate $\operatorname{E}_{p}{[f(x)]}$, while we are only
sampling from $y$?

The answer is yes, as long as we can guarantee that $q(.)$ is non-zero everywhere
$p(.)$ is. In other words if the probability of extracting a particular value
of $x$ is non-zero, $y$ should also be able to have that value with some non-zero
probability. If this condition satisfies we can write
\begin{equation*}
\operatorname{E}_{p}{[f(x)]} = \operatorname{E}_{q}{[w(y) f(y)]}, \qquad w(y) = \frac{p(y)}{q(y)}
\end{equation*}
$w$ is called the importance sampling weighting. Proving this relation is
very simple, we just need to write the definition of the expectation. To do such,
we will start from the right hand side
\begin{align*}
\operatorname{E}_{q}{[w(y) f(y)]} &= \int_{-\infty}^{\infty}{w(y) f(y) q(y) \ dy}  \\
&= \int_{-\infty}^{\infty}{\frac{p(y)}{\cancel{q(y)}} f(y) \cancel{q(y)} \ dy} \\
&= \int_{-\infty}^{\infty}{p(y) f(y) \ dy} = \operatorname{E}_{p}{[f(x)]}
\end{align*}
Note that changing $y$ to $x$ is eligible since they are just dummy variables.

Now that we have introduced the idea of importance sampling, we will go back
to the path integral problem. First we will introduce importance sampling for
path integrals, then we will discuss how it can help to improve the sample
efficiency.

Lets assume that, for calculating the path integral expectations in
equation~\eqref{eq:pi_optimal_controller}, we use the following controlled system
dynamics instead of the uncontrolled dynamics (equation \ref{eq:pi_uc_system}).
\begin{equation} \label{eq:pi_c_system}
\dot{\vx} = \vf(t,\vx) + \vg(t,\vx) \left(u + \varepsilon\right), \qquad \varepsilon \sim \mathcal{N}(\mathbf{0},\mathbf{\Sigma}), \quad \vx(t=s)=\vy
\end{equation}
Now according to the importance sampling method, we need to incorporate the
importance sampling weighting into the expectation. This weighting should be 
defined as
\begin{equation*}
\frac{\mathbb{P}_{uc}(\tau \mid s, \vy)}{\mathbb{P}_{c}(\tau \mid s, \vy)} = 
\frac{\prod\limits_{i=0}^{N-1} {\mathcal{N} \Big(\vx(t_i) + \vf(t_i,\vx(t_i)) dt, \mathbf{\Xi}(t_i,\vx(t_i)) dt \Big)}}{\prod\limits_{i=0}^{N-1} {\mathcal{N} \Big(\vx(t_i) + \vf(t_i,\vx(t_i)) dt + \vg(t_i,\vx(t_i)) \vu(t_i) dt, \mathbf{\Xi}(t_i,\vx(t_i)) dt \Big)}}
\end{equation*}
where $\mathbb{P}_{c}$ is the probability distribution of the trajectory generated by the controlled system in equation~\eqref{eq:pi_c_system}. For the sake of simplicity, we will temporarily drop the time and state dependency of the functions. Therefore we will have
\begin{align*}
\frac{\mathbb{P}_{uc}(\tau \mid s, \vy)}{\mathbb{P}_{c}(\tau \mid s, \vy)} &= 
\prod\limits_{i=0}^{N-1} {\frac{ \mathcal{N} \Big(\vx_i + \vf_i dt, \mathbf{\Xi}_i dt \Big)}{\mathcal{N} \Big(\vx_i + \vf_i dt + \vg_i \vu_i dt, \mathbf{\Xi}_i dt \Big)}} \\
&= \prod\limits_{i=0}^{N-1} { \frac{ \exp\Big(-\frac{1}{2} \norm{\vx_{i+1}-\vx_i - \vf_i dt}^2_{\mathbf{\Xi}_i dt} \Big) }{ \exp\Big(-\frac{1}{2} \norm{\vx_{i+1} - \vx_i - \vf_i dt - \vg_i \vu_i dt}^2_{\mathbf{\Xi}_i dt} \Big) } }
\end{align*}
Since the trajectory is generated through the controlled dynamics, we know $d\vx_{i+1} = \vf_i dt + \vg_i \vu_i dt + \vg_i d\vw_i$. Hence we get
\begin{align*}
\frac{\mathbb{P}_{uc}(\tau \mid s, \vy)}{\mathbb{P}_{c}(\tau \mid s, \vy)} &= 
\prod\limits_{i=0}^{N-1} { \frac{ \exp\Big(-\frac{1}{2} \norm{\vg_i \vu_i dt + \vg_i d\vw_i}^2_{\mathbf{\Xi}_i dt} \Big) }{ \exp\Big(-\frac{1}{2} \norm{ \vg_i d\vw_i}^2_{\mathbf{\Xi}_i dt} \Big) } } \\
&= \prod\limits_{i=0}^{N-1} { \exp\Big( -\frac{1}{2} \norm{\vg_i \vu_i dt + \vg_i d\vw_i}^2_{\mathbf{\Xi}_i dt} + \frac{1}{2} \norm{ \vg_i d\vw_i}^2_{\mathbf{\Xi}_i dt}  \Big) } \\
&= \prod\limits_{i=0}^{N-1} { \exp\Big( -\frac{1}{2} \norm{\vg_i \vu_i dt}^2_{\mathbf{\Xi}_i dt}  -\cancel{\frac{1}{2} \norm{\vg_i d\vw_i}^2_{\mathbf{\Xi}_i dt}} - \vu_i^T \vg_i^T \mathbf{\Xi}_i^{-1} \vg_i d\vw_i + \cancel{\frac{1}{2} \norm{\vg_i d\vw_i}^2_{\mathbf{\Xi}_i dt}}  \Big) } \\
&= \prod\limits_{i=0}^{N-1} { \exp\Big( -\frac{1}{2} \vu_i^T \vg_i^T \mathbf{\Xi}_i^{-1} \vg_i \vu_i dt  - \vu_i^T \vg_i^T \mathbf{\Xi}_i^{-1} \vg_i d\vw_i \Big) } \\
&= \prod\limits_{i=0}^{N-1} { \exp\Big( -\frac{1}{2 \lambda} \vu_i^T \vR \vu_i dt  - \frac{1}{\lambda} \vu_i^T \vR d\vw_i \Big) } \\
&= \exp\Big( -\frac{1}{\lambda} \sum\limits_{i=0}^{N-1} { \frac{1}{2} \vu_i^T \vR \vu_i dt  + \vu_i^T \vR d\vw_i } \Big) \\
&=  \exp \Big( -\frac{1}{\lambda}  \displaystyle\int_{t_0}^{t_f} { \frac{1}{2} \vu^T \vR \vu dt + \vu^T \vR d\vw } \Big)
\end{align*}
Now if we multiply this importance sampling correction term to the expectations in equation~\eqref{eq:pi_optimal_controller}, we get
\begin{equation} 
\vu^*(s,\vy) = \frac{  \operatorname{E}_{\tau_{c}} \left\{ (\vu+\varepsilon) \  \operatorname{e}^{-\frac{1}{\lambda}  \Big( \Phi(\vx(t_f))+ \int_{t_0}^{t_f} {q(t,\vx) + \frac{1}{2} \vu^T \vR \vu \ dt + \vu^T \vR d\vw} \Big) } \right\} }    {\operatorname{E}_{\tau_{c}} \left\{ \operatorname{e}^{-\frac{1}{\lambda}  \Big( \Phi(\vx(t_f))+ \int_{t_0}^{t_f} {q(t,\vx) + \frac{1}{2} \vu^T \vR \vu \ dt + \vu^T \vR d\vw} \Big) } \right\}}
\end{equation} 
We have changed the $\tau_{uc}$ subscript to $\tau_c$ and substituted
$\varepsilon$ by $\vu + \varepsilon$. We can introduce the return of a trajectory
as
\begin{equation}
R(\tau;s,\vy) = \Phi(\vx(t_f))+ \int_{s}^{t_f} {\left(q(t,\vx) + \frac{1}{2} \vu^T \vR \vu\right) dt} + \int_{s}^{t_f} {\vu^T \vR d\vw} 
\end{equation}
Note that since the Brownian motion has a zero mean, the expectation of the
return (at initial time $t_0$ and initial state $\vx_0$) is equal to the cost
function of the policy $\vu$, i.e. $J = E{[R(\tau;t_0,\vx_0)]}$. Using the
return notation, the optimal control equation reduces to the following form
\begin{equation}  \label{eq:pi_is_optimal_controller}
\vu^*(s,\vy) = \vu(s,\vy) + \frac{  \operatorname{E}_{\tau_{c}} \left\{ \vvep(s) \  \operatorname{e}^{-\frac{1}{\lambda}  R(\tau;s,\vy) } \right\} }    {\operatorname{E}_{\tau_{c}} \left\{ \operatorname{e}^{-\frac{1}{\lambda}  R(\tau;s,\vy) } \right\}}
\end{equation}

\section{Path Integral with Function Approximation} \label{sec:PI_FA}
The equation~\eqref{eq:pi_is_optimal_controller} indicates that in order to
calculate the optimal control at a specific time and state we must calculate
two expectations. In practice for a nonlinear system we estimate these
expectations through a Monte Carlo method. This means that we extract a
sufficiently big set of samples by forward simulating the controlled dynamics
from a given time and state. Then we estimate the expectations by a numerical
average over the samples. However in order to find the optimal policy, we need to
repeat this process for all the time steps in the period $[t_0,t_t]$ and at least
a subset of the state space. As you can imagine, this is typically quite a costly process.
On the other hand, we know that the final optimal control manifold is relatively
smooth (or at least we prefer to have a smooth approximation of it), which means
the estimated optimal control for a specific time and state could possibly be used
for its neighboring time steps and states. This idea can be included in the path
integral approach by using function approximation on the optimal control. Lets
assume that want to approximate the optimal control input through the
following linear model (linear w.r.t. the parameter vector)
\begin{equation}
u^*_i(s,\vy) = \vUpsilon_i^T(s,\vy) \vth_i^* + error
\end{equation}
where $u_i$ refers to the \textit{i}th control input. $\vUpsilon_i(s,\vy)$ is a
nonlinear basis function vector for the control input \textit{i} which can have
any arbitrary functionality of time and state. $\vth_i^*$ is the optimal
parameter vector. Calculating the optimal parameter vector for this linear model
introduces a linear regression problem. In order to solve the regression problem,
we must define an appropriate objective function. Here we will consider the
mean squared error criterion. The optimal parameter vector for each of the
$m$-control inputs can be calculated through the following optimization.
\begin{align}
\vth_i^* &= \argmax\limits_{\theta_i}{L(\vth_i)} \notag \\
&= \argmax\limits_{\vth_i}{ \int_{t_0}^{t_f}{{\int \frac{1}{2}\norm{u^*_i(s,\vy) - \vUpsilon_i^T(s,\vy) \vth_i}^2_2 \ p(s,\vy)d\vy} ds} }
\end{align}
where $p(s,\vy)$ is an arbitrary weighting function that has a unit integral
(therefore it can be considered as a probability distribution function). Since
the gradient of the optimal parameter vector should be zero, we can write
\begin{align}
 \frac{\partial L(\vth_i^*)}{\partial \vth_i} = \int_{t_0}^{t_f}{{\int \Big(u^*_i(s,\vy) - \vUpsilon_i^T(s,\vy) \vth_i^* \Big) \vUpsilon_i(s,\vy) \ p(s,\vy)d\vy} ds} = 0
\end{align}
Through multiplying and dividing the integrand by $\operatorname{E}_{\tau_{c}} \left\{ \operatorname{e}^{-\frac{1}{\lambda}  R(\tau;s,\vy) } \right\}$, we get
\begin{align}
 \int_{t_0}^{t_f}{{\int \frac{ \operatorname{E}_{\tau_{c}} \left\{ \operatorname{e}^{-\frac{1}{\lambda}  R(\tau;s,\vy) } \right\} }    {\operatorname{E}_{\tau_{c}} \left\{ \operatorname{e}^{-\frac{1}{\lambda}  R(\tau;s,\vy) } \right\}} \Big(u^*_i(s,\vy) - \vUpsilon_i^T(s,\vy) \vth_i^* \Big) \vUpsilon_i(s,\vy) \ p(s,\vy)d\vy} ds} = 0
\end{align}
We then push all of the terms inside of the nominator's
$\operatorname{E}_{\tau_{c}}\{.\}$. Notice that this operation is eligible
because the initial state $\vy$ is not a part of the expectation and is
assumed to be known.
\begin{equation}
 \int_{t_0}^{t_f}{{\int_{\Omega} \operatorname{E}_{\tau_{c}} \left\{\frac{  \operatorname{e}^{-\frac{1}{\lambda}  R(\tau;s,\vy) } }    {\operatorname{E}_{\tau_{c}} \left\{ \operatorname{e}^{-\frac{1}{\lambda}  R(\tau;s,\vy) } \right\}} \Big(u^*_i(s,\vy) - \vUpsilon_i^T(s,\vy) \vth_i^* \Big) \vUpsilon_i(s,\vy) \ p(s,\vy) \right\} d\vy} ds} = 0
\end{equation}
In order to shorten notation, we will show all of the expectations in integral
form and only use one integral sign, instead of three.
\begin{equation}
 \int {\frac{  \operatorname{e}^{-\frac{1}{\lambda}  R(\tau;s,\vy) } }    {\operatorname{E}_{\tau_{c}} \left\{ \operatorname{e}^{-\frac{1}{\lambda}  R(\tau;s,\vy) } \right\}} \Big(u^*_i(s,\vy) - \vUpsilon_i^T(s,\vy) \vth_i^* \Big) \vUpsilon_i(s,\vy) \ \mathbb{P}_{\tau_c}(\tau\mid s, \vy) p(s,\vy) \  d\tau d\vy ds} = 0
\end{equation}
In the next step, we will add and subtract the terms $u_i(s,\vy) + \varepsilon(s)$.
\begin{align}
 \int \frac{  \operatorname{e}^{-\frac{1}{\lambda}  R(\tau;s,\vy) } }    {\operatorname{E}_{\tau_{c}} \left\{ \operatorname{e}^{-\frac{1}{\lambda}  R(\tau;s,\vy) } \right\}} \Big(u^*_i - u_i - \varepsilon + u_i + \varepsilon - \vUpsilon_i^T \vth_i^* \Big) \vUpsilon_i(s,\vy) \ \mathbb{P}_{\tau_c}(\tau\mid s, \vy) p(s,\vy) d\tau d\vy ds = 0
\end{align}
We break this integral into two integrals as follows
\begin{align}
 \int &{\frac{  \operatorname{e}^{-\frac{1}{\lambda}  R(\tau;s,\vy) } }    {\operatorname{E}_{\tau_{c}} \left\{ \operatorname{e}^{-\frac{1}{\lambda}  R(\tau;s,\vy) } \right\}} \Big(u^*_i(s,\vy) -u_i(s,\vy) - \varepsilon \Big) \vUpsilon_i(s,\vy) \ \mathbb{P}_{\tau_c}(\tau\mid s, \vy) p(s,\vy) d\tau d\vy ds} + \notag\\
 &\int {\frac{  \operatorname{e}^{-\frac{1}{\lambda}  R(\tau;s,\vy) } }    {\operatorname{E}_{\tau_{c}} \left\{ \operatorname{e}^{-\frac{1}{\lambda}  R(\tau;s,\vy) } \right\}} \Big(u_i(s,\vy) + \varepsilon - \vUpsilon_i^T(s,\vy) \vth_i^* \Big) \vUpsilon_i(s,\vy) \ \mathbb{P}_{\tau_c}(\tau\mid s, \vy) p(s,\vy) d\tau d\vy ds} = 0
\end{align}
For the first integral, we will first integrate it with respect to trajectory
$\tau$ and then use equation~\eqref{eq:pi_is_optimal_controller}.
\begin{align*}
 & \int {\frac{  \operatorname{e}^{-\frac{1}{\lambda}  R(\tau;s,\vy) } }    {\operatorname{E}_{\tau_{c}} \left\{ \operatorname{e}^{-\frac{1}{\lambda}  R(\tau;s,\vy) } \right\}} \Big(u^*_i(s,\vy) -u_i(s,\vy) - \varepsilon \Big) \vUpsilon_i(s,\vy) \ \mathbb{P}_{\tau_c}(\tau\mid s, \vy) p(s,\vy) d\tau d\vy ds} = \notag\\
 & \int { \left(u^*_i(s,\vy) - u_i(s,\vy) - \frac{  \operatorname{E}_{\tau_{c}} \left\{\varepsilon \operatorname{e}^{-\frac{1}{\lambda}  R(\tau;s,\vy) } \right\} }    {\operatorname{E}_{\tau_{c}} \left\{ \operatorname{e}^{-\frac{1}{\lambda}  R(\tau;s,\vy) } \right\}} \right)  \vUpsilon_i(s,\vy) \ (\tau\mid s, \vy) p(s,\vy) d\vy ds} = 0
\end{align*}
By substituting this result back into the previous equation we get
\begin{equation}
 \int {\frac{  \operatorname{e}^{-\frac{1}{\lambda}  R(\tau;s,\vy) } }    {\operatorname{E}_{\tau_{c}} \left\{ \operatorname{e}^{-\frac{1}{\lambda}  R(\tau;s,\vy) } \right\}} \Big(u_i(s,\vy) + \varepsilon - \vUpsilon_i^T(s,\vy) \vth_i^* \Big) \vUpsilon_i(s,\vy) \ \mathbb{P}_{\tau_c}(\tau\mid s, \vy) p(s,\vy) d\tau d\vy ds} = 0
\end{equation}
which is equivalent to the following optimization problem
\begin{equation} \label{eq:pi_is_fa_optimal_controller_original}
 \vth_i^* = \argmin_{\vth_i}{\int {\frac{  \operatorname{e}^{-\frac{1}{\lambda}  R(\tau;s,\vy) } }    {\operatorname{E}_{\tau_{c}} \left\{ \operatorname{e}^{-\frac{1}{\lambda}  R(\tau;s,\vy) } \right\}} \norm{\vUpsilon_i^T(s,\vy) \vth_i - u_i(s,\vy) - \varepsilon}^2_2 \ \mathbb{P}_{\tau_c}(\tau\mid s, \vy) p(s,\vy) d\tau d\vy ds}}
\end{equation}
If we use the same function approximation model for $u_i(s,\vy)$ and $u_i(s,\vy) \approx \vUpsilon_i^T(s,\vy) \vth_{i,c}$, we get
\begin{equation} \label{eq:pi_is_fa_optimal_controller}
 \vth_i^* = \vth_{i,c} + \argmin_{\Delta \vth_i}{\int {\frac{  \operatorname{e}^{-\frac{1}{\lambda}  R(\tau;s,\vy) } }    {\operatorname{E}_{\tau_{c}} \left\{ \operatorname{e}^{-\frac{1}{\lambda}  R(\tau;s,\vy) } \right\}} \norm{\vUpsilon_i^T(s,\vy) \Delta \vth_i - \varepsilon}^2_2 \ \mathbb{P}_{\tau_c}(\tau\mid s, \vy) p(s,\vy) d\tau d\vy ds}}
\end{equation}
Equation~\eqref{eq:pi_is_fa_optimal_controller_original} or
\eqref{eq:pi_is_fa_optimal_controller} introduce a method to blend the function
approximation and optimal control problems into a single optimization problem.
In the next section, we will introduce the general path integral  algorithm which
uses importance sampling and function approximation.

\section{General Path Integral Algorithm}
In this section, we will present the path integral general algorithm. In this
algorithm we will use all the components that we already have introduced in the
previous sections. Before putting things all to gather, we will describe the
reason behind each of the component separately.

\paragraph{Importance Sampling} As discussed previously, the original path
integral algorithm requires that the samples are extracted from the uncontrolled
system. Therefore, estimating the optimal control through a Monte Carlo method
will be extremely inefficient. However by using the Importance Sampling scheme
introduced in \sref{sec:PI_SI}, we can implement an incremental method to
estimate the optimal control which is more sample efficient. 

In this approach, we can start the sampling with a sophisticated guess
of the optimal control. Note that this is optional,though, and we could also start from the
uncontrolled system.In each iteration of the algorithm, we use our
latest guess of the optimal control to extract a batch of samples. Using this
batch of samples, we can improve our estimation of the optimal control. Our sampling efficiently inherently improves at the same time, since, as
our estimation improves, we sample an the area with lower accumulated cost and
therefore the sample weighings, $\alpha(\tau;s,\vy)$ (equation
\ref{eq:pi_sample_weighting}), will have more significant values. It can be shown
that if we are in the $\epsilon$-vicinity of the optimal solution the sample
efficiency will improve linearly with $\epsilon$.

\paragraph{Function Approximation} The original path integral theorem introduces
a sampling method for estimating the optimal control at a specific time and
state. Therefore for estimating the optimal control in a given time period and some subspace of the state space, we should repeat the estimation individually
for every point. In \sref{sec:PI_FA}, we introduced a function
approximation scheme in which the estimated optimal control for a single point
generalizes to its neighboring area. Another advantage of function approximation
is the introduction of $p(s,\vy)$ (the weighting function of MSE). For example,
assume we have chosen $p(s,\vy)$ as the probability distribution of the state under
the latest estimation of the optimal control. In this case, the function
approximation will be more precise in the area more likely to be visited
using the optimal policy.

Algorithm~\ref{alg:gpi} shows the General Path Integral Algorithm which uses
importance sampling and function approximation. 

\begin{algorithm}[tpb] \caption{General Path Integral Algorithm} \label{alg:gpi}
\begin{algorithmic} 
\State \textbf{given}
\State \hspace{4mm} The cost function: 
\State \hspace{8mm} $J = \Phi(\vx(t_f))+ \int_{s}^{t_f} {\left(q(t,\vx) + \frac{1}{2} \vu^T \vR \vu\right) dt}$
\State \hspace{4mm} A PDF defining the quality of approximation of optimal control at each time-state pair: $p(t,\vx)$
\State \hspace{4mm} An initial policy and a Linear Model: $\vu(t,\vx) = [u_i(t,\vx)] = [\vUpsilon^T_i(t,\vx) \vth_i]$
\Repeat{}
\State (a) Randomly choose a time-state pair from $p(t,\vx)$: $(s,\vy)$
\State (b) Forward simulate the controlled system for $K$ different rollouts: $\{ \tau^k \}_{k=1}^K$ 
\State \hspace{8mm} $\dot{\vx} = \vf(t,\vx) + \vg(t,\vx) \left(\vu + \vvep \right) $ 
\State \hspace{8mm} $ \vu(t,\vx) = [\vUpsilon^T_i(t,\vx) \vth_i], \ \vvep \sim \mathcal{N}(\mathbf{0},\mathbf{\Sigma}),\ \vx(t=s)=\vy$ 
\State (c) Calculate the return for each rollout: $\{ R^k \}_{k=1}^K$
\State \hspace{8mm} $R(\tau;s,\vy) = \Phi(\vx(t_f))+ \int_{s}^{t_f} {\left(q(t,\vx) + \frac{1}{2} \vu^T \vR \vu\right) dt} + \int_{s}^{t_f} {\vu^T \vR d\vw}$
\State (d) Calculate $\{\alpha^k\}_{k=1}^K$ 
\State \hspace{8mm} $\alpha^k(s,\vy) = \sfrac{ \exp (-\frac{1}{\lambda} R^k) }    {\frac{1}{K} \sum\limits_{j=1}^{K} \exp
(-\frac{1}{\lambda}  R^j) }$
\State (e) Solve the following linear regression problem for each control input $i$:
\State \hspace{8mm} $\Delta \vth_i = \argmin{\sum\limits_{k=1}^{K} {\alpha^k \norm{\vUpsilon_i^T(s,\vy) \Delta \vth_i - \varepsilon^k_i(s)}^2_2 }}$
\State (f) Update the parameter vector for each control input $i$:
\State \hspace{8mm} $\vth_i \leftarrow \vth_i + \omega \Delta \vth_i$
\Until{convergence}
\end{algorithmic} 
\end{algorithm}

\section{PI2 Algorithm: Time-dependent Policy}
The Policy Improvement Path Integral algorithm (PI2) is a variant of General Path
Integral algorithm. PI2 is an instance of a sample-based reinforcement learning
algorithm that can retrieve trajectory samples directly from the real
environment i.e. the real system. In this section, we will derive the PI2
algorithm from the General Path Integral algorithm.

Originally the PI2 algorithm was developed for optimizing parameterized,
time-dependent policies. The time-dependent policy can be either a parameterized
dynamical system like Dynamic Movement Primitives (DMPs) or simply a function
approximation of the control input like the one in
equation~\eqref{eq:pi2_t_policy}. The essential requirements for the algorithm are linearity w.r.t
the parameter vector, and the usage of time-dependent basis functions. In this section,
we will use function-approximation policies since it will later give us the
opportunity to extend the PI2 algorithm to more general polices that are
functions of time and state.
\begin{align} \label{eq:pi2_t_policy}
u_i(t) &= \vUpsilon^T(t) \vth_i \notag \\
\vUpsilon(t) &= \big[ \Upsilon_n(t) \big]_{N \times 1} =
\left[ e^{-\frac{1}{2}\frac{(t-\mu_n)^2}{\sigma_{n}^{2}}} \right]_{N \times 1}
\end{align}
In this equation, $u_i$ refers to the $i$th control input and $\vUpsilon$ is
a time-dependent basis function. Each element of $\Upsilon$ (denoted by
$\Upsilon_n$) is a bell-shape function with a mean and variance, $\mu_n$ and
$\sigma^2_n$ respectively. The dimension of the parameter vector as well as the
basis function vector is $N$.

In order to derive PI2 from the General Path Integral algorithm, we should make
the following assumptions:
\begin{enumerate}
\item The weighting function of MSE, $p(t,\vx)$, is assumed to be the probability
distribution of the state under the latest estimation of the optimal control.
Therefore the rollout trajectory's states are considered to be extracted from
$p(t,\vx)$.
\item The return function, $R(\tau;s,\vy)$, won't be a function of state if the
system has been initialized from a similar initial condition. An immediate result
of this assumption is that $\alpha^k(s,\vy)$ is only a function of time i.e.
$\alpha^k(s)$. Therefore for a batch of trajectories extracted from a similar
initial condition, $\alpha^k(s)$ can be estimated for each time step by the
following.
\begin{equation}
\alpha^k(s) = \frac{ \exp (-\frac{1}{\lambda} R^k(s)) }    {\frac{1}{K} \sum\limits_{j=1}^{K} \exp
(-\frac{1}{\lambda}  R^j(s) ) }
\end{equation}
Notice that in contrast to the general path integral algorithm, where we are
required to estimate $\alpha^k(s,\vy)$ by extracting trajectories starting from
$(s,\vy)$, we can use trajectories with different state values but
the same time indexes to estimate $\alpha$. Therefore, if we have a batch of
trajectories that are generated from similar initial conditions, we can estimate
$\alpha$ for all of the states and over the given time horizon, $[t_0,t_f]$.
\item The basis function vector for all of the control inputs is the same. Hence we
will drop the $i$ subscript of the basis function as $[u_i(t)]=[\vUpsilon^T(t)
\vth_i]$.
\item Instead of adding noise to the control input, the noise is added
directly to the parameter vector. Therefore, the input noise can be expressed as
$\vvep_i = \vUpsilon(t) \vep_i$, where $\vep_i$ is the noise that is added to the
parameter vector of the $i$th control input.
\item The PI2 regression problem should be modified as follows
\begin{equation}
\Delta \vth_i = \argmin{\sum\limits_{s=t_0}^{t_f}{\sum\limits_{k=1}^{K} {\alpha^k(s) \norm{\vUpsilon_i^T(s) \Delta \vth_i - \varepsilon^k_i(s)}^2_2 }}}
\end{equation}
In contrast to the General Path Integral Algorithm, PI2 assumes that samples
are extracted over the entire time horizon, not only for a single time step.
Therefore, the regression problem should be over all of the sampled trajectories as well as all of the sub-trajectories of each individual trajectory, where a sub-trajectory is a portion of a trajectory starting from some time $s$ until the end of the time horizon (as was illustrated in \sref{sec:mc_exploring_starts}).

In order to solve this regression problem, PI2 breaks it into two separate
optimizations. In doing so, it is assumed that the 
the regression error is zero mean over the samples. In this method, the
optimization is first solved for each time step separately. Therefore the first
optimization will find a time-dependent parameter vector increment that has the
minimum error over the rollouts at each time step. In the second optimization, we find a parameter vector increment that approximates the time-dependent one.

The first optimization is defined as follows for each time step $s$
\begin{equation} \label{eq:pi2_regression_inner}
\Delta \vth_i^*(s) = \argmin{\sum\limits_{k=1}^{K} {\alpha^k(s) \norm{\vUpsilon_i^T(s) \Delta \vth_i - \varepsilon^k_i(s)}^2_2 }}
\end{equation}
In this optimization, all of the regressors are the same. Therefore in order to find
the solution to this problem, we should use the right inverse which will give a
solution with the minimum-norm $\Delta \vth_i^*(s)$
\begin{equation} \label{eq:pi2_regression_inner_sol1}
\Delta \vth_i^*(s) = \sum\limits_{k=1}^{K} { \alpha^k(s) \frac{\vUpsilon_i(s)}{\vUpsilon_i^T(s) \vUpsilon_i(s)} \varepsilon^k_i(s) }
\end{equation}
If we use the fourth assumption that the noise is directly added to the parameter
vector, we get
\begin{equation} \label{eq:pi2_regression_inner_sol2}
\Delta \vth_i^*(s) = \sum\limits_{k=1}^{K} { \alpha^k(s) \frac{\vUpsilon_i(s) \vUpsilon_i^T(s)}{\vUpsilon_i^T(s) \vUpsilon_i(s)} \vep^k_i(s) }
\end{equation}
The second optimization for finding the optimal $\Delta\vth_{i}^*$ is defined in
equation~\eqref{eq:pi2_regression_outer}. The index $n$ refers to the $n$th
element of the vector $\Delta\vth_{i}^*=[\Delta\theta_{i,n}^*]$
\begin{equation} \label{eq:pi2_regression_outer}
\Delta \theta_{i,n}^* = \argmin\limits_{\Delta \theta_{i,n}}{\sum\limits_{s=t_0}^{t_f}{  ( \Delta \theta_{i,n} - \Delta \theta_{i,n}^*(s) )^2 \Upsilon_n(s) }}
\end{equation}
$\Upsilon_n(t)$ is the $n$th element of the basis function vector $\vUpsilon(t)$. 
The solution to this optimization can be easily calculated as
\begin{equation}
\Delta \theta_{in}^* = \frac{ \sum\limits_{t=t_0}^{t_f}{\Delta \theta_{in}^*(s) \Upsilon_n(s)} }{ \sum\limits_{t=t_0}^{t_f}{\Upsilon_n(s)} }
\end{equation}
By using element-wise multiplication and division and replacing the summation
with an integral, we can also write
\begin{equation} \label{eq:pi2_regression_outer_sol}
\Delta \vth_{i}^* = \left({ \int\limits_{t_0}^{t_f}{\Delta \vth_{i}^*(s) \circ \vUpsilon(s) ds} }\right)  \cdot\bigg/ { \int\limits_{t_0}^{t_f} {\vUpsilon(s) ds} }
\end{equation}
where $\circ$ and $\cdot/$ are element-wise multiplication and division.
\end{enumerate}
Using these assumptions, we can derive PI2 from the General Path Integral
algorithm. The complete algorithm is given as
Algorithm~\ref{alg:pi2_time_policy}. Notice that in PI2, we do not need
to artificially initialize the system at random points $(s,\vy)$ according to
$p(s,\vy)$. Therefore this algorithm is a sample-based
reinforcement learning algorithm and can be used to learn the optimal policy through
samples that are generated directly from the real physical system.

\newpage

\begin{algorithm}[H] \caption{PI2 Algorithm for time-dependent policy} \label{alg:pi2_time_policy}
\begin{algorithmic}
\State \textbf{given}
\State \hspace{4mm} The cost function: 
\State \hspace{8mm} $J = \Phi(\vx(t_f))+ \int_{t_0}^{t_f} {\left(q(t,\vx) + \frac{1}{2} \vu^T \vR \vu\right) dt}$
\State \hspace{4mm} A linear model for function approximation: $\vu(t) = [u_i(t)] = [\vUpsilon^T(t) \vth_i]$
\State \hspace{4mm} Initialize $[\vth_i]$ with a sophisticated guess 
\State \hspace{4mm} Initialize exploration noise standard deviation: $c$

\Repeat
\State Create $K$ rollouts of the system with the perturbed parameter $[\vth_i] +
[\vep_i], \quad \vep_i \sim \mathcal{N} (\mathbf{0},c^2\vI)$
\For{the $i$th control input}
\For{each time, s}
\State Calculate the Return from starting time $s$ for the $k$th rollout:
\State \hspace{4mm} $R(\tau^k(s)) = \Phi(\vx(t_f)) + \int_{s}^{t_f} {\left(q(t,\vx) + \frac{1}{2} \vu^T \vR \vu\right) dt}$
\State Calculate $\alpha$ from starting time $s$ for the $k$th rollout:
\State \hspace{4mm} $\alpha^k(s) = \sfrac{ \exp ( -\frac{1}{\lambda}R(\tau^k(s)) ) }{
\sum_{k=1}^{K}{ \exp( -\frac{1}{\lambda}R(\tau^k(s)) ) } }$
\State Calculate the time varying parameter increment $\Delta \vth_{i}(s)$:
\State \hspace{4mm} $\Delta \vth_{i}(s) = \sum_{k=1}^{K}{\alpha^k(s) \frac{\vUpsilon(s) \vUpsilon^T(s)}{ \vUpsilon^T(s) \vUpsilon(s)} \vep^k_i(s)}$
\EndFor
\State Time-averaging the parameter vector
\State \hspace{4mm} $\Delta \vth_{i} = \left({ \int\limits_{t_0}^{t_f}{\Delta \vth_{i}(s) \circ \vUpsilon(s) ds} }\right) \cdot\bigg/ { \int\limits_{t_0}^{t_f} {\vUpsilon(s) ds} }$
\State Update parameter vector for control input $i$, $\vth_i$:
\State \hspace{4mm} $\vth_i \leftarrow \vth_i + \Delta \vth_i$
\EndFor
\State Decrease c for noise annealing
\Until{maximum number of iterations}
\end{algorithmic}
\end{algorithm}

\section{PI2 Algorithm: Feedback Gain Learning}
In a first attempt to tailor the PI2 algorithm to more general policies, we will introduce a method to
learn a linear time-varying feedback gain. Consider the following tracking problem
\begin{equation}
\begin{cases}
\dot{\vx} = \vf(t,\vx) + \vg(t,\vx)\vu \\
\vu = \vK^T(t)(\vx-\vx_{ref})
\end{cases}
\end{equation}
For now, we will assume that the control input dimension is one. If we substitute
the linear feedback controller back into the system dynamics, we will get
\begin{equation}
\dot{\vx} = \vf(t,\vx) + \vg(t,\vx)  (\vx-\vx_{ref})^T \vK(t) \\
\end{equation}
According to this equation, we can assume a new system with $\vK(t)$ as its
control input. Considering this new system, we can use
Algorithm~\ref{alg:pi2_time_policy} to learn the time varying feedback gains.
This method can be easily extended to a system with multiple control inputs. The
learning algorithm in this case will assume a system that has as many inputs as
the total number of gains. In other words if the dimensions of the state and control
input vectors are $\dim[\vx]$ and $\dim[\vu]$ respectively, the dimension of
control input from the perspective of PI2 will be $\dim[\vx] \times \dim[\vu]$.

In the next section, we will introduce a more general algorithm which will
learn both time-dependent policies and the feedback gains simultaneously.

\section{PI2 Algorithm: General Algorithm}
In this section, the General PI2 algorithm will be modified to allow
policies with linear state dependency and a nonlinear time dependency to be learned. An
instance of such a policy is as follows
\begin{align} \label{eq:pi2_t_x_policy}
u_i(t,\vx) &= \operatorname{grand\,sum}{\left[\bar{\vUpsilon}(t,\vx) \circ \vth_i\right]}  \notag\\
\bar{\vUpsilon}(t,\vx) &= \vUpsilon(t) \begin{bmatrix} 1 & \vx^T \end{bmatrix} =
\left[ e^{-\frac{1}{2}\frac{(t-\mu_n)^2}{\sigma_{n}^{2}}} \begin{bmatrix} 1 & \vx^T \end{bmatrix}
\right]_{N \times (1+\dim{[\vx]})}
\end{align}
The $\operatorname{grand\,sum}$ operator calculates the sum of all the elements
in a matrix and $\circ$ is the element-wise multiplication.  $\vth_i$ is a $N
\times (1+\dim{[\vx]})$ parameter matrix for $i$th control input approximation.
$\bar{\vUpsilon}(t,\vx)$ is the basis function matrix of the same size as
$\vth_i$. Finally $\vUpsilon(t)$ is defined as in equation~\eqref{eq:pi2_t_policy}.

By setting the parameters which are multiplied by the states to zero, this policy
reduces to the time varying policy primarily introduced in the
equation~\eqref{eq:pi2_t_policy}. Using the same argument from the previous
section we can assume that the state dependent segment of the policy is part of
the system dynamics and PI2 will be used to learn the time-varying gains. Based
on the policy in equation~\eqref{eq:pi2_t_x_policy}, we can assume that these
gains are sufficiently approximated by the using time-dependent basis function, $\vUpsilon(t)$.
The compete algorithm is given in Algorithm~\ref{alg:general_pi2}.

\begin{algorithm}[tpb] \caption{General PI2 Algorithm} \label{alg:general_pi2}
\begin{algorithmic}
\State \textbf{given}
\State \hspace{4mm} The cost function: 
\State \hspace{8mm} $J = \Phi(\vx(t_f))+ \int_{t_0}^{t_f} {\left(q(t,\vx) + \frac{1}{2} \vu^T \vR \vu\right) dt}$
\State \hspace{4mm} A linear model for function approximation: $\vu(t,\vx) = [u_i(t,\vx)] = \left[\operatorname{grand\,sum}{[\bar{\vUpsilon}(t,\vx) \circ \vth_i]} \right]$
\State \hspace{4mm} Initialize $\{\vth_i\}$ with a sophisticated guess 
\State \hspace{4mm} Initialize exploration noise standard deviation: $c$

\Repeat
\State Create $K$ rollouts of the system with the perturbed parameter $ \{\vth_i\} +
\{\vep_i\}, \  \{\vep_{i,j}\} \sim \mathcal{N} (\mathbf{0},c^2\vI)$
\For{the $i$th control input}
\For{each time, s}
\State Calculate the Return from starting time $s$ for the $k$th rollout:
\State \hspace{4mm} $R(\tau^k(s)) = \Phi(\vx(t_f)) + \int_{s}^{t_f} {\left(q(t,\vx) + \frac{1}{2} \vu^T \vR \vu\right) dt}$
\State Calculate $\alpha$ from starting time $s$ for the $k$th rollout:
\State \hspace{4mm} $\alpha^k(s) = \sfrac{ \exp ( -\frac{1}{\lambda}R(\tau^k(s)) ) }{
\sum_{k=1}^{K}{ \exp( -\frac{1}{\lambda}R(\tau^k(s)) ) } }$
\State Calculate the time varying parameter increment $\Delta \vth_{i}(s)$:
\State \hspace{4mm} $\Delta \vth_{i}(s) = \sum_{k=1}^{K}{\alpha^k(s) \frac{\vUpsilon(s) \vUpsilon^T(s)}{ \vUpsilon^T(s) \vUpsilon(s)} \vep^k_i(s)}$
\EndFor
\For{the $j$th column of $\Delta \vth_{i}$ matrix, $\Delta \vth_{i,j}$}
\State Time-averaging the parameter vector
\State \hspace{4mm} $\Delta \vth_{i,j} = \left({ \int\limits_{t_0}^{t_f}{\Delta \vth_{i,j}(s) \circ \vUpsilon(s) ds} }\right) \cdot\bigg/ { \int\limits_{t_0}^{t_f} {\vUpsilon(s) ds} }$
\EndFor
\State Update parameter vector for control input $i$, $\vth_i$:
\State \hspace{4mm} $\vth_i \leftarrow \vth_i + \Delta \vth_i$
\EndFor
\State - Decrease c for noise annealing
\Until{maximum number of iterations}
\end{algorithmic}
\end{algorithm}
\thispagestyle{empty}\cleardoublepage
\chapter{Policy Gradient}
Assume the following optimal control problem with the cost function, $J$ defined
as follows
\begin{equation} \label{eq:gd_cost_function}
J= E \left[ \Phi(\vx(N))+\sum_{k=0}^{N-1} L_k \left(\vx(k),\vu(k)\right) \right]
\end{equation}
and the system dynamics as
\begin{equation} \label{eq:gd_discrete_time_ stochastic}
\vx(n+1)=\vf_n(\vx(n),\vu)+\vw(n)
\end{equation}
where $\vw(n)$ is an arbitrary random process. Finding the optimal
controller for this problem is nearly impossible, except in very special cases. Lets assume that,
instead of finding the actual optimal control, we want to find a parameterized
policy that has the lowest cost among a given class of parametrized policies. If
we express the parameterized policy as $\mu(n,\vx;\vth)$, the optimal
control problem will be defined as follows
\begin{align} \label{eq:gd_optimal_problem}
& \vth^* = \arg\min\limits_{\vth} {J(\vth)}= \arg\min\limits_{\vth} { E \left[ \Phi(\vx(N))+\sum_{k=0}^{N-1} L_k \left(\vx(k),\vu(k)\right) \right]} \notag \\
& \begin{cases}
\vx(n+1)=\vf_n(\vx(n),\vu)+\vw(n) \\
\vu(n,\vx) = \mu(n,\vx;\vth)
\end{cases}
\end{align}
From another perspective, we can look at this problem as a function approximation
problem for the optimal control. As opposed to the functional approximation method used in the PI2 algorthim, in which the policies had to be  linear with respect to the parameters, here we can consider a more general class of policies which are non-linear with respect to $\vth$.
 
The proposed optimal control problem in equation~\eqref{eq:gd_optimal_problem}
can be directly solved if we can calculate the gradient of the cost function with
respect to the parameter vector, $\vth$. Then we can simply use an optimization
algorithm to find the optimal parameter vector, $\vth^*$.
Algorithm~\ref{alg:gradient_descent} introduces an instance of this optimization
algorithm which uses the gradient descent method. In this algorithm, $\omega$ is
the learning rate.

\begin{algorithm}[tpb] \caption{Gradient Descent Algorithm} \label{alg:gradient_descent}
\begin{algorithmic}
\State \textbf{given}
\State \hspace{4mm} A method to compute $\nabla_{\vth}J(\vth)$ for all $\vth$
\State \hspace{4mm} An initial value for the parameter vector: $\vth \leftarrow \vth_0$ 
\Repeat
\State Compute the cost function gradient at $\vth$
\State \hspace{4mm} $\vg = \nabla_{\vth}J(\vth)$
\State Update the parameter vector
\State \hspace{4mm} $\vth \leftarrow \vth - \omega \vg$
\Until{convergence}
\end{algorithmic}
\end{algorithm}

However in order to find a closed form formula for the cost function gradient we
first need to calculate the cost function as a function of the parameter vector. This
requires that the states are computed as a function of the parameter vector by
solving the difference equation of the system. The solution to this problem does not exist, though, for a
general nonlinear system.

In the absence of the closed form solution, we need to use a numerical method to
directly estimate the gradient. In the reinforcement learning and optimal control
literature, numerous methods have been proposed to estimate the gradient (as well
as Hessian matrix if we want to use an optimization method that requires the
Hessian matrix). In this chapter we will introduce a very basic, yet
effective, method to estimate the cost function gradient. This technique is known
as the Finite Difference method.

\section{Finite Difference}
The Finite Difference (FD) method estimates the cost function gradient with respect to the optimal control parameter vector numerically. Before introducing the algorithm in its complete form, we will
start with a slightly simpler problem. First consider a policy that has a single
parameter and assume that there is no stochasticity in the problem i.e.
$\vw(n) = 0$. The gradient of the cost function will be a scaler which is the derivative of the cost with respect to the parameter. Based on the
definitions, the derivative of a function can be calculated through each of the
two following formulas.
\begin{align} \label{eq:1d_single_side_dev}
\frac{d J(\theta)}{d \theta \ \ } &= \lim\limits_{d \theta \to 0} \frac{J(\theta + d \theta) - J(\theta)}{d \theta}   \\ 
\label{eq:1d_double_side_dev} 
\frac{d J(\theta)}{d \theta \ \ } &= \lim\limits_{d \theta \to 0} \frac{J(\theta + d \theta/2) - J(\theta - d \theta/2)}{d \theta}
\end{align} 
The method in equation~\eqref{eq:1d_single_side_dev} is called the single-sided
derivative and the one in equation~\eqref{eq:1d_double_side_dev} is called the
double-sided derivative. For an infinitely small perturbation, $d\theta$, these
methods calculate the derivative of the cost function at $\theta$. However with a
sufficiently small perturbation $\Delta \theta$, we can approximate the
derivative through the followings methods
\begin{align} \label{eq:1d_single_side_dev_app}
\frac{d J(\theta)}{d \theta \ \ } &\approx \frac{J(\theta + \Delta \theta) - J(\theta)} {\Delta \theta}   \\ 
\label{eq:1d_double_side_dev_app} 
\frac{d J(\theta)}{d \theta \ \ } &\approx \frac{J(\theta + \Delta \theta/2) - J(\theta - \Delta \theta/2)} {\Delta \theta}
\end{align} 
Therefore by only calculating the value of the cost function at two different
points ($\theta$ and $\theta + \Delta \theta$ for single-sided approximation and
$\theta - \frac{1}{2} \Delta \theta$ and $\theta + \frac{1}{2} \Delta \theta$ for
the double-sided approximation), we can approximate the derivative at $\theta$.
To compute the cost function, we need to simulate the system dynamics with the
given parameter value and then calculate the cost function for the rollout. Since there is no stochasticity in this simple example, we are guaranteed to get the true value of the cost function from this single rollout.

\begin{figure}[tpb]
\centering
\includegraphics[width=0.3\textwidth]{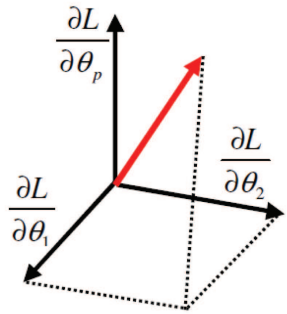}
\caption{Gradient w.r.t. a vector of parameter $\theta$ with the size $p$}
\label{fig:partial_derivative}
\end{figure}

A natural extension to this approach is to assume that we have a vector of parameters
instead of a single parameter. In this case we need to calculate the gradient of
the cost. By definition, in order to calculate one element of the
gradient vector, we should only perturb the parameter associated with that element
(Figure~\ref{fig:partial_derivative}). In this case if we use the single-sided
approximation, we need to calculate the cost function $\dim[\vth] + 1$ times. One for the cost at the non-perturbed parameter $\vth$ and
$\dim[\vth]$ times for each element of the parameter vector. If we use the double-sided method, we need $2 \times \dim[\vth]$
times of the cost function's evaluations, with $2$ perturbations per parameter
element.

In the next section, we will examine a case where the perturbations are not
evaluated at each parameter element separately, but instead are applied in a random
directions.

\subsection{FD with Random Perturbations}
\begin{figure} [tpb]
\centering
\includegraphics[width=0.3\textwidth]{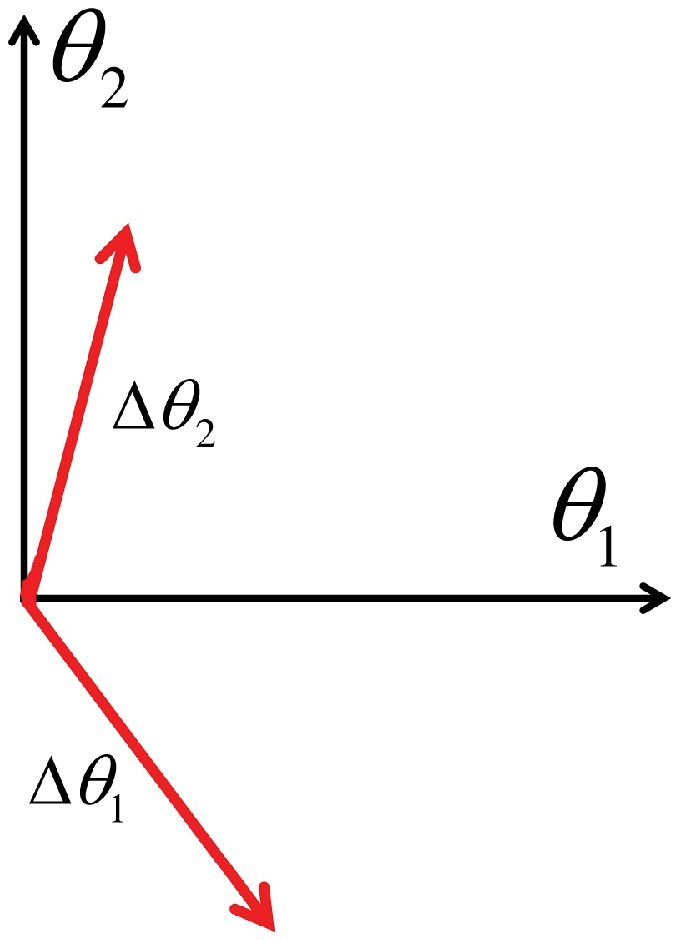}
\caption{FD example with 3 computations of cost function: $J(\vth)$, $J(\vth + \Delta \vth_1)$, $J(\vth + \Delta \vth_2)$}
\label{fig:fd_2_perturbations}
\end{figure}

Before introducing the FD method with random perturbations, we will use two
simple examples as an introduction to the main idea. Both of the examples are
considered to be deterministic. In the first example, we will assume that we have
just two parameters. In this example we will use the single-sided method. Thus we
need three evaluations of the cost function. Lets assume that these measurements
are $J(\vth)$, $J(\vth + \Delta \vth_1)$, $J(\vth + \Delta \vth_2)$ which are the values of the
cost function for the parameter vector without perturbation and two vectors with random perturbations, $\Delta \vth_1$ and
$\Delta \vth_2$ (Figure~\ref{fig:fd_2_perturbations}). In order to estimate the
gradient at $\vth$, we will use the first order Taylor expansion of the cost
function around $\vth$
\begin{align}
J(\vth + \Delta \theta_1)  &\approx J(\vth) + \Delta \vth_1^T \nabla J(\vth) \notag \\
J(\vth + \Delta \theta_2)  &\approx J(\vth) + \Delta \vth_2^T \nabla J(\vth) 
\end{align}

\begin{figure} [tpb]
\centering
\includegraphics[width=0.3\textwidth]{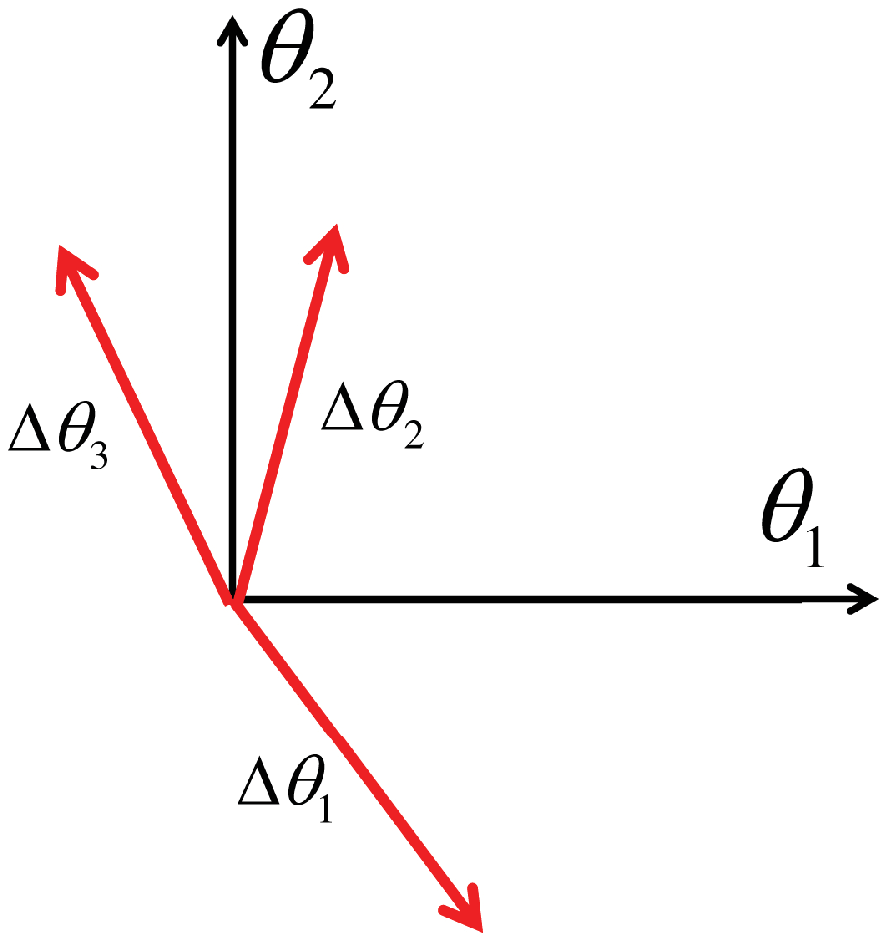}
\caption{FD example with 3 computations of cost function: $J(\vth + \Delta \vth_1)$, $J(\vth + \Delta \vth_2)$, $J(\vth+\vth_3)$}
\label{fig:fd_3_perturbations}
\end{figure}

Using matrix notation, these two equalities can be written in the following form
\begin{equation}
\begin{bmatrix}
\Delta \vth_1^T \\
\Delta \vth_2^T
\end{bmatrix}
\nabla J(\vth) =
\begin{bmatrix}
J(\vth + \Delta \vth_1) - J(\vth) \\
J(\vth + \Delta \vth_2) - J(\vth)
\end{bmatrix}
\end{equation}
In oder to find the gradient of the cost function, we need to solve this equality. If
the perturbations are not parallel (linearly independent), $\begin{bmatrix}
\Delta \vth_1^T \\ \Delta \vth_2^T \end{bmatrix}$ is a 2-by-2 invertible matrix.
Therefore the gradient can be calculated as
\begin{equation}
\nabla J(\vth) =
\begin{bmatrix}
\Delta \vth_1^T \\
\Delta \vth_2^T
\end{bmatrix}^{-1}
\begin{bmatrix}
J(\vth + \Delta \vth_1) - J(\vth) \\
J(\vth + \Delta \vth_2) - J(\vth)
\end{bmatrix}
\end{equation}
In the second example, we will consider the same problem except this time, instead
of computing the cost function twice perturbed and once not perturbed, we will compute the cost function using 
three different perturbed values of the parameter vector (Figure~\ref{fig:fd_3_perturbations}). Therefore we have the
following: $J(\vth + \Delta \vth_1)$, $J(\vth + \Delta \vth_2)$,
$J(\vth+ \Delta \vth_3)$.

In contrast to the previous example, in this case we don't have the cost function's
value at the nominal value of the parameters. We can still estimate both the
gradient of the cost function and the value of the cost function for this point, however.
Again we start with the first order Taylor expansion of the cost function
\begin{align}
J(\vth + \Delta \vth_1)  &= J(\vth) + \Delta \vth_1^T \nabla J(\vth) \notag\\
J(\vth + \Delta \vth_2)  &= J(\vth) + \Delta \vth_2^T \nabla J(\vth) \notag\\
J(\vth + \Delta \vth_3)  &= J(\vth) + \Delta \vth_3^T \nabla J(\vth) 
\end{align}
In this case, the unknown entities are $J(\vth)$ and $\nabla J(\vth)$. We can reformulate these equalities in the following matrix form.
\begin{equation}
\begin{bmatrix}
\Delta \vth_1^T  & 1\\
\Delta \vth_2^T & 1\\
\Delta \vth_3^T & 1
\end{bmatrix}
\begin{bmatrix}
\nabla J(\vth) \\
J(\vth)
\end{bmatrix}
=
\begin{bmatrix}
J(\vth + \Delta \vth_1) \\
J(\vth + \Delta \vth_2) \\
J(\vth + \Delta \vth_3)
\end{bmatrix}
\end{equation}
If the perturbations are \emph{pairwise independent} (ie each pair of vectors are independent from each-other) the left matrix is invertible and we can solve for the desired quantities. Notice that since the dimension of the parameter space is two, we can have
a maximum of two independent vectors. Thus the set of three perturbations will never be completely independent.
\begin{equation}
\begin{bmatrix}
\nabla J(\vth) \\
J(\vth)
\end{bmatrix} = 
\begin{bmatrix}
\Delta \vth_1^T & 1 \\
\Delta \vth_2^T & 1 \\
\Delta \vth_3^T & 1
\end{bmatrix}^{-1}
\begin{bmatrix}
J(\vth + \Delta \vth_1) \\
J(\vth + \Delta \vth_2) \\
J(\vth + \Delta \vth_3)
\end{bmatrix}
\end{equation}
This equation estimates the cost function and the gradient of the cost function
simultaneously. In the general case, for a parameter vector of the length $p$ with
$(p+1)$ perturbations such that each $p$-member subset of perturbations is
independent, we can write the following
  
\begin{minipage}{0.55\textwidth}
\begin{align*}
J(\vth + \Delta \vth_1) &= J(\vth) + \Delta \vth_1^T \nabla J(\vth)\\
J(\vth + \Delta \vth_2) &= J(\vth) + \Delta \vth_2^T \nabla J(\vth)\\
&\vdotswithin{=} \hspace{40mm} \Longrightarrow \\
J(\vth + \Delta \vth_p) &= J(\vth) + \Delta \vth_p^T \nabla J(\vth) \\
J(\vth + \Delta \vth_{p+1})  &= J(\vth) + \Delta \vth_{p+1}^T \nabla J(\vth)
\end{align*}
\end{minipage}
\begin{minipage}{0.45\textwidth}
\begin{equation*}
\underbrace{\begin{bmatrix}
J(\vth + \Delta \vth_1) \\ J(\vth + \Delta \vth_2) \\ \vdots \\ J(\vth + \Delta \vth_{p+1})
\end{bmatrix}}_{\text{\large{$\vJ$}}} =
\underbrace{\begin{bmatrix}
\Delta \vth_1^T & 1 \\
\Delta \vth_2^T & 1 \\
\vdots & \\
\Delta \vth_{p+1}^T & 1
\end{bmatrix}}_{\text{\large{$\Delta \vTh$}}}
\begin{bmatrix}
\nabla J(\vth) \\ J(\vth)
\end{bmatrix}
\end{equation*}
\end{minipage} \\ \\
Then we can estimate the cost function's value and gradient at $\vth$ by the following
\begin{equation}
\begin{bmatrix} \nabla J(\vth) \\ J(\vth) \end{bmatrix} =
\Delta \vTh^{-1} \vJ
\end{equation}
In the next section, we will consider the case that we have more than $p+1$
perturbations. Extra perturbations can be helpful for two main reasons:
compensating for the numerical error and dealing with the stochasticity of the
problem.

\subsection{FD for Stochastic Problems}
In this section, we will consider the original problem in
equation~\eqref{eq:gd_optimal_problem} in which the system dynamics are
stochastic. A similar idea introduced in the previous section can be used here as
well. We should notice that the cost function in this case is defined as an
expectation. Thus for estimating this expectation through a numerical method, we
need to compute the value of the function for many executions of a given policy. Here we will define the
return function, $R$, as the value of the cost function for a single execution of
the policy. Notice that
$J = E[R]$.
\begin{equation*}
R = \Phi(\vx(N))+\sum_{k=0}^{N-1} L_k \left(\vx(k),\vu(k)\right)
\end{equation*} 
If we assume that the we need $K$ samples to evaluate the cost function for a single policy, we will need $K \times (p+1)$
evaluations of the cost function to estimate the cost function's gradient. We can re-write the previous equations as
follows
\begin{equation*}
\underbrace{\begin{bmatrix}
\frac{1}{K}\sum_{k=1}^{K}{R^k(\vth + \Delta \vth_1)} \\ \frac{1}{K}\sum_{k=1}^{K}{R^k(\vth + \Delta \vth_2)} \\ \vdots \\ \frac{1}{K}\sum_{k=1}^{K}{R^k(\vth + \Delta \vth_{p+1})}
\end{bmatrix}}_{\text{\large{$\vJ$}}} =
\underbrace{\begin{bmatrix}
\Delta \vth_1^T & 1 \\
\Delta \vth_2^T & 1 \\
\vdots & \\
\Delta \vth_{p+1}^T & 1
\end{bmatrix}}_{\text{\large{$\Delta \vTh$}}}
\begin{bmatrix}
\nabla J(\vth) \\ J(\vth)
\end{bmatrix}
\end{equation*}
and the cost function's value and gradient can be estimated as
\begin{equation}
\begin{bmatrix} \nabla J(\vth) \\ J(\vth) \end{bmatrix} =
\Delta \vTh^{-1} \vJ
\end{equation}
By extending the idea of random perturbations, we can assume that instead of $K$ evaluations of the cost function for each fixed perturbation, we can evaluate the
cost function for $N \leq K \times (p+1)$ different perturbations. Therefore we
can write
\begin{equation*}
\underbrace{\begin{bmatrix}
R(\vth + \Delta \vth_1) \\ R(\vth + \Delta \vth_2) \\ \vdots \\ R(\vth + \Delta \vth_N)
\end{bmatrix}}_{\text{\large{$\vR$}}} =
\underbrace{\begin{bmatrix}
\Delta \vth_1^T & 1 \\
\Delta \vth_2^T & 1 \\
\vdots & \\
\Delta \vth_N^T & 1
\end{bmatrix}}_{\text{\large{$\Delta \vTh$}}}
\begin{bmatrix}
\nabla J(\vth) \\ J(\vth)
\end{bmatrix}
\end{equation*}
Notice that $\Delta \vTh$ is a $N \times (p+1)$ matrix. If we assume that $N \geq
p+1$, $\Delta \vTh$ will have  rank $p+1$. Therefore, for estimating the cost
function's value and gradient, we should use the left pseudo-inverse of $\Delta
\vTh$.
\begin{equation}
\begin{bmatrix} \nabla J(\vth) \\ J(\vth) \end{bmatrix} = \Delta \vTh^\dagger \vR =
(\Delta \vTh^T \Delta \vTh)^{-1} \Delta \vTh^T \vR
\end{equation}
\subsection{Finite Difference: The General Method}
In this section, we will introduce the general FD method which unifies the
previous sections. Lets assume the general problem introduced in
equation~\eqref{eq:gd_optimal_problem}. The goal is to estimate the value of the
cost function and its gradient. First we generate $N$ ($N \geq p+1$) random
perturbations using an isometric distribution. Then we evaluate the return for
each of these perturbations.   The cost function's value and gradient can be
estimated as follows
\begin{equation} \label{eq:fd_general}
\begin{bmatrix} \nabla J(\vth) \\ J(\vth) \end{bmatrix} = \Delta \vTh^\dagger \vR =
(\Delta \vTh^T \Delta \vTh + \lambda \vI)^{-1} \Delta \vTh^T \vR
\end{equation}
where $\vR$ and $\Delta \vTh$ are defined as
\begin{equation*}
\vR = 
\begin{bmatrix}
R(\vth + \Delta \vth_1) \\ R(\vth + \Delta \vth_2) \\ \vdots \\ R(\vth + \Delta \vth_N)
\end{bmatrix},
\qquad \qquad
\Delta \vTh = 
\begin{bmatrix}
\Delta \vth_1^T & 1 \\
\Delta \vth_2^T & 1 \\
\vdots & \\
\Delta \vth_N^T & 1
\end{bmatrix}
\end{equation*} 
The term $\lambda \vI$ in equation~\eqref{eq:fd_general} is a regularization
term. Although it adds a bias to the estimation of the gradient, it is used to reduce
the covariance of the estimation. Notice that even if we have a
deterministic problem, we still might want to use more perturbations than $p+1$
in order to increase the numerical accuracy of the estimation. This is actually
the case if we want to use the double-sided method for approximating the gradient.
Note also that in deterministic problems, $R$ is equal to $J$.

Algorithm~\ref{alg:general_FD} illustrates an implementation of the gradient
descend algorithm which uses the generalized Finite Difference method for
estimating the gradient of the cost function. Even though the algorithm is
written for a finite-horizon discrete optimal control problem, it can be easily
modified to the continuous time problem by changing the computation of the return
function to an appropriate integral form. For the infinite-time horizon case, we
also need to change the return function to incorporate the decay factor. However
we should notice that generating rollouts for an infinite-horizon problem is
impractical unless the task is episodic (ie the task ends by reaching
a particular set of states after a certain maximum time horizon).

\begin{algorithm}[tpb] \caption{Gradient Descend Algorithm with Finite Difference Method} \label{alg:general_FD}
\begin{algorithmic}
\State \textbf{given} 
\State \hspace{4mm} The cost function: 
\State \hspace{8mm} $J = E \left[ \Phi(\vx(N))+\sum_{k=0}^{N-1} L_k \left(\vx(k),\vu(k)\right) \right]$
\State \hspace{4mm} A policy (function approximation) for the control input: $\vu(n,\vx) = \mu(n,\vx;\vth)$
\State \hspace{4mm} An initial value for the parameter vector: $\vth \leftarrow \vth_0$ 
\State \hspace{4mm} The parameter exploration standard deviation: $c$
\State \hspace{4mm} The regularization coefficient: $\lambda$
\State \hspace{4mm} The learning rate: $\omega$
\Repeat
\State Create $N$ rollouts of the system with the perturbed parameters $ \vth + \Delta \vth, \  \Delta \vth \sim \mathcal{N} (\mathbf{0},c^2\vI)$
\State Calculate the return from the initial time and state for the $n$th rollout:
\State \hspace{4mm} $R(\vth + \Delta \vth_n) = \Phi(\vx^n(N))+\sum_{k=0}^{N-1} L_k \left(\vx^n(k),\vu^n(k)\right) $
\State Construct $\vR$ and $\vTh$ matrices as:
\State \hspace{4mm} $\vR_{N \times 1} = \left[R(\vth + \Delta \vth_n)\right]$, $\vTh_{N \times (p+1)} = \left[ \Delta \vth_n^T \ 1 \right]$
\State Calculate the value and gradient of the cost function at $\vth$
\State \hspace{4mm} $\begin{bmatrix} \nabla J(\vth) \\ J(\vth) \end{bmatrix} = (\Delta \vTh^T \Delta \vTh + \lambda \vI)^{-1} \Delta \vTh^T \vR$
\State Update the parameter vector:
\State \hspace{4mm} $\vth \leftarrow \vth - \omega \nabla J(\vth)$
\Until{convergence}
\end{algorithmic}
\end{algorithm}

\thispagestyle{empty}\cleardoublepage

\appendix
%

%
%
%

%
%
%
%
%
\end{document}